\documentclass[
 reprint,
 amsmath,amssymb,
 aps,
nofootinbib,
superscriptaddress
]{revtex4-2}

\usepackage{graphicx}
\usepackage{dcolumn}
\usepackage{bm}
\usepackage{physics}
\usepackage{bbold}
\usepackage{subcaption}
\usepackage{tikz}
\usepackage{simpler-wick}
\usepackage{enumerate}
\usepackage{mathtools}
\usepackage{hyperref}

\newcommand{\clo}{\mathcal{O}}
\newcommand{\bbi}{\mathbb{1}}

\newcommand\msmall[1]{\mbox{\small\ensuremath{#1}}}
\newcommand{\fker}[6]{
\mathbb{F}_{#1 #2}\!{\msmall{\arraycolsep=0.2\arraycolsep\ensuremath{\begin{bmatrix}#4 & #3 \\ #5 & #6\end{bmatrix}}}}
}
\newcommand{\sker}[3]{
\mathbb{S}_{#1 #2}[#3]
}

\begin{document}

\preprint{APS/123-QED}

\title{Surgery and Statistics in 3d Gravity}

\author{Jan de Boer}
\affiliation{%
Institute for Theoretical Physics (ITFA),\\
University of Amsterdam, \\ Amsterdam, The Netherlands
}%

\author{Joshua Kames-King}%
\affiliation{%
Laboratory for Theoretical Fundamental Physics and Fields and Strings Laboratory,\\ École Polytechnique Fédérale de Lausanne (EPFL),\\
 Lausanne, Switzerland
}%

\author{Boris Post}
\email{b.p.post@uva.nl}
\affiliation{%
Institute for Theoretical Physics (ITFA),\\
University of Amsterdam, \\ Amsterdam, The Netherlands
}%

\date{June 13, 2025}

\begin{abstract}
We extend the correspondence between universal statistical features of large-$c$ 2d CFTs and surgery methods in pure AdS$_3$ quantum gravity. In particular, we introduce a method that we call RMT surgery, which relates a large class of off-shell partition functions in 3d gravity to the spectral statistics of general CFT observables. We apply this method to construct and compute an off-shell Euclidean wormhole whose boundaries are four-punctured spheres, which captures level repulsion in the high-energy sector of the boundary CFT. Using a similar gluing prescription, we also explore a new class of off-shell torus wormholes with trumpet boundaries, contributing to statistical moments of the density of primary states. Lastly, we demonstrate that surgery methods can be used as an intermediate step towards computing Seifert manifolds directly in 3d gravity.
\end{abstract}

\maketitle


\section{\label{sec:intro} Introduction}

The old subject of pure AdS$_3$ quantum gravity has recently received much renewed interest, thanks to conceptual progress in establishing an averaged version of the holographic correspondence \cite{Belin:2020hea, Chandra:2022bqq,Belin:2023efa,Jafferis:2024jkb,deBoer:2023vsm} as well as technical advances in evaluating the Euclidean gravitational path integral \cite{Collier:2023fwi,Collier:2024mgv}. Inspired by the duality between pure JT gravity and a random matrix integral \cite{Saad:2019lba}, strong evidence has been gathered for the case that pure AdS$_3$ gravity, and the sum over bulk geometries, correctly describes the universal \emph{statistical} properties of holographic 2d CFT's \cite{Kraus:2016nwo,Cardy:2017qhl,Collier:2019weq, Belin:2021ryy,Anous:2021caj,Jafferis:2024jkb}, in a way that closely resembles random matrix theory (RMT) \cite{Cotler:2020ugk,Cotler:2020hgz,DiUbaldo:2023qli,Haehl:2023mhf,Haehl:2023tkr, Haehl:2023xys,Boruch:2025ilr} and the eigenstate thermalization hypothesis (ETH) \cite{deBoer:2024mqg, Jafferis:2022uhu}.

In this paper, we draw a close parallel between statistical methods in 2d CFT, familiar from quantum chaos theory, and \emph{surgery} techniques in 3d gravity. By surgery, we mean a set of topological operations on 3-manifolds that, through a cutting-and-gluing prescription of the gravitational path integral, can be used to compute exact partition functions of complicated 3-manifolds from more basic building blocks. Owing to the absence of local degrees of freedom in 3d gravity, these topological methods are an effective framework to study a wide variety of 3-manifolds contributing to the sum over topologies in quantum gravity. In particular, they allow us to describe new \emph{off-shell} topologies, not studied before in 3d gravity, which will turn out to be crucial for understanding the statistical properties of the dual CFT.
\footnote{By `off-shell topology', we mean any topological manifold $M$ that does not admit a metric solving the gravitational equations of motion with the prescribed boundary conditions. Consequently, the Euclidean gravitational path integral on such $M$ does not admit a saddle-point approximation. In Euclidean signature, on-shell solutions correspond to hyperbolic 3-manifolds, so `off-shell' can be taken as synonymous with `non-hyperbolic'.}

In what follows, we will introduce four distinct surgery methods, each of which captures a different aspect of the statistical theory modeling the CFT at high energies. In Section \ref{sec:ETH}, we introduce the notion of \emph{ETH surgery}, as a method of evaluating the partition functions of on-shell Euclidean wormhole geometries by cutting and gluing along higher-genus surfaces. This formalism captures the variance of higher-genus partition functions and relates it to ensembles of OPE data, governed by the eigenstate thermalization hypothesis and its extension to 2d CFT \cite{Belin:2020hea}. Using the machinery of Virasoro TQFT \cite{Collier:2023fwi}, we revisit how wormhole geometries correspond to Wick contractions of OPE coefficients in a simplified Gaussian model and show how non-Gaussianities can be incorporated via the action of the mapping class group \cite{deBoer:2024mqg}.

Next, in Section \ref{sec:RMT}, we introduce \emph{RMT surgery}, as a tool to evaluate the exact gravitational partition function (without saddle-point approximation) on a class of off-shell  3d topologies, by cutting and gluing along embedded tori. This method relies on a set of assumptions about the off-shell path integral that we will  state explicitly. Under these assumptions, we show, both in a simple example and more generally, that the partition functions obtained using RMT surgery match  the universal statistical prediction of level repulsion in each spin sector of the boundary CFT. Moreover, gravity offers a natural modular completion of this RMT prediction by summing over a PSL$(2,\mathbb{Z})$ family of topologies. 
 
In Section \ref{sec:trumpetgluing}, we use the 3d analogue of the ``trumpet'' geometry \cite{Cotler:2020ugk} to construct a class of off-shell 3-manifolds with asymptotically AdS$_3$ torus boundaries and compute their partition functions. Including this class in the sum over topologies gives rise to non-perturbatively small corrections to the spectral density and its higher moments. 

Lastly, in Section \ref{sec:seifert}, we turn to \emph{Dehn surgery}, which we use as an intermediate step in the computation of Seifert manifolds. Seifert manifolds constitute a class of off-shell topologies that, when summed over in the gravitational path integral, have been conjectured to cure the negativity in the average spectral density associated to a PSL$(2,\mathbb{Z})$ family of BTZ black holes \cite{Maloney:2007ud,Keller:2014xba,Maxfield:2020ale}. We reduce the problem of computing the sum over all genus-zero Seifert manifolds to the evaluation of a single topology, namely the $n$-boundary torus wormhole $\Sigma_{0,n}\times S^1$. Even though we do not solve this last step, we make a conjecture for the partition function $Z_\text{grav}(\Sigma_{0,n}\times S^1)$ based on a random matrix \emph{ansatz} for the primary spectrum above the black hole threshold, in fixed spin sectors. Using the formalism of topological recursion \cite{Eynard:2007kz, Chekhov:2006rq}, this ansatz, when combined with Dehn surgery, will be shown to agree with Maxfield and Turiaci's analysis \cite{Maxfield:2020ale} in the large spin near-extremal limit.

\section{\label{sec:ETH} ETH surgery}
 As a first example of the interplay between statistical methods in CFT$_2$ and surgery techniques in AdS$_3$, we will review the duality between OPE statistics and on-shell (i.e.~hyperbolic) topologies, reformulated in the language of 3-manifold surgery using Virasoro TQFT \cite{Collier:2023fwi}. Specifically, we will focus on a simple CFT observable, namely the genus-two partition function:
\begin{align}\label{eq:genustwo}
    &Z_\Sigma(\Omega,\bar\Omega) =\\ &\int \prod_{i=1}^3\dd h_i\,\dd \bar h_i\,\rho(h_i,\bar h_i)\,C_{h_1h_2h_3}C^*_{h_1h_2h_3}\,|\mathcal{F}_\Sigma(\{h_k\};\Omega) |^2.\nonumber
\end{align}
We have expanded into `sunset channel' conformal blocks $\mathcal{F}_\Sigma$, which are labeled by moduli $\Omega=(\omega_1,\omega_2,\omega_3)$ and primary conformal weights $h_1,h_2,h_3$. The spectral density $\rho$ is a distribution supported on the exact spectrum of the CFT$_2$, and $C_{h_1h_2h_3}$ are the OPE coefficients.

The statistical moments of $Z_\Sigma$ can be described by sampling the OPE coefficients of heavy operators from a statistical ensemble. In the most basic ensemble, first considered in \cite{Belin:2020hea,Chandra:2022bqq}, we assume that the spectrum consists of a unique vacuum state, a gap up to the black hole threshold $h_t,\bar h_t = \frac{c-1}{24}$ and a dense (discrete) spectrum of states above the threshold. In this set-up, approximate modular invariance and crossing symmetry dictate the form of the average density of states and OPE density: 
\begin{multline}\label{eq:average}
    \overline{\textstyle\prod_{i=1}^3\rho(h_i,\bar h_i)\,C_{h_1h_2h_3}C^*_{h_1h_2h_3}} \approx  \\[1em] \textstyle\prod_{i=1}^3 \rho_0(h_i,\bar h_i)\left| C_0(h_1,h_2,h_3)\right|^2
\end{multline}
for $h_i \geq h_t$. Here $\rho_0(h,\bar h) = \rho_0(h)\rho_0(\bar h)\delta_{\mathbb{Z}}(h-\bar h)$ is the coarse-grained density of states, including spin quantization, with a universal Cardy growth at high energies. $\rho_0$ and $C_0$ are universal functions of the (holomorphic) conformal weights, derived in \cite{Collier:2019weq}, and $|\bullet |^2$ denotes a product with the anti-holomorphic counterpart. 

Using this average for the heavy states $h_i \geq h_t$ propagating in Eq.~\eqref{eq:genustwo}, and using the selection rules $C_{\mathbb{1}h_1h_2} = \delta_{h_1h_2}\delta_{\bar h_1\bar h_2}$ for the vacuum contributions $h_i = \bbi$, one finds that the average genus-two partition function is a sum of vacuum conformal blocks in different OPE channels (see Section 8.2 of \cite{Chandra:2022bqq}). In the bulk, this coincides with a sum over a set of 3-dimensional \emph{handlebodies},
\begin{equation}
    \overline{Z_\Sigma(\Omega,\bar\Omega)} \approx \sum_M \,Z_\text{grav}[M;\Omega,\bar \Omega]\,,
\end{equation}
where the handlebodies are specified by a set of cycles on the boundary surface that can be contracted in the bulk:
\begin{equation}\label{eq:cyclesset}
    M \in \left\{ 
\begin{array}{c}
\includegraphics[width=0.2\linewidth]{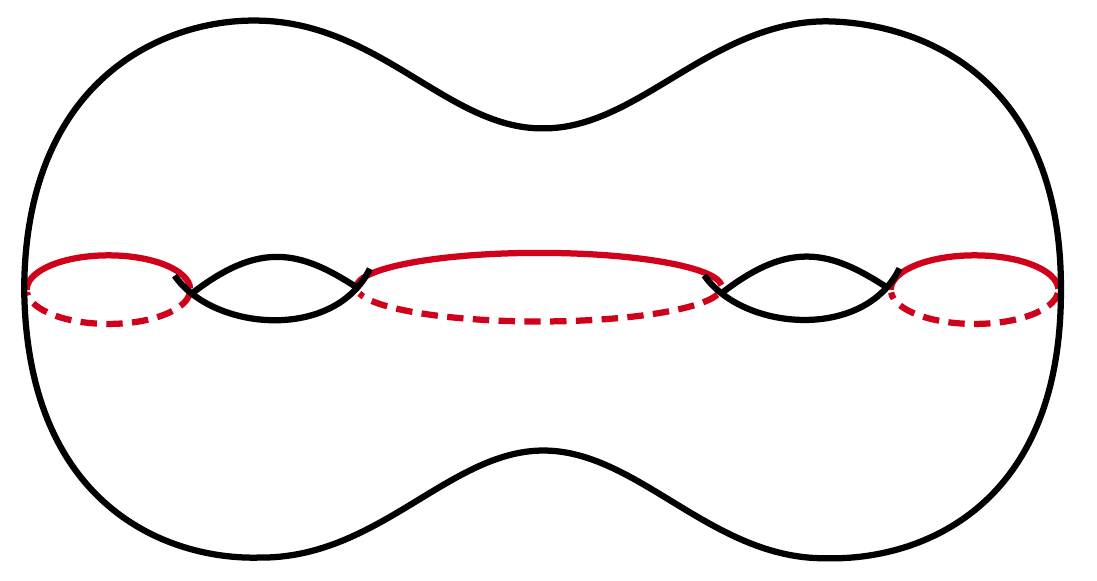} \quad
\includegraphics[width=0.2\linewidth]{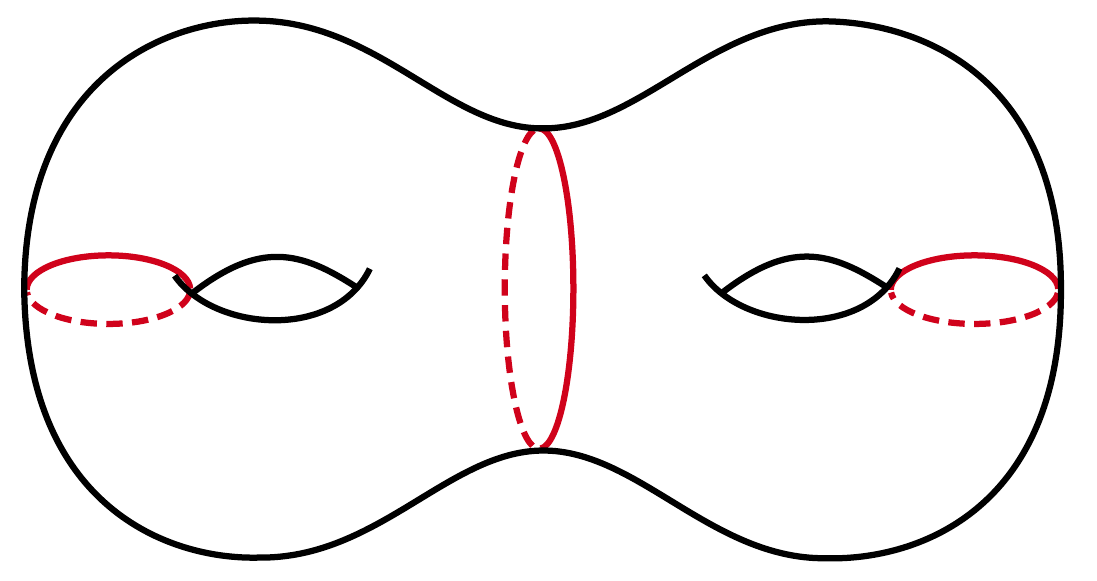} \quad
\includegraphics[width=0.2\linewidth]{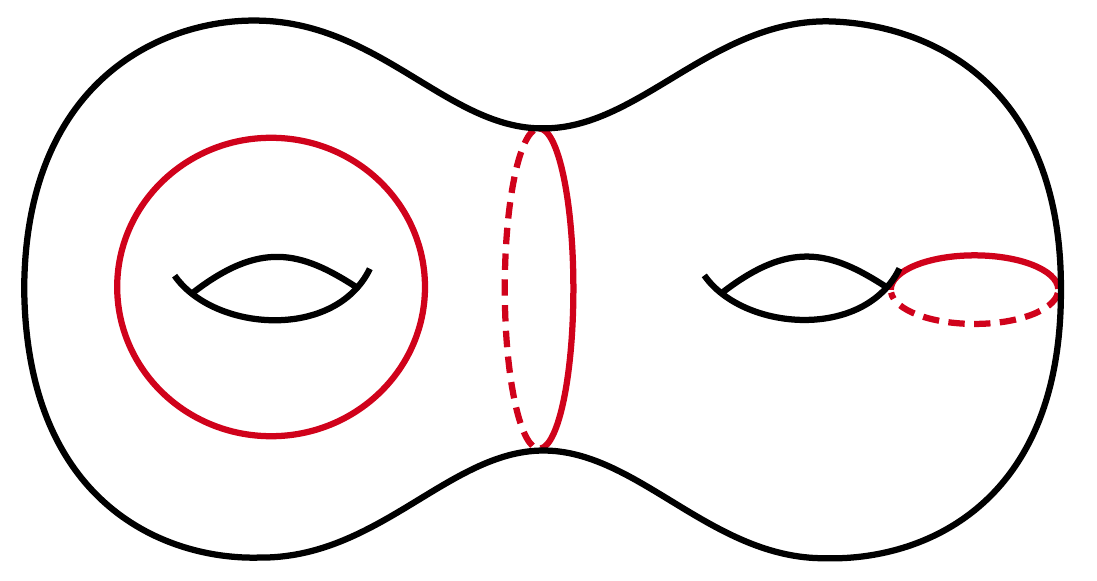} \\
\includegraphics[width=0.2\linewidth]{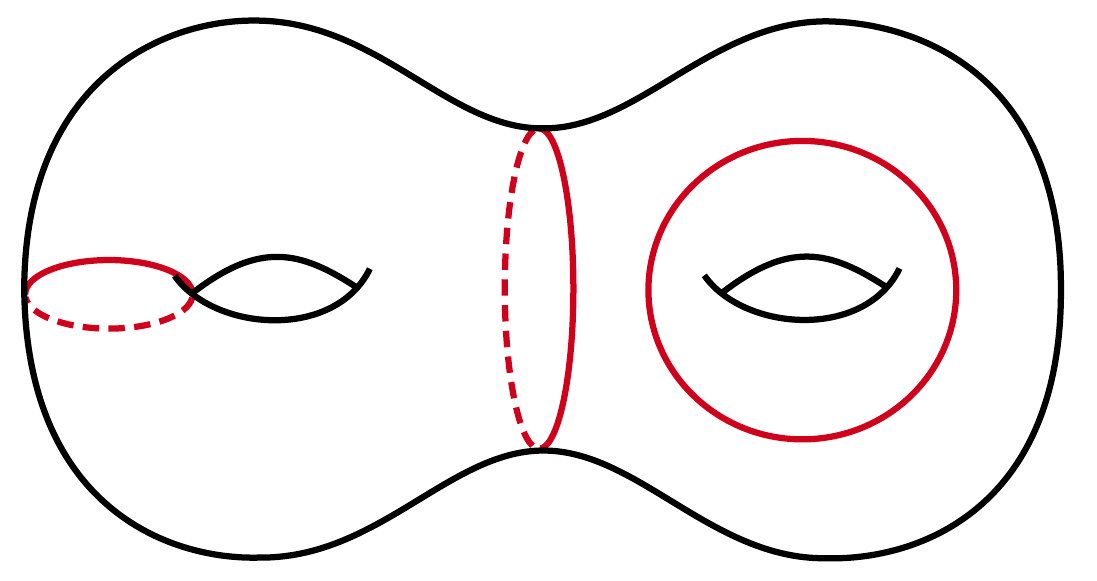} \quad
\includegraphics[width=0.2\linewidth]{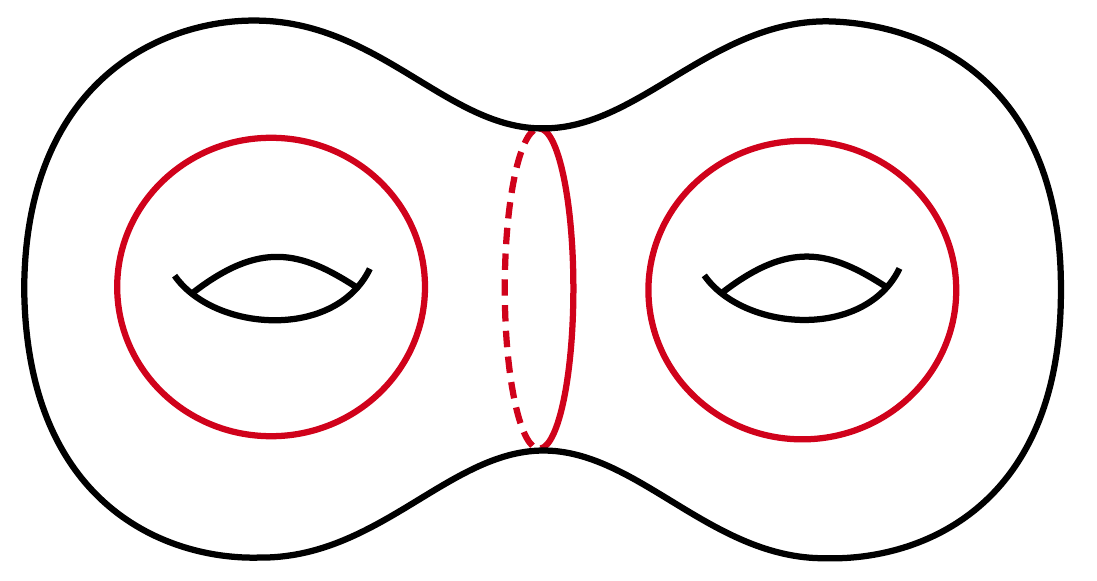}
\end{array}
\right\}.
\end{equation}
The reason why the identity blocks appear in different OPE channels, while we expanded in one channel only, is that the densities $\rho_0$ and $C_0$ appearing in Eq.~\eqref{eq:average} are proportional to the identity modular S-kernel $\mathbb{S}_{\mathbb{1}P}$ and identity fusion kernel $\mathbb{F}_{\mathbb{1}P}$, respectively, which change the channel decomposition of the blocks \cite{Eberhardt:2023mrq}. The reason why a fixed handlebody topology evaluates to the Virasoro identity block, to all orders in $G_\text{N}$, can be understood by quantizing 3d gravity using Virasoro TQFT \cite{Collier:2023fwi}.

Instead of computing the average, one can also compute the \emph{variance} of $Z_\Sigma$ in the OPE ensemble. This is where ETH surgery comes in. First, from the boundary perspective, the `maximally ignorant' statistical theory of $C_{h_1h_2h_3}$ (given only $\overline{Z}_\Sigma$) is a Gaussian random ensemble, analogous to the ETH \cite{deBoer:2023vsm}. In this simplified Gaussian model, the variance of $Z_\Sigma$ is computed by the connected Wick contractions of four OPE coefficients,
\begin{multline}\label{eq:wick}
    \wick{\c1 C_{123} \c2 C^*_{123}\c1 C_{456}\c2 C^*_{456}} =\\ (\delta_{14}\delta_{25}\delta_{36} \pm \text{perm.}) \big | C_0(h_1,h_2,h_3)^2 \big |^2
\end{multline}
which sets three pairs of indices equal. We used the short-hand notation $C_{123}=C_{h_1h_2h_3}$ and $\delta_{12} = \delta_{h_1h_2}\delta_{\bar h_1\bar h_2}$. The signed permutations arise from the CFT property $C_{\sigma(1)\sigma(2)\sigma(3)} = \text{sgn}(\sigma)^{J_1+J_2+J_3}C_{123}$, where $J_i = h_i-\bar h_i$ is the spin. Substituting Eq.~\eqref{eq:wick} into the connected average of $(Z_\Sigma)^2$, and approximating the high-energy spectrum by $\rho_0$, we will show that the variance takes the form\vspace{1mm}
\begin{equation}\label{eq:variance}
    \overline{Z_\Sigma(\Omega,\bar\Omega)Z_\Sigma(\Omega',\bar\Omega')}^{\,c} \approx \sum_W Z_\text{\text{grav}}[W;\Omega,\bar\Omega,\Omega',\bar\Omega']
\end{equation}
where each $W$ is a Euclidean wormhole with the topology $\Sigma \times I$. In the simplest case, its on-shell geometry is the Maldacena-Maoz solution \cite{Maldacena:2004rf,Belin:2020hea}, but more generally there is an infinite sum to account for the phase factors in Eq.~\eqref{eq:wick} and the quantization of the spins $J_i$.

 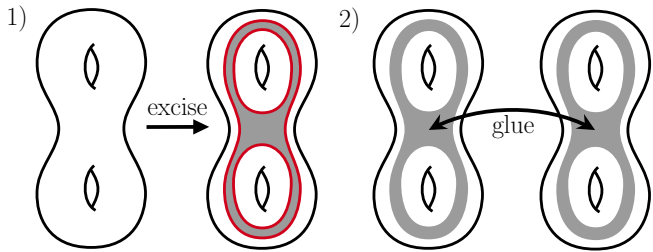
\begin{figure}
    \centering
    \resizebox{\linewidth}{!}{
    \begin{tikzpicture}[x=0.75pt,y=0.75pt,yscale=-1,xscale=1,baseline={([yshift=-0.5ex]current bounding box.center)}]
\draw [color={rgb, 255:red, 208; green, 2; blue, 27 }  ,draw opacity=1 ][fill={rgb, 255:red, 155; green, 155; blue, 155 }  ,fill opacity=1 ][line width=2.25]    (279.6,23.2) .. controls (311.45,23.42) and (320.3,54.95) .. (319.88,75.78) .. controls (319.45,102.56) and (303.58,117.94) .. (303.7,138.78) .. controls (303.55,159.96) and (319.66,176.08) .. (319.78,201.82) .. controls (319.63,223) and (311.54,254.5) .. (279.41,254.28) .. controls (246.75,254.41) and (238.71,222.89) .. (239.13,201.71) .. controls (239.02,176.32) and (255.43,159.89) .. (255.31,138.71) .. controls (255.19,117.88) and (239.08,102.1) .. (239.23,75.67) .. controls (238.85,55.18) and (246.94,23.68) .. (279.6,23.2) -- cycle ;
\draw [color={rgb, 255:red, 208; green, 2; blue, 27 }  ,draw opacity=1 ][fill={rgb, 255:red, 255; green, 255; blue, 255 }  ,fill opacity=1 ][line width=2.25]    (248.79,195.72) .. controls (248.42,174.58) and (261.09,154.77) .. (279.31,155.19) .. controls (295.15,154.77) and (310.06,172.35) .. (310.14,195.82) .. controls (310.03,215.14) and (303.87,243.87) .. (279.43,243.67) .. controls (254.59,243.79) and (248.47,215.04) .. (248.79,195.72) -- cycle ;
\draw [color={rgb, 255:red, 208; green, 2; blue, 27 }  ,draw opacity=1 ][fill={rgb, 255:red, 255; green, 255; blue, 255 }  ,fill opacity=1 ][line width=2.25]    (249.19,73.29) .. controls (248.82,52.16) and (261.49,32.34) .. (279.7,32.76) .. controls (295.54,32.34) and (310.45,49.92) .. (310.54,73.39) .. controls (310.42,92.71) and (305.44,121.03) .. (279.83,121.24) .. controls (253.57,121.45) and (248.87,92.61) .. (249.19,73.29) -- cycle ;
\draw [line width=2.25]    (97.82,11.4) .. controls (143.37,11.64) and (156.02,46.15) .. (155.42,68.96) .. controls (154.81,98.28) and (132.11,115.13) .. (132.28,137.93) .. controls (132.07,161.13) and (155.11,178.78) .. (155.27,206.96) .. controls (155.06,230.15) and (143.49,264.64) .. (97.55,264.4) .. controls (50.85,264.54) and (39.36,230.03) .. (39.96,206.84) .. controls (39.79,179.05) and (63.26,161.05) .. (63.09,137.86) .. controls (62.92,115.05) and (39.88,97.78) .. (40.1,68.84) .. controls (39.55,46.41) and (51.12,11.92) .. (97.82,11.4) -- cycle ;
\draw [line width=2.25]    (100.64,48.1) .. controls (88.62,57.63) and (86.64,87.2) .. (102.13,95.75) ;
\draw [line width=2.25]    (96.86,52.38) .. controls (108.06,68.8) and (106.08,78.66) .. (98.17,92.46) ;
\draw [line width=2.25]    (100.71,177.37) .. controls (88.68,186.89) and (86.71,216.47) .. (102.19,225.01) ;
\draw [line width=2.25]    (96.92,181.64) .. controls (108.12,198.07) and (106.15,207.92) .. (98.24,221.72) ;
\draw [line width=3]    (156.33,136.35) -- (216.96,136.2) ;
\draw [shift={(222.96,136.19)}, rotate = 179.86] [fill={rgb, 255:red, 0; green, 0; blue, 0 }  ][line width=0.08]  [draw opacity=0] (16.97,-8.15) -- (0,0) -- (16.97,8.15) -- cycle    ;
\draw [line width=2.25]    (279.17,12.36) .. controls (324.72,12.6) and (337.37,47.11) .. (336.77,69.92) .. controls (336.16,99.25) and (313.46,116.09) .. (313.63,138.9) .. controls (313.42,162.09) and (336.46,179.75) .. (336.62,207.92) .. controls (336.41,231.11) and (324.84,265.6) .. (278.9,265.36) .. controls (232.2,265.51) and (220.71,230.99) .. (221.31,207.8) .. controls (221.14,180.01) and (244.61,162.02) .. (244.44,138.83) .. controls (244.27,116.02) and (221.23,98.74) .. (221.45,69.8) .. controls (220.9,47.38) and (232.47,12.89) .. (279.17,12.36) -- cycle ;
\draw [line width=2.25]    (281.99,49.07) .. controls (269.97,58.6) and (267.99,88.17) .. (283.48,96.71) ;
\draw [line width=2.25]    (278.21,53.34) .. controls (289.41,69.77) and (287.43,79.62) .. (279.52,93.43) ;
\draw [line width=2.25]    (282.06,178.33) .. controls (270.03,187.86) and (268.06,217.43) .. (283.54,225.97) ;
\draw [line width=2.25]    (278.27,182.6) .. controls (289.47,199.03) and (287.5,208.89) .. (279.59,222.69) ;
\draw [color={rgb, 255:red, 155; green, 155; blue, 155 }  ,draw opacity=1 ][fill={rgb, 255:red, 155; green, 155; blue, 155 }  ,fill opacity=1 ][line width=2.25]    (456.32,23.2) .. controls (488.18,23.42) and (497.02,54.95) .. (496.6,75.78) .. controls (496.18,102.56) and (480.31,117.94) .. (480.42,138.78) .. controls (480.27,159.96) and (496.39,176.08) .. (496.5,201.82) .. controls (496.35,223) and (488.26,254.5) .. (456.14,254.28) .. controls (423.48,254.41) and (415.44,222.89) .. (415.86,201.71) .. controls (415.74,176.32) and (432.15,159.89) .. (432.04,138.71) .. controls (431.92,117.88) and (415.8,102.1) .. (415.96,75.67) .. controls (415.57,55.18) and (423.66,23.68) .. (456.32,23.2) -- cycle ;
\draw [color={rgb, 255:red, 155; green, 155; blue, 155 }  ,draw opacity=1 ][fill={rgb, 255:red, 255; green, 255; blue, 255 }  ,fill opacity=1 ][line width=2.25]    (425.51,195.72) .. controls (425.15,174.58) and (437.82,154.77) .. (456.03,155.19) .. controls (471.87,154.77) and (486.78,172.35) .. (486.87,195.82) .. controls (486.75,215.14) and (480.6,243.87) .. (456.16,243.67) .. controls (431.31,243.79) and (425.19,215.04) .. (425.51,195.72) -- cycle ;
\draw [color={rgb, 255:red, 155; green, 155; blue, 155 }  ,draw opacity=1 ][fill={rgb, 255:red, 255; green, 255; blue, 255 }  ,fill opacity=1 ][line width=2.25]    (425.91,73.29) .. controls (425.54,52.16) and (438.21,32.34) .. (456.43,32.76) .. controls (472.27,32.34) and (487.18,49.92) .. (487.26,73.39) .. controls (487.15,92.71) and (482.16,121.03) .. (456.55,121.24) .. controls (430.29,121.45) and (425.59,92.61) .. (425.91,73.29) -- cycle ;
\draw [line width=2.25]    (455.89,12.36) .. controls (501.44,12.6) and (514.09,47.11) .. (513.49,69.92) .. controls (512.88,99.25) and (490.19,116.09) .. (490.36,138.9) .. controls (490.14,162.09) and (513.18,179.75) .. (513.35,207.92) .. controls (513.13,231.11) and (501.56,265.6) .. (455.63,265.36) .. controls (408.93,265.51) and (397.43,230.99) .. (398.03,207.8) .. controls (397.87,180.01) and (421.33,162.02) .. (421.17,138.83) .. controls (421,116.02) and (397.95,98.74) .. (398.18,69.8) .. controls (397.62,47.38) and (409.19,12.89) .. (455.89,12.36) -- cycle ;
\draw [line width=2.25]    (458.72,49.07) .. controls (446.69,58.6) and (444.72,88.17) .. (460.2,96.71) ;
\draw [line width=2.25]    (454.93,53.34) .. controls (466.13,69.77) and (464.15,79.62) .. (456.25,93.43) ;
\draw [line width=2.25]    (458.78,178.33) .. controls (446.76,187.86) and (444.78,217.43) .. (460.27,225.97) ;
\draw [line width=2.25]    (454.99,182.6) .. controls (466.2,199.03) and (464.22,208.89) .. (456.31,222.69) ;
\draw [color={rgb, 255:red, 155; green, 155; blue, 155 }  ,draw opacity=1 ][fill={rgb, 255:red, 155; green, 155; blue, 155 }  ,fill opacity=1 ][line width=2.25]    (633.81,24.02) .. controls (665.67,24.24) and (674.51,55.76) .. (674.09,76.59) .. controls (673.67,103.38) and (657.79,118.76) .. (657.91,139.59) .. controls (657.76,160.77) and (673.88,176.9) .. (673.99,202.63) .. controls (673.84,223.81) and (665.75,255.31) .. (633.63,255.1) .. controls (600.96,255.23) and (592.93,223.7) .. (593.35,202.52) .. controls (593.23,177.14) and (609.64,160.71) .. (609.53,139.52) .. controls (609.41,118.69) and (593.29,102.92) .. (593.45,76.48) .. controls (593.06,56) and (601.15,24.5) .. (633.81,24.02) -- cycle ;
\draw [color={rgb, 255:red, 155; green, 155; blue, 155 }  ,draw opacity=1 ][fill={rgb, 255:red, 255; green, 255; blue, 255 }  ,fill opacity=1 ][line width=2.25]    (603,196.53) .. controls (602.63,175.4) and (615.31,155.58) .. (633.52,156) .. controls (649.36,155.58) and (664.27,173.16) .. (664.36,196.63) .. controls (664.24,215.96) and (658.09,244.69) .. (633.65,244.49) .. controls (608.8,244.61) and (602.68,215.85) .. (603,196.53) -- cycle ;
\draw [color={rgb, 255:red, 155; green, 155; blue, 155 }  ,draw opacity=1 ][fill={rgb, 255:red, 255; green, 255; blue, 255 }  ,fill opacity=1 ][line width=2.25]    (603.4,74.11) .. controls (603.03,52.97) and (615.7,33.15) .. (633.92,33.58) .. controls (649.75,33.15) and (664.67,50.74) .. (664.75,74.21) .. controls (664.64,93.53) and (659.65,121.85) .. (634.04,122.06) .. controls (607.78,122.27) and (603.08,93.43) .. (603.4,74.11) -- cycle ;
\draw [line width=2.25]    (633.38,13.18) .. controls (678.93,13.42) and (691.58,47.93) .. (690.98,70.74) .. controls (690.37,100.06) and (667.68,116.9) .. (667.84,139.71) .. controls (667.63,162.91) and (690.67,180.56) .. (690.83,208.74) .. controls (690.62,231.93) and (679.05,266.42) .. (633.12,266.18) .. controls (586.41,266.32) and (574.92,231.81) .. (575.52,208.62) .. controls (575.36,180.83) and (598.82,162.83) .. (598.66,139.64) .. controls (598.49,116.83) and (575.44,99.56) .. (575.66,70.62) .. controls (575.11,48.19) and (586.68,13.7) .. (633.38,13.18) -- cycle ;
\draw [line width=2.25]    (636.21,49.88) .. controls (624.18,59.41) and (622.2,88.98) .. (637.69,97.53) ;
\draw [line width=2.25]    (632.42,54.15) .. controls (643.62,70.58) and (641.64,80.44) .. (633.74,94.24) ;
\draw [line width=2.25]    (636.27,179.14) .. controls (624.25,188.67) and (622.27,218.24) .. (637.76,226.79) ;
\draw [line width=2.25]    (632.48,183.42) .. controls (643.69,199.84) and (641.71,209.7) .. (633.8,223.5) ;
\draw [line width=3]    (461.64,136.69) .. controls (510.44,112.52) and (579.22,111.29) .. (628.45,137.1) ;
\draw [shift={(632.96,139.57)}, rotate = 209.8] [fill={rgb, 255:red, 0; green, 0; blue, 0 }  ][line width=0.08]  [draw opacity=0] (18.75,-9.01) -- (0,0) -- (18.75,9.01) -- (12.45,0) -- cycle    ;
\draw [shift={(456.26,139.5)}, rotate = 331.28] [fill={rgb, 255:red, 0; green, 0; blue, 0 }  ][line width=0.08]  [draw opacity=0] (18.75,-9.01) -- (0,0) -- (18.75,9.01) -- (12.45,0) -- cycle    ;
\draw (6.39,5.87) node [anchor=north west][inner sep=0.75pt]  [font=\huge]  {$1)$};
\draw (155.94,104.66) node [anchor=north west][inner sep=0.75pt]  [font=\huge] [align=left] {excise};
\draw (359.85,7.56) node [anchor=north west][inner sep=0.75pt]  [font=\huge]  {$2)$};
\draw (522.2,122.52) node [anchor=north west][inner sep=0.75pt]  [font=\huge] [align=left] {glue};
\end{tikzpicture}}
    \caption{ETH surgery.}
    \label{fig:ETH_surgery}
\end{figure}

 We can describe the wormholes $W$ using the following surgery prescription, shown in Figure \ref{fig:ETH_surgery}:
 \begin{enumerate}[1)]
    \item Take a genus-two handlebody and excise (carve out) another smaller handlebody from its interior. This produces a so-called `compression body'. 
    \item Take a pair of compression bodies, and glue the two inner boundaries with some gluing map $G$. 
\end{enumerate}
These steps can be implemented on the level of the gravitational path integral, using Virasoro TQFT \cite{Collier:2023fwi}. By quantizing the gravitational phase space, Step 1 prepares a \emph{state} in the Hilbert space defined by the path integral:
\begin{multline}\label{eq:state}
    \ket{Z_\text{grav}(\Omega,\bar \Omega)} =\\ \int\dd \mu(\mathbf{P},\bar{\mathbf{P}}) \braket{\mathcal{F}_\Sigma(\mathbf{P},\bar{\mathbf{P}})}{\Omega,\bar \Omega} \,\ket{\mathcal{F}_\Sigma(\mathbf{P},\bar{\mathbf{P}})} \in \mathcal{H}_\Sigma.
\end{multline}
The gravitational Hilbert space $\mathcal{H}_\Sigma$ (which should not be confused with the microscopic CFT Hilbert space) is spanned by a continuous basis of left- times right-moving non-degenerate Virasoro conformal blocks on $\Sigma$, given by $\ket{\mathcal{F}_\Sigma(\mathbf{P},\bar{\mathbf{P}})} = \ket{\mathcal{F}_\Sigma(\mathbf{P})}\otimes \ket{\mathcal{F}_\Sigma(\bar{\mathbf{P}})}$. This basis of conformal blocks is parametrized by real Liouville momenta $\mathbf{P} = (P_1,P_2,P_3)\in \mathbb{R}^3$, which are related to the conformal weights in Eq.~\eqref{eq:genustwo} as $h = \frac{c-1}{24}+P^2$, and their anti-holomorphic counterparts. The integration measure $\dd \mu$ includes a canonical normalization factor,
\begin{equation}
    \dd\mu(\mathbf{P},\bar{\mathbf{P}}) = \prod_{i=1}^3 \dd P_i\dd \bar P_i\, | \rho_\Sigma(P_1,P_2,P_3)|^2
\end{equation}
given by $\rho_\Sigma(P_1,P_2,P_3) = \prod_{i=1}^3\rho_0(P_i)\,C_0(P_1,P_2,P_3)^2$. See Ref.~\cite{Collier:2023fwi} for an explanation of this normalization.

Next, we proceed to Step 2. The gluing map $G$ is represented as an operator acting on the gravitational Hilbert space, and the two-boundary wormhole partition function after gluing is given by the expectation value 
\begin{equation}\label{eq:expval}
    Z_\text{\text{grav}}[W] = \mel{Z_\text{grav}(\Omega,\bar \Omega)}{\,G\,}{Z_\text{grav}(\Omega',\bar \Omega')}.
\end{equation}
If we want to reproduce the prediction for the variance of $Z_\Sigma$ in the Gaussian model \eqref{eq:wick} from a sum over topologies, as in Eq.~\eqref{eq:variance}, then the relevant family of gluing maps is:
  \begin{equation}\label{eq:gluing_map}
      G = \big|\mathbb{T}_1^{n_1} \mathbb{T}_2^{n_2}\mathbb{T}_3^{n_3}\,\mathbb{B}^{n_k}\big|^2.
  \end{equation}
 Here $\mathbb{T}_i$ is a \emph{Dehn twist}, which is a large diffeomorphism acting on the surface $\Sigma$. It is represented in the Virasoro TQFT Hilbert space as multiplication by a phase $e^{-2\pi i h_i}$. Similarly, the braiding transformation $\mathbb{B}$ is also represented by a phase factor $e^{-\pi i (h_{1}-h_{2}-h_{3})}$. Lastly, $n_{1,2,3}$ are integers and $n_k \in \{0,1\}$. 

We can evaluate the expectation value in Eq.~\eqref{eq:expval} by substituting Eq.~\eqref{eq:state} and using the inner product on the space of conformal blocks derived in \cite{Collier:2023fwi}. Importantly, the inner product is delta function orthogonal. So, since a fixed $G$ is essentially the identity map up to a phase, the Dirac delta functions kill half of the integrals in Eq.~\eqref{eq:expval}. This is the bulk analogue of the statistical Wick contraction that set the microscopic CFT energy levels equal in Eq.~\eqref{eq:wick}. Informally, this is summarized by the slogan
  \begin{equation*}
      \text{``ETH surgery = Wick contraction.''}
  \end{equation*}

After evaluating the inner product, what remains is an integral expression for the wormhole path integral, non-perturbatively exact in $1/c$. If we want to match this Virasoro TQFT answer to the actual ensemble prediction for the variance of $Z_\Sigma$, we should still implement the sum over wormholes $W$ in Eq.~\eqref{eq:variance}. We can now make this precise: it is a sum over all integers $(n_1,n_2,n_3)\in \mathbb{Z}^3$, together with a discrete choice of $n_k$ to account for the signed permutations $(-1)^{h_i-\bar h_i}$ in Eq.~\eqref{eq:wick}. The infinite sum over Dehn twists projects onto integer spins, using the Poisson resummation formula:
 \begin{equation}\label{eq:poissonresum}
    \sum_{n\in\mathbb{Z}}|\mathbb{T}^n|^2= \sum_{n\in \mathbb{Z}}e^{-2\pi i n (h-\bar h)} = \sum_{m\in \mathbb{Z}}\delta(J-m).
 \end{equation}
 Moreover, after performing the sum, the answer is no longer holomorphically factorized, as expected in a non-chiral CFT$_2$. In conclusion, there is a match between, on the one hand, a statistical variance of the genus-two partition function in a Gaussian, ETH-like ensemble of OPE coefficients, and, on the other, a sum over bulk topologies constructed using ETH surgery.
 
\subsection{Non-Gaussianities}
 We can refine the above calculation by including the relevant non-Gaussian statistics in the OPE ensemble. As argued for in \cite{Belin:2021ryy, Anous:2021caj}, such non-Gaussianities are necessary to predict average crossing symmetry of higher-genus partition functions. The analogous statistical framework in quantum chaos theory is the generalized ETH \cite{Foini:2018sdb,Jafferis:2022uhu}.
 
  First, let us revisit the average of $Z_\Sigma$. The full sum over topologies in gravity consists of all the handlebodies and all the non-handlebodies whose boundary is $\Sigma$:
\begin{multline}
    \overline{Z_\Sigma(\Omega,\bar\Omega)} = \\ \sum_{\gamma\in \text{MCG}(\Sigma)\setminus H}Z_\text{grav}[M_\gamma;\Omega,\bar \Omega]+ \text{non-handlebodies.}
\end{multline}
We have labeled the sum over handlebodies by elements $\gamma$ in the mapping class group of $\Sigma$: this is the group of large diffeomorphisms of $\Sigma$, generated by acting with arbitrary compositions of the crossing transformations $\{\mathbb{B}, \mathbb{F}, \mathbb{S}\}$ on the cycles shown in Eq.~\eqref{eq:cyclesset} \cite{Moore:1988qv}. We have quotiented by the so-called handlebody group $H$ to avoid overcounting homeomorphic handlebodies \cite{Collier:2023fwi}. 

The statistical theory that reproduces the handlebody part of the sum is defined by the following quadratic moment for the OPE coefficients: 
\begin{equation}\label{eq:opecorrection}
    \overline{C_{h_1h_2h_3}C_{h_1h_2h_3}^*} = \sum_{\gamma \neq \mathbb{1}} \left | \frac{1}{\rho_0(h_1)} \mathbb{K}(\gamma)_{\mathbb{1}; h_1}^{h_2h_3}\right|^2
\end{equation}
while the average spectral density is still given in terms of $\rho_0$, as in Eq.~\eqref{eq:average}. Here $\mathbb{K}(\gamma)$ is the Virasoro crossing kernel that represents the crossing transformation $\gamma$ by an integral kernel acting on a genus-two identity block. More details on the OPE ensemble that incorporates the handlebody sum, as well as a thorough analysis of the non-handlebody contributions, will appear elsewhere \cite{toappear}.

Second, consider again the variance of $Z_\Sigma$, which contains a connected average of four OPE coefficients. In the generalized ensemble, the quartic moment of $C_{h_1h_2h_3}$ now contains non-Gaussianities such as the ``6j-contraction'' studied in Refs.~\cite{deBoer:2024mqg,Collier:2024mgv,Belin:2023efa,Jafferis:2024jkb}. As shown in Section 4 of \cite{deBoer:2024mqg}, such non-Gaussian contractions correspond, in 3d gravity, to Euclidean wormholes obtained by the surgery of Fig.~\ref{fig:ETH_surgery} with a non-trivial gluing map
 \begin{equation}
     G = |\mathbb{K}(\gamma)|^2, \quad \gamma\in \text{MCG}(\Sigma).
 \end{equation}
 This gluing map is specified by a generic genus-two mapping class $\gamma$, generalizing the $\mathbb{Z}^3$ subgroup generated by Eq.~\eqref{eq:gluing_map}. Topologically, these wormholes are a twisted product $\Sigma \times_\gamma I$, and can be computed exactly using Virasoro TQFT (VTQFT). Summing over all choices of $\gamma$ gives the variance the structure of a relative mapping class group sum, similar to the relative Poincar\'e sum obtained for the torus wormhole in \cite{Cotler:2020ugk}:
 \begin{equation}\label{eq:wormholesum}
     \overline{Z_\Sigma(\Omega,\bar\Omega)Z_\Sigma(\Omega',\bar\Omega')}^c \approx \sum_{\gamma \in \text{MCG}(\Sigma)}Z_\text{grav}[\Sigma \times_\gamma I].
 \end{equation}

An even more refined analysis of the variance of $Z_\Sigma$ (which has not been spelled out in the literature before) goes as follows. The most general CFT consistency condition on the product of four OPE coefficients comes from crossing symmetry of the genus-3 partition function. By tuning the moduli to a regime in which the identity block dominates in a dual channel (à la Cardy), each genus-3 crossing equation gives a universal contribution to the quartic moment \cite{Collier:2019weq,Belin:2021ryy}. The set of these contributions is larger than the set of wormhole topologies in Eq.~\eqref{eq:wormholesum}, since the genus-3 mapping class group contains more elements than the genus-2 mapping class group. So, a more general class of wormholes, not of the Maldacena-Maoz type, is needed to match the ensemble prediction. This class of manifolds can also be described by surgery, but now we glue along an intermediate surface of genus three. The general technique of building 3-manifolds by gluing along surfaces of genus $g \geq 2$ is called \emph{Heegaard splitting}. The minimal genus along which we have to glue in order to produce a given manifold $W$ is called the \emph{Heegaard genus} \cite{Farb2012}. Combining these facts, we conclude that a more general class of wormhole topologies contributing to the variance of $Z_\Sigma$ (consistent with genus-3 crossing) consists of all hyperbolic 3-manifolds with Heegaard genus $\leq 3$ and boundary $\Sigma \sqcup \Sigma'$.

\section{\label{sec:RMT} RMT surgery}
While the previous section was mainly a review, we have highlighted the role of surgery in computing higher statistical moments of CFT observables.
In this section, we will apply this philosophy to a new calculation of an off-shell wormhole, contributing to the \emph{spectral statistics} of the CFT. The main technical tool, which we will call ``RMT surgery'', is conceptually similar to the cutting-and-gluing prescription of the previous section, but there is one main difference:
While in ETH surgery we cut 3-manifolds along surfaces of $g\geq 2$, in RMT surgery we should instead cut along a torus.

A consequence of Thurston's Geometrization Theorem \cite{Thurston1997} is that any 3-manifold that can be cut along an embedded incompressible torus (which cannot be deformed to the boundary) is automatically off-shell (i.e.~non-hyperbolic). Off-shell topologies in the gravitational path integral are expected to contribute to the statistics of the high-energy spectrum of primary states, but this expectation has so far only been corroborated by a calculation of the two-boundary torus wormhole $Z_\text{grav}(T\times I)$ \cite{Cotler:2020ugk} and its spectral interpretation \cite{Cotler:2020hgz, Haehl:2023tkr, Haehl:2023xys, Haehl:2023mhf,DiUbaldo:2023qli}.  

From the boundary perspective, the key feature of the torus wormhole is that it captures the correlation between two copies of the density of primary states $\rho(h,\bar h)$, similar to the spectral form factor in many-body quantum chaos. But in CFT$_2$, there are other observables that also contain such a ``$\rho$-$\rho$ correlation''.  In this article, we will focus on one of these, namely the \emph{variance} of the sphere 4-point function. Let $\mathcal{G}$ be
 \begin{equation}\label{eq:fourpointfunction}
    \mathcal{G}(z,\bar z) = \langle \clo_1(0)\clo_1(1) \clo_2(z,\bar z)\clo_2(\infty) \rangle_{S^2},
\end{equation}
where $\clo_1$ and $\clo_2$ are scalar primaries inserted on the 2-sphere, with cross-ratios $z,\bar z$, and consider the variance
\begin{equation}
    \Delta \mathcal{G}^2 \coloneqq \overline{\mathcal{G}(z,\bar z) \mathcal{G}(z',\bar z') }^c.
\end{equation}
Let us for the moment be agnostic about the specific ensemble that is used to compute the average. We will expand $\mathcal{G}$ into 4-point conformal blocks in the $t$-channel: 
\begin{equation}
   \mathcal{G}(z,\bar z) = \int \dd h \dd \bar h \,\rho(h,\bar h) |C_{\clo_1\clo_2 h}|^2 |\mathcal{F}_t(h;z)|^2.
\end{equation}
As before, $\rho(h,\bar h)$ is the exact spectral density. We temporarily used the notation $|C_{\clo_1\clo_2 h}|^2 \equiv C_{\clo_1\clo_2 h}C_{\clo_1\clo_2 h}^*$. To discuss the four-point function in the first place, we assume that the CFT has $\clo_1$ and $\clo_2$ in its (sparse) sub-threshold spectrum, which are taken in the conical defect regime $h_\clo,\bar h_\clo \sim c$, staying below the threshold value $h_t = \bar h_t =\frac{c-1}{24}$. This is a slight extension of `pure' 3d gravity, where one adds heavy matter that backreacts to create trajectories of conical defects in the bulk \cite{Benjamin:2020mfz}.

In this conformal block decomposition, we see that the variance $\Delta \mathcal{G}^2$ is computed by the following statistical moment, which we will call $\mathcal{M}$:
\begin{equation}\label{eq:moment}
   \mathcal{M}\coloneqq \overline{\rho(h,\bar h)\rho(h',\bar h')|C_{\clo_1\clo_2 h}|^2|C_{\clo_1\clo_2h'}|^2}.
\end{equation}
We can enumerate the various contributions that the moment $\mathcal{M}$ receives in a statistical model of the CFT data:
\begin{enumerate}[1)]
    \item \textbf{OPE statistics.} This is similar to the previous section. In the Gaussian ensemble of \cite{Chandra:2022bqq}, there is a connected Wick contraction setting $(h,\bar h)=(h',\bar h')$, which in the bulk corresponds to a Euclidean wormhole with conical defects stretching between the two spherical boundaries. There are also non-Gaussian corrections to the quartic moment of $C_{\clo_1\clo_2h}$, which are captured by wormholes where the conical defect Wilson lines can braid, link or tangle \cite{Collier:2024mgv}. 
    \item \textbf{Spectral statistics.} There is a further contribution from the correlation between the two factors of the spectral density in $\mathcal{M}$. We assume that the spectrum of primaries above the threshold is \emph{chaotic}, which by the BGS conjecture \cite{bohigas1984characterization} implies a universal form for the leading correlation between two energy levels at fixed spin, dictated by random matrix theory (RMT):
\begin{multline}\label{eq:spectralcorr}
    \quad\overline{\rho(E_1,J_1)\rho(E_2,J_2)}^c \approx \\[0.7em] \delta_{J_1J_2}\,\rho_{\text{RMT}}(E_1-|J_1|,E_2-|J_2|).\quad
\end{multline}
Here we introduced energy $E = h+\bar h - \frac{c-1}{12}$ and spin $J=h-\bar h$ variables. In the present context, we take as input $\bar\rho \approx \rho_0(E,J)$ for the averaged spectrum above the black hole threshold, and we will model the spectral statistics by a random matrix with edge $E \geq |J|$. Then the leading spectral variance is
\begin{equation}
    \rho_\text{RMT}(t_1,t_2) = -\frac{\mathsf{C}_{\text{RMT}}}{4\pi^2}\frac{1}{\sqrt{t_1t_2}} \frac{t_1+t_2}{(t_1-t_2)^2},
\end{equation}
where $\mathsf{C}_{\text{RMT}}$ is 1 for GUE statistics and 2 for the GOE \cite{Jafferis:2024jkb}.
Eq.~\eqref{eq:spectralcorr} should be seen as an extra assumption, expressing maximal randomness of the high-energy primary spectrum, on top of the usual HKS criteria for a holographic CFT with an AdS$_3$ gravity dual \cite{Hartman:2014oaa}. We can further refine it to a fully modular invariant ansatz \cite{DiUbaldo:2023qli}, as will be discussed below.

    \item \textbf{Mixed correlations.} In a quantum chaotic system, there are generically also cross-correlations between the spectral density and energy eigenstates \cite{Altland:2021rqn}, but not much is known about such correlations in the ensemble description of AdS$_3$/CFT$_2$. For a recent proposal for the cubic cumulant $\overline{\rho C^2}$, see Ref.~\cite{Post:2024itb}.
\end{enumerate}

Of these classes of contributions to the statistical moment $\mathcal{M}$, point 2 will be the focus of this section. In other words, we compute the contribution to the variance $\Delta\mathcal{G}^2$ from the \emph{connected} spectral correlator and \emph{disconnected} Gaussian OPE contraction, schematically $\overline{\rho\rho}^c\times  \overline{C^2}\times \overline{C^2}$. As shown in Appendix \ref{app:fourpoint}, doing so gives
\begin{align}\label{eq:bdy_prediction}
    &\Delta\mathcal{G}^2 \supset \int_0^\infty \dd^4 P \,\Big[\rho_{0,2}(P,\bar P,P',\bar P')
\\ &\big|C_0(P_{\clo_1},P_{\clo_2},P)C_0(P_{\clo_1},P_{\clo_2},P')\mathcal{F}_t(P;z)\mathcal{F}_t(P';z')\big|^2\Big]\nonumber
\end{align}
where $d^4P\coloneqq \dd P\dd\bar P\dd P'\dd \bar P'$. Here $C_0$ is the universal OPE density derived in \cite{Collier:2019weq} and $\rho_{0,2}$ is essentially the random matrix kernel $\rho_\text{RMT}$ in Eq.~\eqref{eq:spectralcorr}, but expressed in Liouville momenta $h = \frac{c-1}{24}+P^2$; see Eq.~\eqref{eq:rho02}.

The goal is to reproduce this boundary prediction---based on a random matrix ansatz---from a bulk calculation of an off-shell Euclidean wormhole with two four-punctured sphere boundaries. In the rest of this section, we will show that this can be achieved by applying \emph{RMT surgery}, which constructs the relevant wormhole topology in two steps, shown in Figure \ref{fig:RMT_surgery}:
 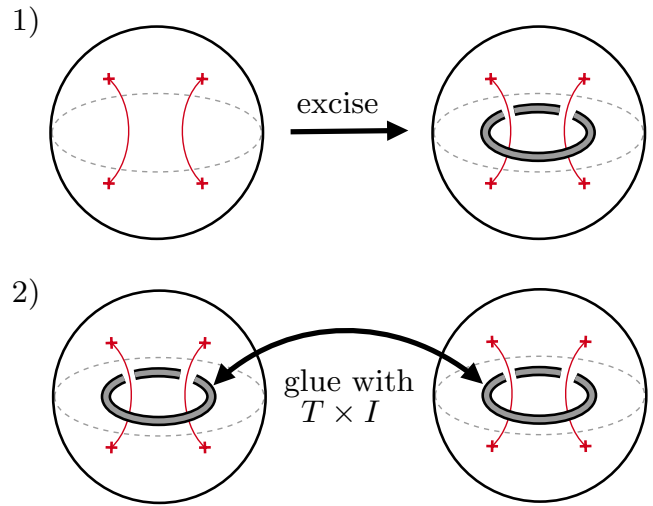
\begin{figure}
    \centering
    \resizebox{\linewidth}{!}{
    \begin{tikzpicture}[x=0.75pt,y=0.75pt,yscale=-1,xscale=1,baseline={([yshift=-0.5ex]current bounding box.center)}]
\draw  [color={rgb, 255:red, 155; green, 155; blue, 155 }  ,draw opacity=1 ][dash pattern={on 1.5pt off 1.5pt on 1.5pt off 1.5pt}] (169.6,60.6) .. controls (169.6,52.32) and (187.51,45.6) .. (209.6,45.6) .. controls (231.69,45.6) and (249.6,52.32) .. (249.6,60.6) .. controls (249.6,68.88) and (231.69,75.6) .. (209.6,75.6) .. controls (187.51,75.6) and (169.6,68.88) .. (169.6,60.6) -- cycle ;
\draw  [line width=0.75]  (169,60.6) .. controls (169,38.18) and (187.18,20) .. (209.6,20) .. controls (232.02,20) and (250.2,38.18) .. (250.2,60.6) .. controls (250.2,83.02) and (232.02,101.2) .. (209.6,101.2) .. controls (187.18,101.2) and (169,83.02) .. (169,60.6) -- cycle ;
\draw [color={rgb, 255:red, 208; green, 2; blue, 27 }  ,draw opacity=1 ]   (191.29,40) .. controls (201.41,49.38) and (201.41,70.38) .. (191.29,80) ;
\draw [shift={(191.29,80)}, rotate = 181.45] [color={rgb, 255:red, 208; green, 2; blue, 27 }  ,draw opacity=1 ][line width=0.75]    (-2.24,0) -- (2.24,0)(0,2.24) -- (0,-2.24)   ;
\draw [shift={(191.29,40)}, rotate = 87.8] [color={rgb, 255:red, 208; green, 2; blue, 27 }  ,draw opacity=1 ][line width=0.75]    (-2.24,0) -- (2.24,0)(0,2.24) -- (0,-2.24)   ;
\draw [color={rgb, 255:red, 208; green, 2; blue, 27 }  ,draw opacity=1 ]   (227,40) .. controls (216.83,49.38) and (216.83,70.38) .. (227,80) ;
\draw [shift={(227,80)}, rotate = 88.43] [color={rgb, 255:red, 208; green, 2; blue, 27 }  ,draw opacity=1 ][line width=0.75]    (-2.24,0) -- (2.24,0)(0,2.24) -- (0,-2.24)   ;
\draw [shift={(227,40)}, rotate = 182.32] [color={rgb, 255:red, 208; green, 2; blue, 27 }  ,draw opacity=1 ][line width=0.75]    (-2.24,0) -- (2.24,0)(0,2.24) -- (0,-2.24)   ;
\draw  [draw opacity=0][line width=3]  (223.12,53.78) .. controls (227.02,55.48) and (229.46,57.89) .. (229.46,60.56) .. controls (229.46,65.69) and (220.44,69.86) .. (209.31,69.86) .. controls (198.19,69.86) and (189.17,65.69) .. (189.17,60.56) .. controls (189.17,57.76) and (191.85,55.24) .. (196.1,53.54) -- (209.31,60.56) -- cycle ; \draw  [line width=3]  (223.12,53.78) .. controls (227.02,55.48) and (229.46,57.89) .. (229.46,60.56) .. controls (229.46,65.69) and (220.44,69.86) .. (209.31,69.86) .. controls (198.19,69.86) and (189.17,65.69) .. (189.17,60.56) .. controls (189.17,57.76) and (191.85,55.24) .. (196.1,53.54) ;  
\draw  [draw opacity=0][line width=3]  (200.2,52.26) .. controls (202.93,51.62) and (206.03,51.26) .. (209.31,51.26) .. controls (212.24,51.26) and (215.03,51.55) .. (217.54,52.06) -- (209.31,60.56) -- cycle ; \draw  [line width=3]  (200.2,52.26) .. controls (202.93,51.62) and (206.03,51.26) .. (209.31,51.26) .. controls (212.24,51.26) and (215.03,51.55) .. (217.54,52.06) ;  
\draw  [color={rgb, 255:red, 155; green, 155; blue, 155 }  ,draw opacity=1 ][dash pattern={on 1.5pt off 1.5pt on 1.5pt off 1.5pt}] (23.6,60.6) .. controls (23.6,52.32) and (41.51,45.6) .. (63.6,45.6) .. controls (85.69,45.6) and (103.6,52.32) .. (103.6,60.6) .. controls (103.6,68.88) and (85.69,75.6) .. (63.6,75.6) .. controls (41.51,75.6) and (23.6,68.88) .. (23.6,60.6) -- cycle ;
\draw  [line width=0.75]  (23,60.6) .. controls (23,38.18) and (41.18,20) .. (63.6,20) .. controls (86.02,20) and (104.2,38.18) .. (104.2,60.6) .. controls (104.2,83.02) and (86.02,101.2) .. (63.6,101.2) .. controls (41.18,101.2) and (23,83.02) .. (23,60.6) -- cycle ;
\draw [color={rgb, 255:red, 208; green, 2; blue, 27 }  ,draw opacity=1 ]   (45.29,40) .. controls (55.41,49.38) and (55.41,70.38) .. (45.29,80) ;
\draw [shift={(45.29,80)}, rotate = 181.45] [color={rgb, 255:red, 208; green, 2; blue, 27 }  ,draw opacity=1 ][line width=0.75]    (-2.24,0) -- (2.24,0)(0,2.24) -- (0,-2.24)   ;
\draw [shift={(45.29,40)}, rotate = 87.8] [color={rgb, 255:red, 208; green, 2; blue, 27 }  ,draw opacity=1 ][line width=0.75]    (-2.24,0) -- (2.24,0)(0,2.24) -- (0,-2.24)   ;
\draw [color={rgb, 255:red, 208; green, 2; blue, 27 }  ,draw opacity=1 ]   (81,40) .. controls (70.83,49.38) and (70.83,70.38) .. (81,80) ;
\draw [shift={(81,80)}, rotate = 88.43] [color={rgb, 255:red, 208; green, 2; blue, 27 }  ,draw opacity=1 ][line width=0.75]    (-2.24,0) -- (2.24,0)(0,2.24) -- (0,-2.24)   ;
\draw [shift={(81,40)}, rotate = 182.32] [color={rgb, 255:red, 208; green, 2; blue, 27 }  ,draw opacity=1 ][line width=0.75]    (-2.24,0) -- (2.24,0)(0,2.24) -- (0,-2.24)   ;
\draw  [draw opacity=0][line width=1.5]  (223.16,53.8) .. controls (227.04,55.5) and (229.46,57.9) .. (229.46,60.56) .. controls (229.46,65.69) and (220.44,69.86) .. (209.31,69.86) .. controls (198.19,69.86) and (189.17,65.69) .. (189.17,60.56) .. controls (189.17,57.74) and (191.88,55.22) .. (196.15,53.52) -- (209.31,60.56) -- cycle ; \draw  [color={rgb, 255:red, 155; green, 155; blue, 155 }  ,draw opacity=1 ][line width=1.5]  (223.16,53.8) .. controls (227.04,55.5) and (229.46,57.9) .. (229.46,60.56) .. controls (229.46,65.69) and (220.44,69.86) .. (209.31,69.86) .. controls (198.19,69.86) and (189.17,65.69) .. (189.17,60.56) .. controls (189.17,57.74) and (191.88,55.22) .. (196.15,53.52) ;  
\draw  [draw opacity=0][line width=1.5]  (200.28,52.18) .. controls (202.96,51.57) and (205.97,51.23) .. (209.15,51.23) .. controls (212.13,51.23) and (214.97,51.53) .. (217.52,52.07) -- (209.15,60.53) -- cycle ; \draw  [color={rgb, 255:red, 155; green, 155; blue, 155 }  ,draw opacity=1 ][line width=1.5]  (200.28,52.18) .. controls (202.96,51.57) and (205.97,51.23) .. (209.15,51.23) .. controls (212.13,51.23) and (214.97,51.53) .. (217.52,52.07) ;  
\draw [line width=1.5]    (114.75,60.25) -- (155.75,59.79) ;
\draw [shift={(159.75,59.75)}, rotate = 179.36] [fill={rgb, 255:red, 0; green, 0; blue, 0 }  ][line width=0.08]  [draw opacity=0] (8.13,-3.9) -- (0,0) -- (8.13,3.9) -- cycle    ;
\draw  [color={rgb, 255:red, 155; green, 155; blue, 155 }  ,draw opacity=1 ][dash pattern={on 1.5pt off 1.5pt on 1.5pt off 1.5pt}] (24.6,161.6) .. controls (24.6,153.32) and (42.51,146.6) .. (64.6,146.6) .. controls (86.69,146.6) and (104.6,153.32) .. (104.6,161.6) .. controls (104.6,169.88) and (86.69,176.6) .. (64.6,176.6) .. controls (42.51,176.6) and (24.6,169.88) .. (24.6,161.6) -- cycle ;
\draw  [line width=0.75]  (24,161.6) .. controls (24,139.18) and (42.18,121) .. (64.6,121) .. controls (87.02,121) and (105.2,139.18) .. (105.2,161.6) .. controls (105.2,184.02) and (87.02,202.2) .. (64.6,202.2) .. controls (42.18,202.2) and (24,184.02) .. (24,161.6) -- cycle ;
\draw [color={rgb, 255:red, 208; green, 2; blue, 27 }  ,draw opacity=1 ]   (46.29,141) .. controls (56.41,150.38) and (56.41,171.38) .. (46.29,181) ;
\draw [shift={(46.29,181)}, rotate = 181.45] [color={rgb, 255:red, 208; green, 2; blue, 27 }  ,draw opacity=1 ][line width=0.75]    (-2.24,0) -- (2.24,0)(0,2.24) -- (0,-2.24)   ;
\draw [shift={(46.29,141)}, rotate = 87.8] [color={rgb, 255:red, 208; green, 2; blue, 27 }  ,draw opacity=1 ][line width=0.75]    (-2.24,0) -- (2.24,0)(0,2.24) -- (0,-2.24)   ;
\draw [color={rgb, 255:red, 208; green, 2; blue, 27 }  ,draw opacity=1 ]   (82,141) .. controls (71.83,150.38) and (71.83,171.38) .. (82,181) ;
\draw [shift={(82,181)}, rotate = 88.43] [color={rgb, 255:red, 208; green, 2; blue, 27 }  ,draw opacity=1 ][line width=0.75]    (-2.24,0) -- (2.24,0)(0,2.24) -- (0,-2.24)   ;
\draw [shift={(82,141)}, rotate = 182.32] [color={rgb, 255:red, 208; green, 2; blue, 27 }  ,draw opacity=1 ][line width=0.75]    (-2.24,0) -- (2.24,0)(0,2.24) -- (0,-2.24)   ;
\draw  [draw opacity=0][line width=3]  (78.12,154.78) .. controls (82.02,156.48) and (84.46,158.89) .. (84.46,161.56) .. controls (84.46,166.69) and (75.44,170.86) .. (64.31,170.86) .. controls (53.19,170.86) and (44.17,166.69) .. (44.17,161.56) .. controls (44.17,158.76) and (46.85,156.24) .. (51.1,154.54) -- (64.31,161.56) -- cycle ; \draw  [line width=3]  (78.12,154.78) .. controls (82.02,156.48) and (84.46,158.89) .. (84.46,161.56) .. controls (84.46,166.69) and (75.44,170.86) .. (64.31,170.86) .. controls (53.19,170.86) and (44.17,166.69) .. (44.17,161.56) .. controls (44.17,158.76) and (46.85,156.24) .. (51.1,154.54) ;  
\draw  [draw opacity=0][line width=3]  (55.2,153.26) .. controls (57.93,152.62) and (61.03,152.26) .. (64.31,152.26) .. controls (67.24,152.26) and (70.03,152.55) .. (72.54,153.06) -- (64.31,161.56) -- cycle ; \draw  [line width=3]  (55.2,153.26) .. controls (57.93,152.62) and (61.03,152.26) .. (64.31,152.26) .. controls (67.24,152.26) and (70.03,152.55) .. (72.54,153.06) ;  
\draw  [draw opacity=0][line width=1.5]  (78.16,154.8) .. controls (82.04,156.5) and (84.46,158.9) .. (84.46,161.56) .. controls (84.46,166.69) and (75.44,170.86) .. (64.31,170.86) .. controls (53.19,170.86) and (44.17,166.69) .. (44.17,161.56) .. controls (44.17,158.74) and (46.88,156.22) .. (51.15,154.52) -- (64.31,161.56) -- cycle ; \draw  [color={rgb, 255:red, 155; green, 155; blue, 155 }  ,draw opacity=1 ][line width=1.5]  (78.16,154.8) .. controls (82.04,156.5) and (84.46,158.9) .. (84.46,161.56) .. controls (84.46,166.69) and (75.44,170.86) .. (64.31,170.86) .. controls (53.19,170.86) and (44.17,166.69) .. (44.17,161.56) .. controls (44.17,158.74) and (46.88,156.22) .. (51.15,154.52) ;  
\draw  [draw opacity=0][line width=1.5]  (55.28,153.18) .. controls (57.96,152.57) and (60.97,152.23) .. (64.15,152.23) .. controls (67.13,152.23) and (69.97,152.53) .. (72.52,153.07) -- (64.15,161.53) -- cycle ; \draw  [color={rgb, 255:red, 155; green, 155; blue, 155 }  ,draw opacity=1 ][line width=1.5]  (55.28,153.18) .. controls (57.96,152.57) and (60.97,152.23) .. (64.15,152.23) .. controls (67.13,152.23) and (69.97,152.53) .. (72.52,153.07) ;  
\draw  [color={rgb, 255:red, 155; green, 155; blue, 155 }  ,draw opacity=1 ][dash pattern={on 1.5pt off 1.5pt on 1.5pt off 1.5pt}] (170.1,161.1) .. controls (170.1,152.82) and (188.01,146.1) .. (210.1,146.1) .. controls (232.19,146.1) and (250.1,152.82) .. (250.1,161.1) .. controls (250.1,169.38) and (232.19,176.1) .. (210.1,176.1) .. controls (188.01,176.1) and (170.1,169.38) .. (170.1,161.1) -- cycle ;
\draw  [line width=0.75]  (169.5,161.1) .. controls (169.5,138.68) and (187.68,120.5) .. (210.1,120.5) .. controls (232.52,120.5) and (250.7,138.68) .. (250.7,161.1) .. controls (250.7,183.52) and (232.52,201.7) .. (210.1,201.7) .. controls (187.68,201.7) and (169.5,183.52) .. (169.5,161.1) -- cycle ;
\draw [color={rgb, 255:red, 208; green, 2; blue, 27 }  ,draw opacity=1 ]   (191.79,140.5) .. controls (201.91,149.88) and (201.91,170.88) .. (191.79,180.5) ;
\draw [shift={(191.79,180.5)}, rotate = 181.45] [color={rgb, 255:red, 208; green, 2; blue, 27 }  ,draw opacity=1 ][line width=0.75]    (-2.24,0) -- (2.24,0)(0,2.24) -- (0,-2.24)   ;
\draw [shift={(191.79,140.5)}, rotate = 87.8] [color={rgb, 255:red, 208; green, 2; blue, 27 }  ,draw opacity=1 ][line width=0.75]    (-2.24,0) -- (2.24,0)(0,2.24) -- (0,-2.24)   ;
\draw [color={rgb, 255:red, 208; green, 2; blue, 27 }  ,draw opacity=1 ]   (227.5,140.5) .. controls (217.33,149.88) and (217.33,170.88) .. (227.5,180.5) ;
\draw [shift={(227.5,180.5)}, rotate = 88.43] [color={rgb, 255:red, 208; green, 2; blue, 27 }  ,draw opacity=1 ][line width=0.75]    (-2.24,0) -- (2.24,0)(0,2.24) -- (0,-2.24)   ;
\draw [shift={(227.5,140.5)}, rotate = 182.32] [color={rgb, 255:red, 208; green, 2; blue, 27 }  ,draw opacity=1 ][line width=0.75]    (-2.24,0) -- (2.24,0)(0,2.24) -- (0,-2.24)   ;
\draw  [draw opacity=0][line width=3]  (223.62,154.28) .. controls (227.52,155.98) and (229.96,158.39) .. (229.96,161.06) .. controls (229.96,166.19) and (220.94,170.36) .. (209.81,170.36) .. controls (198.69,170.36) and (189.67,166.19) .. (189.67,161.06) .. controls (189.67,158.26) and (192.35,155.74) .. (196.6,154.04) -- (209.81,161.06) -- cycle ; \draw  [line width=3]  (223.62,154.28) .. controls (227.52,155.98) and (229.96,158.39) .. (229.96,161.06) .. controls (229.96,166.19) and (220.94,170.36) .. (209.81,170.36) .. controls (198.69,170.36) and (189.67,166.19) .. (189.67,161.06) .. controls (189.67,158.26) and (192.35,155.74) .. (196.6,154.04) ;  
\draw  [draw opacity=0][line width=3]  (200.7,152.76) .. controls (203.43,152.12) and (206.53,151.76) .. (209.81,151.76) .. controls (212.74,151.76) and (215.53,152.05) .. (218.04,152.56) -- (209.81,161.06) -- cycle ; \draw  [line width=3]  (200.7,152.76) .. controls (203.43,152.12) and (206.53,151.76) .. (209.81,151.76) .. controls (212.74,151.76) and (215.53,152.05) .. (218.04,152.56) ;  
\draw  [draw opacity=0][line width=1.5]  (223.66,154.3) .. controls (227.54,156) and (229.96,158.4) .. (229.96,161.06) .. controls (229.96,166.19) and (220.94,170.36) .. (209.81,170.36) .. controls (198.69,170.36) and (189.67,166.19) .. (189.67,161.06) .. controls (189.67,158.24) and (192.38,155.72) .. (196.65,154.02) -- (209.81,161.06) -- cycle ; \draw  [color={rgb, 255:red, 155; green, 155; blue, 155 }  ,draw opacity=1 ][line width=1.5]  (223.66,154.3) .. controls (227.54,156) and (229.96,158.4) .. (229.96,161.06) .. controls (229.96,166.19) and (220.94,170.36) .. (209.81,170.36) .. controls (198.69,170.36) and (189.67,166.19) .. (189.67,161.06) .. controls (189.67,158.24) and (192.38,155.72) .. (196.65,154.02) ;  
\draw  [draw opacity=0][line width=1.5]  (200.78,152.68) .. controls (203.46,152.07) and (206.47,151.73) .. (209.65,151.73) .. controls (212.63,151.73) and (215.47,152.03) .. (218.02,152.57) -- (209.65,161.03) -- cycle ; \draw  [color={rgb, 255:red, 155; green, 155; blue, 155 }  ,draw opacity=1 ][line width=1.5]  (200.78,152.68) .. controls (203.46,152.07) and (206.47,151.73) .. (209.65,151.73) .. controls (212.63,151.73) and (215.47,152.03) .. (218.02,152.57) ;  
\draw [line width=1.5]    (88.54,154.14) .. controls (114.71,130.95) and (155.06,129.62) .. (185.49,154.35) ;
\draw [shift={(188.29,156.71)}, rotate = 221.47] [fill={rgb, 255:red, 0; green, 0; blue, 0 }  ][line width=0.08]  [draw opacity=0] (8.13,-3.9) -- (0,0) -- (8.13,3.9) -- cycle    ;
\draw [shift={(85.25,157.25)}, rotate = 314.82] [fill={rgb, 255:red, 0; green, 0; blue, 0 }  ][line width=0.08]  [draw opacity=0] (8.13,-3.9) -- (0,0) -- (8.13,3.9) -- cycle    ;
\draw (115.86,43.88) node [anchor=north west][inner sep=0.75pt]  [font=\small] [align=left] {excise};
\draw (7,10.9) node [anchor=north west][inner sep=0.75pt]    {$1)$};
\draw (7,114.9) node [anchor=north west][inner sep=0.75pt]    {$2)$};
\draw (110.86,149.88) node [anchor=north west][inner sep=0.75pt]  [font=\small] [align=left] {glue with};
\draw (117.5,161.9) node [anchor=north west][inner sep=0.75pt]    {$T \times I$};
\end{tikzpicture}}
    \caption{RMT surgery.}
    \label{fig:RMT_surgery}
\end{figure}
\begin{enumerate}[1)]
    \item Take a four-punctured sphere and fill it in by a ball, with bulk Wilson lines connecting the punctures in pairs. Excise from its interior a solid torus (shown in grey) linking both Wilson lines.
    \item Take two copies of the manifold constructed in Step 1 and glue their torus boundaries with the torus wormhole $T\times I$, through some gluing map $G_\text{T}$.
\end{enumerate}

The resulting topology, which we will call $\mathcal{W}$, is of the wormhole type $S^2\times I$, where $I=[0,1]$, but it contains a non-trivial linking of the two pairs of Wilson lines. In particular, it is not homeomorphic to the four-point wormhole $\Sigma_{0,4} \times I$ that has been computed in Ref.~\cite{Chandra:2022bqq}. The topology of $\mathcal{W}$ is visualized in Fig.~\ref{fig:ETHvsRMT}(b), with more details given in Appendix \ref{app:topology}.

By construction, the wormhole $\mathcal{W}$ is off-shell, because the torus along which we have glued the topologies in Step 2 becomes an incompressible torus embedded into the manifold after gluing, hence violating one of Thurston's criteria for hyperbolicity \cite{Thurston1982}. This torus is incompressible because it has been non-trivially linked with the Wilson lines. Note that \emph{before} gluing, each piece is separately hyperbolic: the filled-in four-punctured sphere admits a hyperbolic metric due to the conical defect lines, and excising the torus creates a \emph{cusp} boundary component. The fact that surgery may change the allowed metrics on a manifold is well known in geometric topology, and also occurs in Dehn surgery \cite{Thurston1997}.

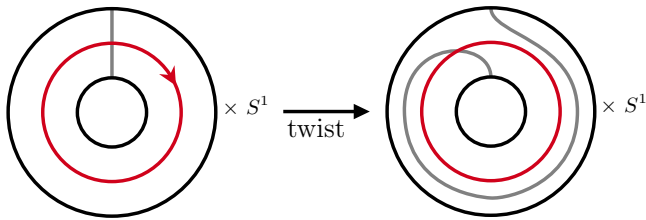
\begin{figure}
    \centering
    \resizebox{\linewidth}{!}{
    \begin{tikzpicture}[x=0.75pt,y=0.75pt,yscale=-1,xscale=1,baseline={([yshift=-0.5ex]current bounding box.center)}]
\draw [color={rgb, 255:red, 128; green, 128; blue, 128 }  ,draw opacity=1 ][line width=1.5]    (339,39.5) .. controls (338.6,55.3) and (389,58.1) .. (389,99.5) .. controls (389,140.9) and (349,150.1) .. (339,149.5) .. controls (329,148.9) and (289.8,142.9) .. (289,99.5) .. controls (288.2,56.1) and (338.6,57.3) .. (339,79.5) ;
\draw [color={rgb, 255:red, 128; green, 128; blue, 128 }  ,draw opacity=1 ][line width=1.5]    (120,40) -- (120,80) ;
\draw  [line width=1.5]  (100,100) .. controls (100,88.95) and (108.95,80) .. (120,80) .. controls (131.05,80) and (140,88.95) .. (140,100) .. controls (140,111.05) and (131.05,120) .. (120,120) .. controls (108.95,120) and (100,111.05) .. (100,100) -- cycle ;
\draw  [line width=1.5]  (60,100) .. controls (60,66.86) and (86.86,40) .. (120,40) .. controls (153.14,40) and (180,66.86) .. (180,100) .. controls (180,133.14) and (153.14,160) .. (120,160) .. controls (86.86,160) and (60,133.14) .. (60,100) -- cycle ;
\draw  [color={rgb, 255:red, 208; green, 2; blue, 27 }  ,draw opacity=1 ][line width=1.5]  (80,100) .. controls (80,77.91) and (97.91,60) .. (120,60) .. controls (142.09,60) and (160,77.91) .. (160,100) .. controls (160,122.09) and (142.09,140) .. (120,140) .. controls (97.91,140) and (80,122.09) .. (80,100) -- cycle ;
\draw [color={rgb, 255:red, 208; green, 2; blue, 27 }  ,draw opacity=1 ]   (156.39,82.83) .. controls (156.39,83.59) and (153.62,78.68) .. (155.52,81.97) ;
\draw [shift={(157,84.56)}, rotate = 240.26] [fill={rgb, 255:red, 208; green, 2; blue, 27 }  ,fill opacity=1 ][line width=0.08]  [draw opacity=0] (10.72,-5.15) -- (0,0) -- (10.72,5.15) -- (7.12,0) -- cycle    ;
\draw  [line width=1.5]  (319,99.5) .. controls (319,88.45) and (327.95,79.5) .. (339,79.5) .. controls (350.05,79.5) and (359,88.45) .. (359,99.5) .. controls (359,110.55) and (350.05,119.5) .. (339,119.5) .. controls (327.95,119.5) and (319,110.55) .. (319,99.5) -- cycle ;
\draw  [line width=1.5]  (279,99.5) .. controls (279,66.36) and (305.86,39.5) .. (339,39.5) .. controls (372.14,39.5) and (399,66.36) .. (399,99.5) .. controls (399,132.64) and (372.14,159.5) .. (339,159.5) .. controls (305.86,159.5) and (279,132.64) .. (279,99.5) -- cycle ;
\draw  [color={rgb, 255:red, 208; green, 2; blue, 27 }  ,draw opacity=1 ][line width=1.5]  (299,99.5) .. controls (299,77.41) and (316.91,59.5) .. (339,59.5) .. controls (361.09,59.5) and (379,77.41) .. (379,99.5) .. controls (379,121.59) and (361.09,139.5) .. (339,139.5) .. controls (316.91,139.5) and (299,121.59) .. (299,99.5) -- cycle ;
\draw [line width=1.5]    (219,99.5) -- (265,99.5) ;
\draw [shift={(269,99.5)}, rotate = 180] [fill={rgb, 255:red, 0; green, 0; blue, 0 }  ][line width=0.08]  [draw opacity=0] (9.29,-4.46) -- (0,0) -- (9.29,4.46) -- cycle    ;
\draw (182.53,89.4) node [anchor=north west][inner sep=0.75pt]  [font=\small]  {$\times \ S^{1}$};
\draw (220,102.5) node [anchor=north west][inner sep=0.75pt] [font=\large]  [align=left] {twist};
\draw (401,88.4) node [anchor=north west][inner sep=0.75pt]  [font=\small]  {$\times \ S^{1}$};
\end{tikzpicture}}
    \caption{Bulk mapping class acting on an annulus$\times S^1$.}
    \label{fig:dehn_twists}
\end{figure}

Having discussed the topology of the four-point wormhole $\mathcal{W}$, let us discuss its bulk mapping class group (MCG). By this we mean the group of orientation-preserving large diffeomorphisms of $\mathcal{W}$ that act trivially on the boundary. This is an important notion in quantum gravity because diffeomorphisms are gauged, meaning that we should ``divide out'' the bulk MCG in the gravitational path integral. For hyperbolic 3-manifolds, this is easy, because their MCG is finite (even trivial for manifolds with boundaries of $g\geq 2$) \cite{Collier:2023fwi}. However, for non-hyperbolic manifolds, the MCG may be infinite, and dividing it out requires some care. In the case at hand, we claim that the bulk MCG of $\mathcal{W}$ has the simple form
\begin{equation}\label{eq:MCG}
    \text{MCG}(\mathcal{W}) = \mathbb{Z}\times \mathbb{Z}.
\end{equation}
To see why, note that there is an incompressible torus $T$ embedded in $\mathcal{W}$, meaning that the map $\pi_1(T) \to \pi_1(\mathcal{W})$ is injective. So there are two homotopy classes of closed curves, $[\alpha]$ and $[\beta]$, that are non-contractible in $\mathcal{W}$. A small neighborhood of the embedded torus is $T \times I$, and the cycles $\alpha$ and $\beta$ can be homotoped to non-intersecting curves (shrink one cycle slightly along the interval $I$). Taking the surfaces $\alpha\times S^1$ and $S^1\times \beta$ and rotating them by $2\pi$ along $\alpha$ and $\beta$ generates the commuting group $\mathbb{Z}\times \mathbb{Z}$. In Figure \ref{fig:dehn_twists}, we have drawn a single such twist along a cycle $\alpha$ (in red), together with the effect that this diffeomorphism has on a line segment (in gray) inside the torus neighborhood $T\times I \cong A \times S^1$.  

The diffeo's described above only act on the `internal' embedded torus, leaving the boundaries of the torus neighborhood in Fig. \ref{fig:dehn_twists} invariant and acting trivially on the complement $\mathcal{W}\setminus (T\times I)$. This is the same as the mapping class group action on the internal torus of the ``double trumpet'' studied by Cotler and Jensen \cite{Cotler:2020ugk,Jafferis:2024jkb}.

\subsection{Wormhole partition function}\label{sec:whpartitionfunction}

Now we will compute the exact gravitational partition function of the wormhole topology $\mathcal{W}$. Since we argued that $\mathcal{W}$ is off-shell, we should pause to say what we mean by an off-shell partition function in 3d gravity. In general, the Euclidean  path integral on a  topology $M$ is
\begin{equation}\label{eq:GPI}
    Z_\text{grav}[M;\Sigma] = \int_{g\mid_\Sigma} \frac{\mathcal{D}g}{\text{Diff}(M)} e^{-S_E[g]}
\end{equation}
where $S_E[g]$ is the Einstein-Hilbert action with an appropriate boundary action for the conformal boundary $\Sigma = \partial M$. The group $\text{Diff}(M)$ contains both large and small diffeomorphisms of $M$. Since 3d gravity in first-order variables is related to Chern-Simons theory (for which small diffeomorphisms are gauge redundancies) \cite{Witten:1988hc}, the only non-trivial step is to quotient by the mapping class group, as discussed before.  

 In the sum over topologies $M$ (where each $M$ has the same asymptotic boundary $\Sigma$), there are both on-shell and off-shell contributions. The main difference is that for on-shell topologies, the path integral \eqref{eq:GPI} admits a saddle-point approximation, while for off-shell topologies it does not. Indeed, for on-shell topologies we can compute the partition function exactly using Virasoro TQFT, and verify explicitly that it is well-approximated by a saddle-point evaluation at large $c$ \cite{Collier:2024mgv}. In contrast, for the off-shell partition function computed in this section, we will demonstrate that the resulting integral expression does not have a saddle-point. This lack of a saddle-point approximation in the large-$c$ limit implies that, for off-shell topologies, we should compute the path integral over metrics in \eqref{eq:GPI} by other means.\footnote{We could also choose to not sum over off-shell topologies, but, as shown in Ref.~\cite{Maloney:2007ud}, this leads to pathologies in the (putative) holographic boundary theory, such as a continuous spectrum with an exponentially large negativity near extremality \cite{Keller:2014xba}.} 

Returning to the wormhole $\mathcal{W}$ constructed using RMT surgery, we will take inspiration from Section \ref{sec:ETH} and use the Hilbert space formalism defined by the path integral. In this formalism, gluing 3-manifolds along common surfaces amounts to taking inner products of states defined by the path integral, such as in Eq.~\eqref{eq:expval}. This method relies on the assumption that the bulk mapping class group acts separately on each of the components that are being glued, which is not always true for generic off-shell topologies. But, since we argued that the bulk MCG of $\mathcal{W}$ only acts locally on the torus neighborhood $T\times I$, this assumption is justified in the case at hand.

Following Step 1 of the RMT surgery that was illustrated in Figure \ref{fig:RMT_surgery}, the path integral  defines a state in the gravitational Hilbert space of the torus:
\begin{multline}\label{eq:state2}
    \ket{Z_\text{grav}(z,\bar z)} =\\ \int_0^\infty \dd P\,\dd \bar P \braket{P,\bar P}{Z_\text{grav}(z,\bar z)}\ket{P,\bar P}\in \mathcal{H}_\text{T}.
\end{multline}
By quantizing the gravitational phase space, it can be shown that $\mathcal{H}_\text{T}$ is spanned (as a vector space) by a basis of holomorphic times anti-holomorphic non-degenerate torus conformal blocks $\ket{P,\bar P}\coloneqq \ket{\mathcal{F}_{1,0}(P)}\otimes \ket{\mathcal{F}_{1,0}(\bar P)}$, with $P,\bar P\in \mathbb{R}_{\geq 0}$. These basis states are related to the Virasoro characters in the wavefunction basis as 
\begin{equation}
    \braket{\tau,\bar\tau}{P,\bar P}= |\chi_P(\tau)|^2
\end{equation}
where $\chi_P(\tau) = \frac{1}{\eta(\tau)}e^{2\pi i \tau P^2}$.  To make $\mathcal{H}_\text{T}$ into a Hilbert space, we assume that the inner product is delta function orthogonal with a flat measure in $P$ and $\bar P$:
\begin{equation}\label{eq:innerprodassumption}
    \braket{P,\bar P}{P',\bar P'} = \delta(P-P')\delta(\bar P-\bar P').
\end{equation}
This inner product can be derived from Verlinde's quantization of Teichm\"uller space \cite{Verlinde:1989ua}, if one discards an infinite volume factor similar to the infinite prefactor of the Liouville partition function \cite{Collier:2023fwi}. Eq.~\eqref{eq:innerprodassumption} can also be justified by a regularization procedure from the inner product on the once-punctured torus \cite{toappear3}.

The state in Eq.~\eqref{eq:state2} has expansion coefficients (in the basis of Virasoro characters) given by evaluating the gravitational path integral on the following 3-manifold:
\tikzset{every picture/.style={line width=0.75pt}}
\begin{equation}
    \braket{P,\bar P}{Z_\text{grav}(z,\bar z)} = Z_{\text{grav}}\left(\begin{tikzpicture}[x=0.75pt,y=0.75pt,yscale=-1.3,xscale=1.3,baseline={([yshift=-0.5ex]current bounding box.center)}]
\draw  [color={rgb, 255:red, 155; green, 155; blue, 155 }  ,draw opacity=1 ][dash pattern={on 1.5pt off 1.5pt on 1.5pt off 1.5pt}] (110.6,200.6) .. controls (110.6,192.32) and (128.51,185.6) .. (150.6,185.6) .. controls (172.69,185.6) and (190.6,192.32) .. (190.6,200.6) .. controls (190.6,208.88) and (172.69,215.6) .. (150.6,215.6) .. controls (128.51,215.6) and (110.6,208.88) .. (110.6,200.6) -- cycle ;
\draw  [line width=1.5]  (110,200.6) .. controls (110,178.18) and (128.18,160) .. (150.6,160) .. controls (173.02,160) and (191.2,178.18) .. (191.2,200.6) .. controls (191.2,223.02) and (173.02,241.2) .. (150.6,241.2) .. controls (128.18,241.2) and (110,223.02) .. (110,200.6) -- cycle ;
\draw [color={rgb, 255:red, 208; green, 2; blue, 27 }  ,draw opacity=1 ]   (132.29,180) .. controls (142.41,189.38) and (142.41,210.38) .. (132.29,220) ;
\draw [shift={(132.29,220)}, rotate = 181.45] [color={rgb, 255:red, 208; green, 2; blue, 27 }  ,draw opacity=1 ][line width=0.75]    (-2.24,0) -- (2.24,0)(0,2.24) -- (0,-2.24)   ;
\draw [shift={(132.29,180)}, rotate = 87.8] [color={rgb, 255:red, 208; green, 2; blue, 27 }  ,draw opacity=1 ][line width=0.75]    (-2.24,0) -- (2.24,0)(0,2.24) -- (0,-2.24)   ;
\draw [color={rgb, 255:red, 208; green, 2; blue, 27 }  ,draw opacity=1 ]   (168,180) .. controls (157.83,189.38) and (157.83,210.38) .. (168,220) ;
\draw [shift={(168,220)}, rotate = 88.43] [color={rgb, 255:red, 208; green, 2; blue, 27 }  ,draw opacity=1 ][line width=0.75]    (-2.24,0) -- (2.24,0)(0,2.24) -- (0,-2.24)   ;
\draw [shift={(168,180)}, rotate = 182.32] [color={rgb, 255:red, 208; green, 2; blue, 27 }  ,draw opacity=1 ][line width=0.75]    (-2.24,0) -- (2.24,0)(0,2.24) -- (0,-2.24)   ;
\draw  [draw opacity=0][line width=1.5]  (164.12,193.78) .. controls (168.02,195.48) and (170.46,197.89) .. (170.46,200.56) .. controls (170.46,205.69) and (161.44,209.86) .. (150.31,209.86) .. controls (139.19,209.86) and (130.17,205.69) .. (130.17,200.56) .. controls (130.17,197.76) and (132.85,195.24) .. (137.1,193.54) -- (150.31,200.56) -- cycle ; \draw  [line width=1.5]  (164.12,193.78) .. controls (168.02,195.48) and (170.46,197.89) .. (170.46,200.56) .. controls (170.46,205.69) and (161.44,209.86) .. (150.31,209.86) .. controls (139.19,209.86) and (130.17,205.69) .. (130.17,200.56) .. controls (130.17,197.76) and (132.85,195.24) .. (137.1,193.54) ;  
\draw  [draw opacity=0][line width=1.5]  (143.1,191.87) .. controls (145.34,191.47) and (147.77,191.26) .. (150.31,191.26) .. controls (152.86,191.26) and (155.29,191.47) .. (157.53,191.87) -- (150.31,200.56) -- cycle ; \draw  [line width=1.5]  (143.1,191.87) .. controls (145.34,191.47) and (147.77,191.26) .. (150.31,191.26) .. controls (152.86,191.26) and (155.29,191.47) .. (157.53,191.87) ;  
\draw (118.54,176.4) node [anchor=north west][inner sep=0.75pt]  [font=\footnotesize]  {$\clo_1$};
\draw (118.54,214.79) node [anchor=north west][inner sep=0.75pt]  [font=\footnotesize]  {$\clo_1$};
\draw (169.54,175.44) node [anchor=north west][inner sep=0.75pt]  [font=\footnotesize]  {$\clo_2$};
\draw (169.54,214.44) node [anchor=north west][inner sep=0.75pt]  [font=\footnotesize]  {$\clo_2$};
\draw (140.29,210.54) node [anchor=north west][inner sep=0.75pt]  [font=\small]  {$P,\bar P$};
\end{tikzpicture}\right).\label{eq:step1pathintegral}
\end{equation}
Here $P$ labels the holonomy of the Wilson loop wrapping the torus. The identification between this holonomy and the Liouville momentum $P$ follows from the quantization of the Teichm\"uller component of the moduli space of flat $SL(2,\mathbb{R})$ connections, in the first order formulation of 3d gravity \cite{Collier:2023fwi}. Similarly, $\bar P$ parametrizes the right-moving $SL(2,\mathbb{R})$ holonomy. The dependence on the cross-ratio's $z,\bar z$ is implicit in the insertion points of the operators on the boundary sphere. 

To evaluate the path integral in Eq.~\eqref{eq:step1pathintegral}, we use Virasoro TQFT for the left- and right-movers separately. We follow the same steps as \cite{Post:2024itb} to unlink the Wilson loop:
\begin{align}
&Z_{\text{Vir}}\!\left[\begin{tikzpicture}[x=0.75pt,y=0.75pt,yscale=-0.7,xscale=0.7,baseline={([yshift=-0.5ex]current bounding box.center)}]
\draw [color={rgb, 255:red, 208; green, 2; blue, 27 }  ,draw opacity=1 ]   (132.29,180) .. controls (142.41,189.38) and (142.41,210.38) .. (132.29,220) ;
\draw [shift={(132.29,220)}, rotate = 181.45] [color={rgb, 255:red, 208; green, 2; blue, 27 }  ,draw opacity=1 ][line width=0.75]    (-2.24,0) -- (2.24,0)(0,2.24) -- (0,-2.24)   ;
\draw [shift={(132.29,180)}, rotate = 87.8] [color={rgb, 255:red, 208; green, 2; blue, 27 }  ,draw opacity=1 ][line width=0.75]    (-2.24,0) -- (2.24,0)(0,2.24) -- (0,-2.24)   ;
\draw [color={rgb, 255:red, 208; green, 2; blue, 27 }  ,draw opacity=1 ]   (168,180) .. controls (157.83,189.38) and (157.83,210.38) .. (168,220) ;
\draw [shift={(168,220)}, rotate = 88.43] [color={rgb, 255:red, 208; green, 2; blue, 27 }  ,draw opacity=1 ][line width=0.75]    (-2.24,0) -- (2.24,0)(0,2.24) -- (0,-2.24)   ;
\draw [shift={(168,180)}, rotate = 182.32] [color={rgb, 255:red, 208; green, 2; blue, 27 }  ,draw opacity=1 ][line width=0.75]    (-2.24,0) -- (2.24,0)(0,2.24) -- (0,-2.24)   ;
\draw  [draw opacity=0][line width=1.5]  (164.12,193.78) .. controls (168.02,195.48) and (170.46,197.89) .. (170.46,200.56) .. controls (170.46,205.69) and (161.44,209.86) .. (150.31,209.86) .. controls (139.19,209.86) and (130.17,205.69) .. (130.17,200.56) .. controls (130.17,197.76) and (132.85,195.24) .. (137.1,193.54) -- (150.31,200.56) -- cycle ; \draw  [line width=1.5]  (164.12,193.78) .. controls (168.02,195.48) and (170.46,197.89) .. (170.46,200.56) .. controls (170.46,205.69) and (161.44,209.86) .. (150.31,209.86) .. controls (139.19,209.86) and (130.17,205.69) .. (130.17,200.56) .. controls (130.17,197.76) and (132.85,195.24) .. (137.1,193.54) ;  
\draw  [draw opacity=0][line width=1.5]  (143.1,191.87) .. controls (145.34,191.47) and (147.77,191.26) .. (150.31,191.26) .. controls (152.86,191.26) and (155.29,191.47) .. (157.53,191.87) -- (150.31,200.56) -- cycle ; \draw  [line width=1.5]  (143.1,191.87) .. controls (145.34,191.47) and (147.77,191.26) .. (150.31,191.26) .. controls (152.86,191.26) and (155.29,191.47) .. (157.53,191.87) ;  
\draw (110.25,176.4) node [anchor=north west][inner sep=0.75pt]  [font=\footnotesize]  {$\clo_1$};
\draw (110.54,214.79) node [anchor=north west][inner sep=0.75pt]  [font=\footnotesize]  {$\clo_1$};
\draw (169.54,175.44) node [anchor=north west][inner sep=0.75pt]  [font=\footnotesize]  {$\clo_2$};
\draw (169.29,214.44) node [anchor=north west][inner sep=0.75pt]  [font=\footnotesize]  {$\clo_2$};
\draw (143.29,213.54) node [anchor=north west][inner sep=0.75pt]  [font=\footnotesize]  {$P$};
\end{tikzpicture}\right]\nonumber \\[0.6em] &=\int_0^\infty\!\!\dd P_1 \, \fker{\bbi}{P_1}{\clo_2}{\clo_1}{\clo_1}{\clo_2} Z_{\text{Vir}}\!\left[\begin{tikzpicture}[x=0.75pt,y=0.75pt,yscale=-0.8,xscale=0.8,baseline={([yshift=-0.5ex]current bounding box.center)}]
\draw [color={rgb, 255:red, 208; green, 2; blue, 27 }  ,draw opacity=1 ]   (170.36,203.06) -- (170.27,238.06) ;
\draw  [draw opacity=0][line width=1.5]  (175.83,211.61) .. controls (184.27,212.72) and (190.46,216.3) .. (190.46,220.56) .. controls (190.46,225.69) and (181.44,229.86) .. (170.31,229.86) .. controls (159.19,229.86) and (150.17,225.69) .. (150.17,220.56) .. controls (150.17,216.11) and (156.95,212.38) .. (166,211.47) -- (170.31,220.56) -- cycle ; \draw  [line width=1.5]  (175.83,211.61) .. controls (184.27,212.72) and (190.46,216.3) .. (190.46,220.56) .. controls (190.46,225.69) and (181.44,229.86) .. (170.31,229.86) .. controls (159.19,229.86) and (150.17,225.69) .. (150.17,220.56) .. controls (150.17,216.11) and (156.95,212.38) .. (166,211.47) ;  
\draw [color={rgb, 255:red, 208; green, 2; blue, 27 }  ,draw opacity=1 ]   (152.29,200) .. controls (163.55,203.73) and (176.64,204.27) .. (188,200) ;
\draw [shift={(188,200)}, rotate = 24.39] [color={rgb, 255:red, 208; green, 2; blue, 27 }  ,draw opacity=1 ][line width=0.75]    (-2.24,0) -- (2.24,0)(0,2.24) -- (0,-2.24)   ;
\draw [shift={(152.29,200)}, rotate = 63.32] [color={rgb, 255:red, 208; green, 2; blue, 27 }  ,draw opacity=1 ][line width=0.75]    (-2.24,0) -- (2.24,0)(0,2.24) -- (0,-2.24)   ;
\draw [color={rgb, 255:red, 208; green, 2; blue, 27 }  ,draw opacity=1 ]   (152.29,240) .. controls (165.73,236.27) and (178.82,238.09) .. (188,240) ;
\draw [shift={(188,240)}, rotate = 56.75] [color={rgb, 255:red, 208; green, 2; blue, 27 }  ,draw opacity=1 ][line width=0.75]    (-2.24,0) -- (2.24,0)(0,2.24) -- (0,-2.24)   ;
\draw [shift={(152.29,240)}, rotate = 29.5] [color={rgb, 255:red, 208; green, 2; blue, 27 }  ,draw opacity=1 ][line width=0.75]    (-2.24,0) -- (2.24,0)(0,2.24) -- (0,-2.24)   ;
\draw (130.58,196.07) node [anchor=north west][inner sep=0.75pt]  [font=\footnotesize]  {$\clo_1$};
\draw (130.54,234.46) node [anchor=north west][inner sep=0.75pt]  [font=\footnotesize]  {$\clo_1$};
\draw (189.54,196.07) node [anchor=north west][inner sep=0.75pt]  [font=\footnotesize]  {$\clo_2$};
\draw (189.29,234.44) node [anchor=north west][inner sep=0.75pt]  [font=\footnotesize]  {$\clo_2$};
\draw (191.29,215.18) node [anchor=north west][inner sep=0.75pt]  [font=\footnotesize]  {$P$};
\draw (153.73,214.66) node [anchor=north west][inner sep=0.75pt]  [font=\scriptsize]  {$P_{1}$};
\end{tikzpicture}\right] \\[1em]
&= \int_0^\infty \!\!\dd P_1 \, \fker{\bbi}{P_1}{\clo_2}{\clo_1}{\clo_1}{\clo_2} \frac{\sker{P}{P_1}{\bbi}}{\sker{\bbi}{P_1}{\bbi}}\,Z_{\text{Vir}}\!\left[\begin{tikzpicture}[x=0.75pt,y=0.75pt,yscale=-0.8,xscale=0.8,baseline={([yshift=-0.5ex]current bounding box.center)}]
\draw [color={rgb, 255:red, 208; green, 2; blue, 27 }  ,draw opacity=1 ]   (190.36,223.06) -- (190.27,258.06) ;
\draw [color={rgb, 255:red, 208; green, 2; blue, 27 }  ,draw opacity=1 ]   (172.29,220) .. controls (183.55,223.73) and (196.64,224.27) .. (208,220) ;
\draw [shift={(208,220)}, rotate = 24.39] [color={rgb, 255:red, 208; green, 2; blue, 27 }  ,draw opacity=1 ][line width=0.75]    (-2.24,0) -- (2.24,0)(0,2.24) -- (0,-2.24)   ;
\draw [shift={(172.29,220)}, rotate = 63.32] [color={rgb, 255:red, 208; green, 2; blue, 27 }  ,draw opacity=1 ][line width=0.75]    (-2.24,0) -- (2.24,0)(0,2.24) -- (0,-2.24)   ;
\draw [color={rgb, 255:red, 208; green, 2; blue, 27 }  ,draw opacity=1 ]   (172.29,260) .. controls (185.73,256.27) and (198.82,258.09) .. (208,260) ;
\draw [shift={(208,260)}, rotate = 56.75] [color={rgb, 255:red, 208; green, 2; blue, 27 }  ,draw opacity=1 ][line width=0.75]    (-2.24,0) -- (2.24,0)(0,2.24) -- (0,-2.24)   ;
\draw [shift={(172.29,260)}, rotate = 29.5] [color={rgb, 255:red, 208; green, 2; blue, 27 }  ,draw opacity=1 ][line width=0.75]    (-2.24,0) -- (2.24,0)(0,2.24) -- (0,-2.24)   ;
\draw (150.92,216.73) node [anchor=north west][inner sep=0.75pt]  [font=\footnotesize]  {$\clo_1$};
\draw (150.92,250.46) node [anchor=north west][inner sep=0.75pt]  [font=\footnotesize]  {$\clo_1$};
\draw (209.54,216.73) node [anchor=north west][inner sep=0.75pt]  [font=\footnotesize]  {$\clo_2$};
\draw (209.29,250.44) node [anchor=north west][inner sep=0.75pt]  [font=\footnotesize]  {$\clo_2$};
\draw (170.73,234.66) node [anchor=north west][inner sep=0.75pt]  [font=\footnotesize]  {$P_{1}$};
\end{tikzpicture}\right].\label{eq:tblock}
\end{align}
In the first equality, we inserted an identity line and used the Virasoro fusion kernel to go to the $t$-channel. In the second equality, we used the ``Verlinde loop'' operator to unlink the Wilson loop \cite{Post:2024itb}:
\begin{equation}\label{eq:TQFTrule}
    \begin{tikzpicture}[x=0.75pt,y=0.75pt,yscale=-1.3,xscale=1.3,baseline={([yshift=-.5ex]current bounding box.center)}] 
\draw  [line width=1.2] (100.1,105) -- (119.11,105) ;
\draw  [line width=1.2] [draw opacity=0] (146.06,109.65) .. controls (145.08,121.23) and (140.5,130) .. (135,130) .. controls (128.79,130) and (123.75,118.81) .. (123.75,105) .. controls (123.75,91.19) and (128.79,80) .. (135,80) .. controls (140.68,80) and (145.38,89.35) .. (146.14,101.5) -- (135,105) -- cycle ; \draw [line width=1.2]  (146.06,109.65) .. controls (145.08,121.23) and (140.5,130) .. (135,130) .. controls (128.79,130) and (123.75,118.81) .. (123.75,105) .. controls (123.75,91.19) and (128.79,80) .. (135,80) .. controls (140.68,80) and (145.38,89.35) .. (146.14,101.5) ;  
\draw  [line width=1.2]  (128.11,105) -- (168.75,105) ;
\draw (155,93.4) node [anchor=north west][inner sep=0.75pt]  [font=\small]  {$P_1$};
\draw (115,82.4) node [anchor=north west][inner sep=0.75pt]  [font=\small]  {$P$};
\end{tikzpicture} \;=\, \frac{\mathbb{S}_{PP_1}[\bbi]}{\mathbb{S}_{\bbi P_1}[\bbi]}\,\,
\begin{tikzpicture}[x=0.75pt,y=0.75pt,yscale=-1.2,xscale=1.2,baseline={([yshift=-1.7ex]current bounding box.center)}]
\draw  [line width=1.2]  (99.9,105) -- (168.75,105) ;
\draw (126.2,92.2) node [anchor=north west][inner sep=0.75pt]  [font=\small]  {$P_1$};
\end{tikzpicture}
\end{equation}
where $\sker{\bbi}{P_1}{\bbi} = \rho_0(P_1)$ and $\sker{P}{P_1}{\bbi}$ are Virasoro modular S-kernels (see Appendix \ref{app:kernels}). After unlinking the Wilson loop, the VTQFT partition function in Eq.~\eqref{eq:tblock} evaluates to the 4-point conformal block in the $t$-channel: 
\begin{equation}
    Z_{\text{Vir}}\!\left[\begin{tikzpicture}[x=0.75pt,y=0.75pt,yscale=-0.8,xscale=0.8,baseline={([yshift=-0.5ex]current bounding box.center)}]
\draw [color={rgb, 255:red, 208; green, 2; blue, 27 }  ,draw opacity=1 ]   (190.36,223.06) -- (190.27,258.06) ;
\draw [color={rgb, 255:red, 208; green, 2; blue, 27 }  ,draw opacity=1 ]   (172.29,220) .. controls (183.55,223.73) and (196.64,224.27) .. (208,220) ;
\draw [shift={(208,220)}, rotate = 24.39] [color={rgb, 255:red, 208; green, 2; blue, 27 }  ,draw opacity=1 ][line width=0.75]    (-2.24,0) -- (2.24,0)(0,2.24) -- (0,-2.24)   ;
\draw [shift={(172.29,220)}, rotate = 63.32] [color={rgb, 255:red, 208; green, 2; blue, 27 }  ,draw opacity=1 ][line width=0.75]    (-2.24,0) -- (2.24,0)(0,2.24) -- (0,-2.24)   ;
\draw [color={rgb, 255:red, 208; green, 2; blue, 27 }  ,draw opacity=1 ]   (172.29,260) .. controls (185.73,256.27) and (198.82,258.09) .. (208,260) ;
\draw [shift={(208,260)}, rotate = 56.75] [color={rgb, 255:red, 208; green, 2; blue, 27 }  ,draw opacity=1 ][line width=0.75]    (-2.24,0) -- (2.24,0)(0,2.24) -- (0,-2.24)   ;
\draw [shift={(172.29,260)}, rotate = 29.5] [color={rgb, 255:red, 208; green, 2; blue, 27 }  ,draw opacity=1 ][line width=0.75]    (-2.24,0) -- (2.24,0)(0,2.24) -- (0,-2.24)   ;
\draw (150.92,216.73) node [anchor=north west][inner sep=0.75pt]  [font=\footnotesize]  {$\clo_1$};
\draw (150.92,250.46) node [anchor=north west][inner sep=0.75pt]  [font=\footnotesize]  {$\clo_1$};
\draw (209.54,216.73) node [anchor=north west][inner sep=0.75pt]  [font=\footnotesize]  {$\clo_2$};
\draw (209.29,250.44) node [anchor=north west][inner sep=0.75pt]  [font=\footnotesize]  {$\clo_2$};
\draw (170.73,234.66) node [anchor=north west][inner sep=0.75pt]  [font=\footnotesize]  {$P_{1}$};
\end{tikzpicture}\right] = \mathcal{F}_t(P_1;z)
\end{equation}
with an exchange of a heavy operator of weight $h_1 = \frac{c-1}{24}+P_1^2$. So Eq.~\eqref{eq:tblock} is a superposition of 4-point blocks, and the gravitational partition function in Eq.~\eqref{eq:step1pathintegral} is the product with the anti-holomorphic counterpart:
\begin{multline}
    \braket{P,\bar P}{Z_\text{grav}(z,\bar z)} = \\ \left|\int_0^\infty \!\dd P_1 \,C_0(P_1,P_{\clo_1},P_{\clo_2}) \mathbb{S}_{PP_1}[\bbi] \mathcal{F}_t(P_1;z)\right|^2\,.
\end{multline}
Here we used the relation between the universal OPE density $C_0$ and the Virasoro fusion kernel \cite{Collier:2019weq}:
\begin{equation}
   \frac{1}{\sker{\bbi}{P_1}{\bbi}} \fker{\bbi}{P_1}{\clo_2}{\clo_1}{\clo_1}{\clo_2} = C_0(P_{\clo_1},P_{\clo_2},P_1).
\end{equation}

Now we proceed to Step 2 of the RMT surgery shown in Fig.~\ref{fig:RMT_surgery}. That is, we take two copies of the state prepared in Eq.~\eqref{eq:state2}, and glue the torus boundaries with opposite orientation using the gluing map $G_\text{T} : \mathcal{H}_\text{T}\to \mathcal{H}_\text{T}$. In the Hilbert space formalism of the path integral, this means that we should compute the expectation value
\begin{equation}\label{eq:Zwormhole}
    Z_\text{grav}[\mathcal{W}] = \mel{Z_\text{grav}(z,\bar z)}{\,G_\text{T}\,}{Z_\text{grav}(z',\bar z')}.
\end{equation}
Inserting the expansion in torus blocks, Eq.~\eqref{eq:state2}, this step amounts to computing the matrix element
\begin{equation}\label{eq:microcan}
    \mel{P,\bar P}{\hspace{0.1mm}G_\text{T}}{P',\bar P'}.
\end{equation}
We can interpret this matrix element as representing a `microcanonical' version of the AdS$_3$ double trumpet computed by Cotler and Jensen, where we fix the holonomies of two Wilson loops on the boundary tori of $T\times I$, instead of complex structures $\tau,\tau'$. Pictorially,
\begin{equation}\label{eq:microcan2}
 \begin{tikzpicture}[x=0.75pt,y=0.75pt,yscale=-0.7,xscale=0.7,baseline={([yshift=-0.5ex]current bounding box.center)}]
\draw  [line width=1.2]  (70,125) .. controls (70,100.15) and (83.43,80) .. (100,80) .. controls (116.57,80) and (130,100.15) .. (130,125) .. controls (130,149.85) and (116.57,170) .. (100,170) .. controls (83.43,170) and (70,149.85) .. (70,125) -- cycle ;
\draw [line width=1.2]    (95.6,110.63) .. controls (107.6,113.63) and (109.2,135.63) .. (96,140.83) ;
\draw [line width=1.2]    (100,113.08) .. controls (93.88,114.96) and (93.63,134.71) .. (99.88,137.46) ;
\draw  [line width=1.2]  (250,125) .. controls (250,100.15) and (236.57,80) .. (220,80) .. controls (203.43,80) and (190,100.15) .. (190,125) .. controls (190,149.85) and (203.43,170) .. (220,170) .. controls (236.57,170) and (250,149.85) .. (250,125) -- cycle ;
\draw [line width=1.2]    (223.83,110.17) .. controls (211.83,113.17) and (210.63,134.63) .. (223.83,139.83) ;
\draw [line width=1.2]    (220,113.08) .. controls (226.13,114.96) and (226.38,134.71) .. (220.13,137.46) ;
\draw [line width=1.2]    (107.45,81.36) .. controls (140.55,94.64) and (180.36,94.45) .. (213.45,81) ;
\draw [line width=1.2]    (105.82,169.18) .. controls (141.82,154) and (177.64,154.82) .. (214.55,169.36) ;
\draw [color={rgb, 255:red, 208; green, 2; blue, 27 }]  (84.22,125) .. controls (84.22,108.55) and (91.29,95.22) .. (100,95.22) .. controls (108.71,95.22) and (115.78,108.55) .. (115.78,125) .. controls (115.78,141.45) and (108.71,154.78) .. (100,154.78) .. controls (91.29,154.78) and (84.22,141.45) .. (84.22,125) -- cycle ;
\draw  [color={rgb, 255:red, 208; green, 2; blue, 27 }] (204.22,125) .. controls (204.22,108.55) and (211.29,95.22) .. (220,95.22) .. controls (228.71,95.22) and (235.78,108.55) .. (235.78,125) .. controls (235.78,141.45) and (228.71,154.78) .. (220,154.78) .. controls (211.29,154.78) and (204.22,141.45) .. (204.22,125) -- cycle ;
\draw (29.09,115.87) node [anchor=north west][inner sep=0.75pt]  [font=\small]  {$P,\bar{P}$};
\draw (252.04,113.24) node [anchor=north west][inner sep=0.75pt]  [font=\small]  {$P'\!,\bar{P}'$};
\end{tikzpicture}.
\end{equation}

In any TQFT, the topologically trivial manifold $T\times I$ evaluates to the identity map. However, 3d gravity is \emph{not} a TQFT, because we still have to implement the quotient by the bulk mapping class group, recall Eq.~\eqref{eq:MCG}. When implementing Step 1 in the path integral, we were able to use Virasoro TQFT because the bulk MCG acts trivially on the building block in Eq.~\eqref{eq:step1pathintegral}. However, in Step 2 the bulk MCG acts non-trivially on the torus wormhole $T\times I$ that we are `gluing in', so we cannot use Virasoro TQFT to compute the matrix element \eqref{eq:microcan}. 

Instead, we make the following observation. Since the bulk MCG acts in the same way as on the AdS$_3$ double trumpet of Cotler and Jensen, with asymptotically AdS$_3$ boundary conditions labeled by complex structure moduli $\tau,\bar \tau$ and $\tau',\bar\tau'$ on the boundary tori, we can simply borrow the result of \cite{Cotler:2020ugk} and transform to the microcanonical ensemble:
\begin{multline}\label{eq:canonical}
    Z_{T\times I}(\tau,\bar\tau;\tau',\bar\tau') = \mel{\tau,\bar\tau}{\,G_\text{T}\,}{\tau',\bar\tau'} \\[1em] =\int_0^\infty \!\dd^4 P\,\mel{P,\bar P}{\hspace{0.1mm}G_\text{T}}{P',\bar P'} |\chi_{P}(\tau)|^2|\chi_{P'}(\tau')|^2.
\end{multline}
In the second line we have inserted resolutions of the identity on the torus Hilbert space and used the relation between the wavefunctions $\braket{\tau}{P}$ and the Virasoro characters. This lets us read off the microcanonical wormhole amplitude Eq.~\eqref{eq:microcan} from the known result of the canonical torus wormhole. Here we will assume that the path integral computation of Cotler and Jensen \cite{Cotler:2020ugk}, including the treatment of the integration measure and zero-mode factors, is correct. Since their result has passed some stringent consistency checks \cite{Cotler:2020hgz, Haehl:2023tkr, Haehl:2023xys, Haehl:2023mhf,DiUbaldo:2023qli}, we believe that this assumption is well justified.

In Appendix \ref{app:doubletrumpet}, we review the AdS$_3$ double trumpet partition function, with a discussion of the bulk MCG, and show that the integrand in Eq.~\eqref{eq:canonical} takes the form
\begin{equation}\label{eq:mel}
    \mel{P,\bar P}{\hspace{0.1mm}G_\text{T}}{P',\bar P'} = \rho_{0,2}(P,\bar P,P',\bar P'),
\end{equation}
where $\rho_{0,2}$ is precisely the random matrix theory kernel appearing in Eq.~\eqref{eq:bdy_prediction} (as was already noted in \cite{Cotler:2020hgz}). We can insert this result into Eq.~\eqref{eq:Zwormhole}, together with our previous answer for the overlap in Eq.~\eqref{eq:step1pathintegral}, to obtain the four-punctured sphere wormhole partition function $Z_\text{grav}[\mathcal{W}]$ that we sought to compute. 

In order to match to our boundary prediction in Eq.~\eqref{eq:bdy_prediction}, we still need to realize an important property satisfied by the density $\rho_{0,2}$. Namely, the `seed' partition function of Cotler and Jensen is invariant under simultaneous modular transformations \cite{Cotler:2020ugk}, 
\begin{equation}\label{eq:Sinvariance}
    Z_{T\times I}(S\tau, S\bar\tau; S\tau',S\bar\tau') =  Z_{T\times I}(\tau, \bar\tau;\tau'\bar\tau').
\end{equation}
This can be translated to a statement about the microcanonical torus wormhole, if we plug in Eq.~\eqref{eq:canonical} and use the modular transformation rules of the Virasoro characters, by modular S-kernels (described in App.~\ref{app:kernels}). Doing so, we obtain the following identity for the microcanonical partition function, equivalent to Eq.~\eqref{eq:Sinvariance}:
\begin{widetext}
\begin{equation}
    \int_0^\infty \dd P \dd \bar P \dd P'\dd \bar P' \,\sker{P_1}{P}{\bbi}\sker{\bar P_1}{\bar P}{\bbi} \sker{P_2}{P'}{\bbi} \sker{\bar P_2}{\bar P'}{\bbi} \,\rho_{0,2}(P,\bar P, P',\bar P')= \rho_{0,2}(P_1,\bar P_1, P_2,\bar P_2).
\end{equation}
\end{widetext}
This identity, which expresses the topological fact that a boundary modular S-transformation can be deformed through the bulk to the other torus boundary, allows us to perform the integrals over $P,\bar P$ and $P',\bar P'$ in our expression for $Z_\text{grav}[\mathcal{W}]$ in Eq.~\eqref{eq:Zwormhole}.

Putting everything together, we find an exact match between the wormhole partition function for $\mathcal{W}$ and the boundary prediction (on the right-hand side of Eq.~\eqref{eq:bdy_prediction}) for the variance of the sphere four-point function $\Delta \mathcal{G}^2$ that arises from the random matrix statistics of the primary spectrum:
\begin{equation}
    Z_\text{grav}[\mathcal{W};z,\bar z,z',\bar z'] = \overline{\mathcal{G}(z,\bar z) \mathcal{G}(z',\bar z') } \mid_{\text{RMT}}.\vspace{1mm}
\end{equation}

This is the main result of this section. It has been derived from the cutting-and-gluing rules of the gravitational path integral, combining the technology of the Virasoro TQFT with the off-shell calculation of the torus wormhole. Our method is exact to all orders in $1/c$ and does not rely on a saddle-point approximation. 

We stress the close parallel between the RMT surgery computation of the four-point wormhole $\mathcal{W}$ and the ETH surgery described in the previous section. Both started with a disconnected product of handlebodies/punctured balls, forming a connected topology by drilling out handles and gluing along the boundaries of the handles. The main new ingredient was a prescription for how to quotient the bulk mapping class group, for which we used a microcanonical version of the torus wormhole computation of Cotler and Jensen.

The off-shell partition function that we found can be compared to the \emph{on-shell} partition function of the $\Sigma_{0,4}\times I$ wormhole, shown in Fig.~\ref{fig:ETHvsRMT}(a), which accounts for the ETH-like Wick contraction of the OPE coefficients in Eq.~\eqref{eq:moment}. Since this topology is on-shell, we can use Virasoro TQFT to write down its partition function:
\begin{align}\label{eq:OPEavg}
 &\overline{\mathcal{G}(z,\bar z) \mathcal{G}(z',\bar z') } \mid_{\text{ETH}} \,= \int_0^\infty \dd P \dd \bar P \Big[\rho_0(P,\bar P)
\\ &\big|C_0(P_{\clo_1},P_{\clo_2},P)C_0(P_{\clo_1},P_{\clo_2},P)\mathcal{F}_t(P;z)\mathcal{F}_t(P;z')\big|^2\Big].\nonumber
\end{align}
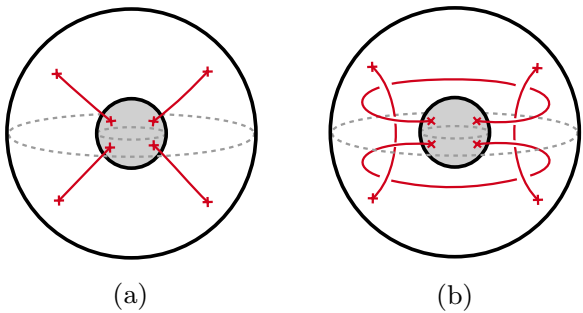
\begin{figure}
    \centering
    \resizebox{0.9\linewidth}{!}{
    \begin{tikzpicture}[x=0.75pt,y=0.75pt,yscale=-1,xscale=1,baseline={([yshift=-0.5ex]current bounding box.center)}]
\draw  [draw opacity=0][dash pattern={on 1.5pt off 1.5pt on 1.5pt off 1.5pt}] (273.06,107.17) .. controls (295.13,108.04) and (311.91,111.09) .. (314.11,114.83) -- (260.17,115.67) -- cycle ; \draw  [color={rgb, 255:red, 155; green, 155; blue, 155 }  ,draw opacity=1 ][dash pattern={on 1.5pt off 1.5pt on 1.5pt off 1.5pt}] (273.06,107.17) .. controls (295.13,108.04) and (311.91,111.09) .. (314.11,114.83) ;  
\draw  [draw opacity=0][dash pattern={on 1.5pt off 1.5pt on 1.5pt off 1.5pt}] (206.45,114.5) .. controls (209.41,110.91) and (225.88,108) .. (247.34,107.16) -- (260.17,115.67) -- cycle ; \draw  [color={rgb, 255:red, 155; green, 155; blue, 155 }  ,draw opacity=1 ][dash pattern={on 1.5pt off 1.5pt on 1.5pt off 1.5pt}] (206.45,114.5) .. controls (209.41,110.91) and (225.88,108) .. (247.34,107.16) ;  
\draw  [color={rgb, 255:red, 155; green, 155; blue, 155 }  ,draw opacity=1 ][dash pattern={on 1.5pt off 1.5pt on 1.5pt off 1.5pt}] (244.81,115.67) .. controls (244.81,114.16) and (251.69,112.93) .. (260.17,112.93) .. controls (268.65,112.93) and (275.52,114.16) .. (275.52,115.67) .. controls (275.52,117.18) and (268.65,118.4) .. (260.17,118.4) .. controls (251.69,118.4) and (244.81,117.18) .. (244.81,115.67) -- cycle ;
\draw  [fill={rgb, 255:red, 155; green, 155; blue, 155 }  ,fill opacity=0.47 ][line width=1.2]  (244.81,115.67) .. controls (244.81,107.19) and (251.69,100.31) .. (260.17,100.31) .. controls (268.65,100.31) and (275.52,107.19) .. (275.52,115.67) .. controls (275.52,124.15) and (268.65,131.02) .. (260.17,131.02) .. controls (251.69,131.02) and (244.81,124.15) .. (244.81,115.67) -- cycle ;
\draw [color={rgb, 255:red, 208; green, 2; blue, 27 }  ,draw opacity=1 ]   (296.63,87.27) .. controls (282.86,99.97) and (283.13,132.49) .. (296.9,145.53) ;
\draw [shift={(296.9,145.53)}, rotate = 88.43] [color={rgb, 255:red, 208; green, 2; blue, 27 }  ,draw opacity=1 ][line width=0.75]    (-2.24,0) -- (2.24,0)(0,2.24) -- (0,-2.24)   ;
\draw [shift={(296.63,87.27)}, rotate = 182.32] [color={rgb, 255:red, 208; green, 2; blue, 27 }  ,draw opacity=1 ][line width=0.75]    (-2.24,0) -- (2.24,0)(0,2.24) -- (0,-2.24)   ;
\draw [color={rgb, 255:red, 255; green, 255; blue, 255 }  ,draw opacity=1 ][line width=2.25]    (284.9,120.53) -- (288.63,120.87) ;
\draw [color={rgb, 255:red, 208; green, 2; blue, 27 }  ,draw opacity=1 ]   (270.03,120.92) .. controls (300.41,117.83) and (311.12,129.39) .. (293.48,134.81) ;
\draw [shift={(270.03,120.92)}, rotate = 39.18] [color={rgb, 255:red, 208; green, 2; blue, 27 }  ,draw opacity=1 ][line width=0.75]    (-2.24,0) -- (2.24,0)(0,2.24) -- (0,-2.24)   ;
\draw [color={rgb, 255:red, 208; green, 2; blue, 27 }  ,draw opacity=1 ]   (223.7,86.73) .. controls (237.42,99.43) and (237.68,131.83) .. (223.97,144.87) ;
\draw [shift={(223.97,144.87)}, rotate = 181.45] [color={rgb, 255:red, 208; green, 2; blue, 27 }  ,draw opacity=1 ][line width=0.75]    (-2.24,0) -- (2.24,0)(0,2.24) -- (0,-2.24)   ;
\draw [shift={(223.7,86.73)}, rotate = 87.8] [color={rgb, 255:red, 208; green, 2; blue, 27 }  ,draw opacity=1 ][line width=0.75]    (-2.24,0) -- (2.24,0)(0,2.24) -- (0,-2.24)   ;
\draw [color={rgb, 255:red, 255; green, 255; blue, 255 }  ,draw opacity=1 ][line width=1.5]    (232.77,121.13) -- (236.1,121.27) ;
\draw [color={rgb, 255:red, 208; green, 2; blue, 27 }  ,draw opacity=1 ]   (249.87,120.9) .. controls (244.31,120.54) and (239.49,120.64) .. (235.43,121.07) .. controls (217.61,122.94) and (214.45,131.15) .. (227.01,135.17) ;
\draw [shift={(249.87,120.9)}, rotate = 228.66] [color={rgb, 255:red, 208; green, 2; blue, 27 }  ,draw opacity=1 ][line width=0.75]    (-2.24,0) -- (2.24,0)(0,2.24) -- (0,-2.24)   ;
\draw  [draw opacity=0][dash pattern={on 1.5pt off 1.5pt on 1.5pt off 1.5pt}] (314.19,115.24) .. controls (314.3,115.5) and (314.35,115.76) .. (314.35,116.02) .. controls (314.35,121.51) and (290.09,125.97) .. (260.17,125.97) .. controls (230.39,125.97) and (206.23,121.56) .. (205.98,116.1) -- (260.17,116.02) -- cycle ; \draw  [color={rgb, 255:red, 155; green, 155; blue, 155 }  ,draw opacity=1 ][dash pattern={on 1.5pt off 1.5pt on 1.5pt off 1.5pt}] (314.19,115.24) .. controls (314.3,115.5) and (314.35,115.76) .. (314.35,116.02) .. controls (314.35,121.51) and (290.09,125.97) .. (260.17,125.97) .. controls (230.39,125.97) and (206.23,121.56) .. (205.98,116.1) ;  
\draw  [line width=1.2]  (205.17,115.67) .. controls (205.17,85.29) and (229.79,60.67) .. (260.17,60.67) .. controls (290.54,60.67) and (315.17,85.29) .. (315.17,115.67) .. controls (315.17,146.04) and (290.54,170.67) .. (260.17,170.67) .. controls (229.79,170.67) and (205.17,146.04) .. (205.17,115.67) -- cycle ;
\draw [color={rgb, 255:red, 208; green, 2; blue, 27 }  ,draw opacity=1 ]   (232.31,94.21) .. controls (244.5,91.83) and (274.5,91.17) .. (287.05,94.86) ;
\draw [color={rgb, 255:red, 255; green, 255; blue, 255 }  ,draw opacity=1 ][line width=2.25]    (232.64,110.19) -- (236.28,110.56) ;
\draw [color={rgb, 255:red, 208; green, 2; blue, 27 }  ,draw opacity=1 ]   (249.74,110.69) .. controls (244.76,111.02) and (240.38,110.98) .. (236.61,110.65) .. controls (236.17,110.61) and (235.73,110.56) .. (235.3,110.52) .. controls (217.48,108.57) and (214.31,100.02) .. (226.88,95.83) ;
\draw [shift={(249.74,110.69)}, rotate = 221.19] [color={rgb, 255:red, 208; green, 2; blue, 27 }  ,draw opacity=1 ][line width=0.75]    (-2.24,0) -- (2.24,0)(0,2.24) -- (0,-2.24)   ;
\draw [color={rgb, 255:red, 255; green, 255; blue, 255 }  ,draw opacity=1 ][line width=2.25]    (284.5,110.72) -- (288.14,111.09) ;
\draw [color={rgb, 255:red, 208; green, 2; blue, 27 }  ,draw opacity=1 ]   (269.89,110.67) .. controls (300.28,113.89) and (310.98,101.85) .. (293.35,96.21) ;
\draw [shift={(269.89,110.67)}, rotate = 51.05] [color={rgb, 255:red, 208; green, 2; blue, 27 }  ,draw opacity=1 ][line width=0.75]    (-2.24,0) -- (2.24,0)(0,2.24) -- (0,-2.24)   ;
\draw [color={rgb, 255:red, 208; green, 2; blue, 27 }  ,draw opacity=1 ]   (232.1,137) .. controls (251.05,141.1) and (268.44,140.82) .. (287.59,136.43) ;
\draw  [draw opacity=0][dash pattern={on 1.5pt off 1.5pt on 1.5pt off 1.5pt}] (130.23,107.67) .. controls (152.3,108.54) and (169.07,111.59) .. (171.28,115.33) -- (117.33,116.17) -- cycle ; \draw  [color={rgb, 255:red, 155; green, 155; blue, 155 }  ,draw opacity=1 ][dash pattern={on 1.5pt off 1.5pt on 1.5pt off 1.5pt}] (130.23,107.67) .. controls (152.3,108.54) and (169.07,111.59) .. (171.28,115.33) ;  
\draw  [draw opacity=0][dash pattern={on 1.5pt off 1.5pt on 1.5pt off 1.5pt}] (63.62,115) .. controls (66.58,111.41) and (83.04,108.5) .. (104.51,107.66) -- (117.33,116.17) -- cycle ; \draw  [color={rgb, 255:red, 155; green, 155; blue, 155 }  ,draw opacity=1 ][dash pattern={on 1.5pt off 1.5pt on 1.5pt off 1.5pt}] (63.62,115) .. controls (66.58,111.41) and (83.04,108.5) .. (104.51,107.66) ;  
\draw  [color={rgb, 255:red, 155; green, 155; blue, 155 }  ,draw opacity=1 ][dash pattern={on 1.5pt off 1.5pt on 1.5pt off 1.5pt}] (101.98,116.17) .. controls (101.98,114.66) and (108.85,113.43) .. (117.33,113.43) .. controls (125.81,113.43) and (132.69,114.66) .. (132.69,116.17) .. controls (132.69,117.68) and (125.81,118.9) .. (117.33,118.9) .. controls (108.85,118.9) and (101.98,117.68) .. (101.98,116.17) -- cycle ;
\draw  [fill={rgb, 255:red, 155; green, 155; blue, 155 }  ,fill opacity=0.47 ][line width=1.2]  (101.98,116.17) .. controls (101.98,107.69) and (108.85,100.81) .. (117.33,100.81) .. controls (125.81,100.81) and (132.69,107.69) .. (132.69,116.17) .. controls (132.69,124.65) and (125.81,131.52) .. (117.33,131.52) .. controls (108.85,131.52) and (101.98,124.65) .. (101.98,116.17) -- cycle ;
\draw [color={rgb, 255:red, 208; green, 2; blue, 27 }  ,draw opacity=1 ]   (151,88.83) .. controls (143.87,96.23) and (138.67,101.83) .. (127.06,110.83) ;
\draw [shift={(127.06,110.83)}, rotate = 187.21] [color={rgb, 255:red, 208; green, 2; blue, 27 }  ,draw opacity=1 ][line width=0.75]    (-2.24,0) -- (2.24,0)(0,2.24) -- (0,-2.24)   ;
\draw [shift={(151,88.83)}, rotate = 178.95] [color={rgb, 255:red, 208; green, 2; blue, 27 }  ,draw opacity=1 ][line width=0.75]    (-2.24,0) -- (2.24,0)(0,2.24) -- (0,-2.24)   ;
\draw [color={rgb, 255:red, 255; green, 255; blue, 255 }  ,draw opacity=1 ][line width=2.25]    (142.07,121.03) -- (145.8,121.37) ;
\draw [color={rgb, 255:red, 208; green, 2; blue, 27 }  ,draw opacity=1 ]   (84.33,89.83) .. controls (92.67,97.5) and (99,103.17) .. (108.33,110.75) ;
\draw [shift={(108.33,110.75)}, rotate = 84.09] [color={rgb, 255:red, 208; green, 2; blue, 27 }  ,draw opacity=1 ][line width=0.75]    (-2.24,0) -- (2.24,0)(0,2.24) -- (0,-2.24)   ;
\draw [shift={(84.33,89.83)}, rotate = 87.61] [color={rgb, 255:red, 208; green, 2; blue, 27 }  ,draw opacity=1 ][line width=0.75]    (-2.24,0) -- (2.24,0)(0,2.24) -- (0,-2.24)   ;
\draw  [draw opacity=0][dash pattern={on 1.5pt off 1.5pt on 1.5pt off 1.5pt}] (171.36,115.74) .. controls (171.47,116) and (171.52,116.26) .. (171.52,116.52) .. controls (171.52,122.01) and (147.26,126.47) .. (117.33,126.47) .. controls (87.56,126.47) and (63.4,122.06) .. (63.15,116.6) -- (117.33,116.52) -- cycle ; \draw  [color={rgb, 255:red, 155; green, 155; blue, 155 }  ,draw opacity=1 ][dash pattern={on 1.5pt off 1.5pt on 1.5pt off 1.5pt}] (171.36,115.74) .. controls (171.47,116) and (171.52,116.26) .. (171.52,116.52) .. controls (171.52,122.01) and (147.26,126.47) .. (117.33,126.47) .. controls (87.56,126.47) and (63.4,122.06) .. (63.15,116.6) ;  
\draw  [line width=1.2]  (62.33,116.17) .. controls (62.33,85.79) and (86.96,61.17) .. (117.33,61.17) .. controls (147.71,61.17) and (172.33,85.79) .. (172.33,116.17) .. controls (172.33,146.54) and (147.71,171.17) .. (117.33,171.17) .. controls (86.96,171.17) and (62.33,146.54) .. (62.33,116.17) -- cycle ;
\draw [color={rgb, 255:red, 255; green, 255; blue, 255 }  ,draw opacity=1 ][line width=2.25]    (141.67,111.22) -- (145.31,111.59) ;
\draw [color={rgb, 255:red, 208; green, 2; blue, 27 }  ,draw opacity=1 ]   (151,146.5) .. controls (143.83,139.75) and (135.67,129.5) .. (127.19,121.42) ;
\draw [shift={(127.19,121.42)}, rotate = 268.63] [color={rgb, 255:red, 208; green, 2; blue, 27 }  ,draw opacity=1 ][line width=0.75]    (-2.24,0) -- (2.24,0)(0,2.24) -- (0,-2.24)   ;
\draw [shift={(151,146.5)}, rotate = 268.29] [color={rgb, 255:red, 208; green, 2; blue, 27 }  ,draw opacity=1 ][line width=0.75]    (-2.24,0) -- (2.24,0)(0,2.24) -- (0,-2.24)   ;
\draw [color={rgb, 255:red, 208; green, 2; blue, 27 }  ,draw opacity=1 ]   (85.33,145.83) .. controls (94.67,136.17) and (100.33,130.17) .. (107.86,122.42) ;
\draw [shift={(107.86,122.42)}, rotate = 359.19] [color={rgb, 255:red, 208; green, 2; blue, 27 }  ,draw opacity=1 ][line width=0.75]    (-2.24,0) -- (2.24,0)(0,2.24) -- (0,-2.24)   ;
\draw [shift={(85.33,145.83)}, rotate = 358.99] [color={rgb, 255:red, 208; green, 2; blue, 27 }  ,draw opacity=1 ][line width=0.75]    (-2.24,0) -- (2.24,0)(0,2.24) -- (0,-2.24)   ;
\draw (107.5,180.5) node [anchor=north west][inner sep=0.75pt]   [align=left] {(a)};
\draw (250.5,181) node [anchor=north west][inner sep=0.75pt]   [align=left] {(b)};
\end{tikzpicture}}
    \caption{The ETH and RMT contributions to the variance $\Delta\mathcal{G}^2$ are associated to distinct bulk topologies.}
    \label{fig:ETHvsRMT}
\end{figure}
This on-shell partition function is superficially similar to the off-shell partition function that we found in Eq.~\eqref{eq:bdy_prediction} for the wormhole $\mathcal{W}$ shown in Fig.~\ref{fig:ETHvsRMT}(b). However, there are some important differences. Firstly, Eq.~\eqref{eq:OPEavg} has half as many integrals as Eq.~\eqref{eq:bdy_prediction}. Secondly, the factor of the density of states $\rho_0(P,\bar P)$ has a Cardy growth
\begin{equation}
   \rho_0(P,\bar P) \sim e^{2\pi Q(P+\bar P)}, \quad P, \bar P \to \infty,
\end{equation}
where $Q^2 = \frac{1}{6}(c-1)$, whereas $\rho_{0,2}(P,\bar P, P',\bar P')$ in the integrand of Eq.~\eqref{eq:bdy_prediction} is independent of the central charge and decays when $P\to \infty$ (for fixed $P'$). This means that the on-shell partition function admits a saddle-point approximation (in fact, in this case the saddle is the Maldacena-Maoz solution \cite{Chandra:2022bqq}), while the off-shell integral expression does not have a saddle-point. Moreover, our comparison demonstrates that the off-shell partition function is non-perturbatively small, at large $c$, compared to the on-shell contribution.

However, there is a kinematical regime in cross-ratio space where the RMT-contribution can become relevant. Namely, we can write the 4-point conformal blocks $\mathcal{F}_t(P;z)$ in the so-called ``pillow'' coordinates $\tau(z),\bar\tau(\bar z)$ \cite{Maldacena:2015iua}. If we write $\tau = x+iy$, and complexify $y,y' = \beta\pm iT$, then the regime of large $T \gg \beta \gg 1$ picks out those contributions to the integral in Eq.~\eqref{eq:bdy_prediction} where $\omega = E-E'$ is small, of the order of the Thouless energy\footnote{We thank Mosh\'e Rozali and Diego Li\v{s}ka for discussions on this.}. For small energy differences, 
 $C_0$ is approximately constant while $\rho_{0,2}$ diverges as $\omega^{-2}$. The net effect is a version of the ``ramp'', which is well-studied for the spectral form factor \cite{Cotler:2016fpe} and the thermal two-point function \cite{Saad:2019pqd}, but now found in the new context of the variance of the sphere four-point function.

\subsection{Modular completion}\label{sec:modcompl}

In parallel with Section \ref{sec:ETH}, we can refine the calculation of the four-point wormhole $\mathcal{W}$ by summing over modular images. In this case, there are two types of modular sums that are relevant. First, there is the \emph{boundary} mapping class group of the four-punctured spheres that constitute the boundary $\partial \mathcal{W} = \Sigma_{0,4}\sqcup \Sigma_{0,4}$. The internal topology of $\mathcal{W}$ is such that we cannot bring one boundary MCG transformation to the other boundary component, so we should sum over two sets of independent images:\vspace{1mm}
\begin{equation}\label{eq:MCGsphere}
    \sum_{\gamma_1,\gamma_2\in \text{MCG}(\Sigma_{0,4})}Z_\text{grav}(\mathcal{W};\gamma_1 z_1,\gamma_1 \bar z_1,\gamma_2 z_2,\gamma_2 \bar z_2).\vspace{-1mm}
\end{equation}
Here $\text{MCG}(\Sigma_{0,4})$ is generated by acting with the fusion and braiding moves $\{\mathbb{F}, \mathbb{B}\}$ on a non-contractible cycle of the four-punctured sphere. The statistical ensemble that reproduces the sum over MCG images in Eq.~\eqref{eq:MCGsphere} has spectral statistics that are still determined by $\rho_{0,2}$, but OPE statistics that modify $\overline{|C_{\clo_1\clo_2h}|^2}$ by a sum over crossing kernels, of the same form as Eq.~\eqref{eq:opecorrection}.

The second type of modular images that are summed over in the gravitational path integral can be obtained by acting with modular transformations on the bulk incompressible torus of $\mathcal{W}$. These modular images will be the focus of this subsection. Although they are not really `images' in a strict sense from the point of view of the boundary 4-point function, these topologies are needed for the spectral average $\overline{\rho\rho}^c$ to be consistent with modular invariance of the torus partition function. We can think of such modular images as providing a modular invariant uplift of the random matrix ansatz in Eq.~\eqref{eq:spectralcorr}, as was also proposed in \cite{DiUbaldo:2023qli,Boruch:2025ilr}.

This second type of topologies can be explicitly computed using the path integral formalism of the previous subsection. Consider a modular transformation
\begin{equation}\label{eq:modtrans}
    \gamma = \begin{pmatrix}
        a & b \\ s & r
    \end{pmatrix} \in\text{PSL}(2,\mathbb{Z})
\end{equation}
labeled by a pair of coprime integers $(r,s)$, and form the manifold $\mathcal{W}_{r,s}$ by following Step 1 of the RMT surgery shown Figure \ref{fig:RMT_surgery}, and then gluing the torus boundaries in Step 2 after a relative modular transformation $\gamma$. The geometric effect of this modular transformation is to identify the meridian $[m]$ of one of the tori with the closed curve $r[m]+s[l]$ on the second torus ($[l]$ is the longitude). On the level of the gravitational path integral, this second step amounts to computing the expectation value\vspace{1mm}
\begin{equation}
    Z_\text{grav}[\mathcal{W}_{r,s}] = \mel{Z_\text{grav}(z,\bar z)}{G_\text{T}\circ \mathbb{U}_{r,s}}{Z_\text{grav}(z',\bar z')}.
\end{equation}
Here $\mathbb{U}_{r,s}:\mathcal{H}_\text{T} \to \mathcal{H}_\text{T}$ is the unitary operator representing the modular transformation $\gamma$ on the torus Hilbert space. Its matrix elements are given by the generalized modular kernels (see App.~\ref{app:kernels} for their functional form):
\begin{equation}\label{eq:genmodker}
    \mel{P,\bar P}{\mathbb{U}_{r,s}}{P',\bar P'} = \mathbb{U}(r,s)_{PP'}\mathbb{U}(r,s)^*_{\bar P\bar P'}.
\end{equation}

If we had instead acted with $\gamma$ on the second torus boundary, and then glued, we would have obtained the same manifold $\mathcal{W}_{r,s}$. This fact is expressed as an operator statement in  $\mathcal{H}_\text{T}$ by the commutation relation\vspace{1mm}
\begin{equation}\label{eq:commute}
     [G_\text{T}, \mathbb{U}_{r,s}] = 0,\vspace{1mm}
\end{equation}
as operators acting by integral transformations on the torus Hilbert space. This commutation relation can also be derived explicitly from the following property satisfied by the AdS$_3$ double trumpet partition function \cite{Cotler:2020ugk}:
\begin{equation}\label{eq:CJproperty}
    Z_{T\times I}(\gamma\tau_1,\gamma\bar\tau_1;\tau_2,\bar\tau_2) = Z_{T\times I}(\tau_1,\bar\tau_1;\tilde\gamma\tau_2,\tilde\gamma\bar\tau_2).
\end{equation}
Here $\gamma\in\text{PSL}(2,\mathbb{Z})$ and $\tilde\gamma = M \gamma^{-1}M$, with the involution $M=\text{diag}(-1,1)$ taking into account the opposite orientation. See App.~\ref{app:doubletrumpet} for a verification of this statement. Inserting resolutions of the identity as in Eq.~\eqref{eq:canonical}, and using the action of $\mathbb{U}_{r,s}$ on the Virasoro characters, the equivalence of Eqs.~\eqref{eq:CJproperty} and \eqref{eq:commute} follows.

As a special case of the above construction, we note that $(r,s) = (1,n)$ corresponds to performing a relative Dehn twist before gluing the tori in Step 2. Summing over $n\in\mathbb{Z}$ implements spin quantization, as in Section \ref{sec:ETH}, using the Poisson resummation formula Eq.~\eqref{eq:poissonresum}. Indeed, one can verify that (see also App.~\ref{app:doubletrumpet}):
\begin{multline}\label{eq:partialsum}
    \sum_{n\in\mathbb{Z}} Z_\text{grav}[\mathcal{W}_{1,n}] = \int_0^\infty \!\!\dd^4 P \delta_{\mathbb{Z}}(J)\Big[\rho_{0,2}(P,\bar P,P',\bar P')
\\ \big|C_0(\clo_1,\clo_2,P)C_0(\clo_1,\clo_2,P')\mathcal{F}_t(P;z)\mathcal{F}_t(P';z')\big|^2\Big]
\end{multline}
where the Dirac comb $\delta_{\mathbb{Z}}(J) = \sum_{n\in \mathbb{Z}}\delta(P^2-\bar P^2 - n)$ enforces integrality of the spin $J$. Note that we only need a single Dirac comb, because $\rho_{0,2}$ is spin-diagonal, setting $J=J'$ through the delta function $\delta(J-J')$. So, in the same way as in Section \ref{sec:ETH}, spin quantization follows from a sum over a subclass of topologies.

In conclusion, the most general modular completion of the wormhole $\mathcal{W}$, consistent with crossing symmetry on the sphere and modular invariance of the torus, is the combined sum
\begin{equation}
     \sum_{\gamma_1,\gamma_2}\sum_{(r,s)=1}Z_\text{grav}(\mathcal{W}_{r,s};\gamma_1 z_1,\gamma_1 \bar z_1,\gamma_2 z_2,\gamma_2 \bar z_2).
\end{equation}
While this looks somewhat daunting, the effect of the Poincar\'e sum over $(r,s)$ is simply to replace the factor of $\rho_{0,2}$ in the off-shell partition function Eq.~\eqref{eq:bdy_prediction} by the \emph{modular completed} spectral correlator 
\begin{multline}\label{eq:modcompleted}
    \rho_{0,2}^{\text{m.c.}}(P_1,\bar P_1,P_2,\bar P_2) \coloneqq \\\sum_{(r,s)=1} \,\mel{P_1,\bar P_1}{\mathbb{U}_{r,s}\circ G_\text{T}}{P_2,\bar P_2}.
\end{multline}
 This function, while written in a different form, is precisely the Poincar\'e sum that was regularized in Ref.~\cite{Cotler:2020ugk}. The double sum over the mapping class group of the four-punctured sphere has not been shown to be convergent or regularizable as of yet, but it is closely related to the conformal block Farey tail of Ref.~\cite{Maloney:2016kee}.

\subsection{Generalization to other topologies}

The new framework of RMT surgery has a much wider applicability than the example of the four-point wormhole studied in the previous subsection. In fact, take any CFT observable with at least one heavy operator exchange in some channel. In the diagrammatic approach to Virasoro TQFT \cite{Collier:2024mgv}, such an observable can be represented by a threevalent graph $\Gamma$, where the vertices represent OPE coefficients and edges represent heavy primary exchanges with Liouville momentum $P_i$. The graph $\Gamma$ should be thought of as embedded in $S^3$, defining the non-contractible cycles of a 3-manifold, which we assume to be hyperbolic (hyperbolic 3-manifolds are fully specified by their fundamental group \cite{Thurston1997}). 

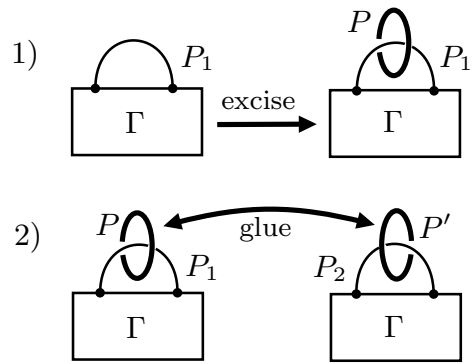
\begin{figure}
    \centering
    \resizebox{0.8\linewidth}{!}{
    \begin{tikzpicture}[x=0.75pt,y=0.75pt,yscale=-1,xscale=1,baseline={([yshift=-0.5ex]current bounding box.center)}]
\draw    (79.32,69.32) .. controls (79.46,44.04) and (109.18,44.89) .. (109.32,69.32) ;
\draw [shift={(109.32,69.32)}, rotate = 89.66] [color={rgb, 255:red, 0; green, 0; blue, 0 }  ][fill={rgb, 255:red, 0; green, 0; blue, 0 }  ][line width=0.75]      (0, 0) circle [x radius= 1.34, y radius= 1.34]   ;
\draw [shift={(79.32,69.32)}, rotate = 270.32] [color={rgb, 255:red, 0; green, 0; blue, 0 }  ][fill={rgb, 255:red, 0; green, 0; blue, 0 }  ][line width=0.75]      (0, 0) circle [x radius= 1.34, y radius= 1.34]   ;
\draw [line width=1.5]    (126.57,83.33) -- (161.17,83.18) ;
\draw [shift={(165.17,83.17)}, rotate = 179.75] [fill={rgb, 255:red, 0; green, 0; blue, 0 }  ][line width=0.08]  [draw opacity=0] (6.97,-3.35) -- (0,0) -- (6.97,3.35) -- cycle    ;
\draw    (180.65,70.25) .. controls (180.88,45.36) and (210.88,45.36) .. (210.65,70.25) ;
\draw [shift={(210.65,70.25)}, rotate = 90.51] [color={rgb, 255:red, 0; green, 0; blue, 0 }  ][fill={rgb, 255:red, 0; green, 0; blue, 0 }  ][line width=0.75]      (0, 0) circle [x radius= 1.34, y radius= 1.34]   ;
\draw [shift={(180.65,70.25)}, rotate = 270.51] [color={rgb, 255:red, 0; green, 0; blue, 0 }  ][fill={rgb, 255:red, 0; green, 0; blue, 0 }  ][line width=0.75]      (0, 0) circle [x radius= 1.34, y radius= 1.34]   ;
\draw [color={rgb, 255:red, 255; green, 255; blue, 255 }  ,draw opacity=1 ][line width=2.25]    (200.24,53.55) -- (201.4,51.21) ;
\draw  [draw opacity=0][line width=1.5]  (189.27,50.12) .. controls (189.62,43.82) and (192.06,38.95) .. (195.03,38.95) .. controls (198.23,38.95) and (200.83,44.63) .. (200.83,51.65) .. controls (200.83,58.66) and (198.23,64.35) .. (195.03,64.35) .. controls (192.41,64.35) and (190.19,60.53) .. (189.47,55.3) -- (195.03,51.65) -- cycle ; \draw  [draw opacity=1 ][line width=1.5]  (189.27,50.12) .. controls (189.62,43.82) and (192.06,38.95) .. (195.03,38.95) .. controls (198.23,38.95) and (200.83,44.63) .. (200.83,51.65) .. controls (200.83,58.66) and (198.23,64.35) .. (195.03,64.35) .. controls (192.41,64.35) and (190.19,60.53) .. (189.47,55.3) ;  
\draw    (81.17,148.6) .. controls (81.39,123.71) and (111.39,123.71) .. (111.17,148.6) ;
\draw [shift={(111.17,148.6)}, rotate = 90.51] [color={rgb, 255:red, 0; green, 0; blue, 0 }  ][fill={rgb, 255:red, 0; green, 0; blue, 0 }  ][line width=0.75]      (0, 0) circle [x radius= 1.34, y radius= 1.34]   ;
\draw [shift={(81.17,148.6)}, rotate = 270.51] [color={rgb, 255:red, 0; green, 0; blue, 0 }  ][fill={rgb, 255:red, 0; green, 0; blue, 0 }  ][line width=0.75]      (0, 0) circle [x radius= 1.34, y radius= 1.34]   ;
\draw [color={rgb, 255:red, 255; green, 255; blue, 255 }  ,draw opacity=1 ][line width=2.25]    (99.88,130.18) -- (100.87,130.68) -- (102.75,131.58) ;
\draw    (210.5,149.1) .. controls (210.28,123.14) and (180.28,123.14) .. (180.5,149.1) ;
\draw [shift={(180.5,149.1)}, rotate = 89.51] [color={rgb, 255:red, 0; green, 0; blue, 0 }  ][fill={rgb, 255:red, 0; green, 0; blue, 0 }  ][line width=0.75]      (0, 0) circle [x radius= 1.34, y radius= 1.34]   ;
\draw [shift={(210.5,149.1)}, rotate = 269.51] [color={rgb, 255:red, 0; green, 0; blue, 0 }  ][fill={rgb, 255:red, 0; green, 0; blue, 0 }  ][line width=0.75]      (0, 0) circle [x radius= 1.34, y radius= 1.34]   ;
\draw [color={rgb, 255:red, 255; green, 255; blue, 255 }  ,draw opacity=1 ][line width=2.25]    (191.65,129.61) -- (189.34,131.1) ;
\draw  [draw opacity=0][line width=1.5]  (201.89,128.48) .. controls (201.55,121.89) and (199.1,116.77) .. (196.13,116.77) .. controls (192.92,116.77) and (190.33,122.7) .. (190.33,130.01) .. controls (190.33,137.33) and (192.92,143.26) .. (196.13,143.26) .. controls (198.77,143.26) and (201,139.21) .. (201.7,133.68) -- (196.13,130.01) -- cycle ; \draw  [draw opacity=1 ][line width=1.5]  (201.89,128.48) .. controls (201.55,121.89) and (199.1,116.77) .. (196.13,116.77) .. controls (192.92,116.77) and (190.33,122.7) .. (190.33,130.01) .. controls (190.33,137.33) and (192.92,143.26) .. (196.13,143.26) .. controls (198.77,143.26) and (201,139.21) .. (201.7,133.68) ;  
\draw [line width=1.5]    (109.87,121.6) .. controls (137.82,114.23) and (156.77,114.56) .. (183.79,121.26) ;
\draw [shift={(187.67,122.25)}, rotate = 194.7] [fill={rgb, 255:red, 0; green, 0; blue, 0 }  ][line width=0.08]  [draw opacity=0] (6.97,-3.35) -- (0,0) -- (6.97,3.35) -- cycle    ;
\draw [shift={(105.67,122.75)}, rotate = 344.27] [fill={rgb, 255:red, 0; green, 0; blue, 0 }  ][line width=0.08]  [draw opacity=0] (6.97,-3.35) -- (0,0) -- (6.97,3.35) -- cycle    ;
\draw  [draw opacity=0][line width=1.5]  (89.78,128.77) .. controls (90.13,122.48) and (92.58,117.6) .. (95.54,117.6) .. controls (98.74,117.6) and (101.34,123.29) .. (101.34,130.3) .. controls (101.34,137.31) and (98.74,143) .. (95.54,143) .. controls (92.92,143) and (90.7,139.19) .. (89.98,133.95) -- (95.54,130.3) -- cycle ; \draw  [draw opacity=1 ][line width=1.5]  (89.78,128.77) .. controls (90.13,122.48) and (92.58,117.6) .. (95.54,117.6) .. controls (98.74,117.6) and (101.34,123.29) .. (101.34,130.3) .. controls (101.34,137.31) and (98.74,143) .. (95.54,143) .. controls (92.92,143) and (90.7,139.19) .. (89.98,133.95) ;  
\draw   (70.33,69.33) -- (120.67,69.33) -- (120.67,95.17) -- (70.33,95.17) -- cycle ;
\draw   (170.33,70) -- (220.67,70) -- (220.67,95.83) -- (170.33,95.83) -- cycle ;
\draw   (70.67,148.67) -- (121,148.67) -- (121,174.5) -- (70.67,174.5) -- cycle ;
\draw   (170.67,149.13) -- (221,149.13) -- (221,174.97) -- (170.67,174.97) -- cycle ;
\draw (89.43,76.93) node [anchor=north west][inner sep=0.75pt]    {$\Gamma $};
\draw (110.87,51.17) node [anchor=north west][inner sep=0.75pt]  [font=\small]  {$P_{1}$};
\draw (45.13,48.9) node [anchor=north west][inner sep=0.75pt]    {$1)$};
\draw (211.13,51.2) node [anchor=north west][inner sep=0.75pt]  [font=\small]  {$P_{1}$};
\draw (175.29,37.86) node [anchor=north west][inner sep=0.75pt]  [font=\small]  {$P$};
\draw (46.3,119.4) node [anchor=north west][inner sep=0.75pt]    {$2)$};
\draw (112.63,132.53) node [anchor=north west][inner sep=0.75pt]  [font=\small]  {$P_{1}$};
\draw (77.79,116.52) node [anchor=north west][inner sep=0.75pt]  [font=\small]  {$P$};
\draw (162.13,132.82) node [anchor=north west][inner sep=0.75pt]  [font=\small]  {$P_{2}$};
\draw (203.12,116.19) node [anchor=north west][inner sep=0.75pt]  [font=\small]  {$P'$};
\draw (189.43,76.6) node [anchor=north west][inner sep=0.75pt]    {$\Gamma $};
\draw (89.77,156.27) node [anchor=north west][inner sep=0.75pt]    {$\Gamma $};
\draw (189.77,156) node [anchor=north west][inner sep=0.75pt]    {$\Gamma$};
\draw (133.67,117.67) node [anchor=north west][inner sep=0.75pt]  [font=\footnotesize] [align=left] {glue};
\draw (126.67,68.67) node [anchor=north west][inner sep=0.75pt]  [font=\footnotesize] [align=left] {excise};
\end{tikzpicture}\hspace{4mm}}
    \caption{RMT surgery in the more general case. The three-valent graph $\Gamma$ is symbolized by a box.}
    \label{fig:generalizedRMT}
\end{figure}

Now suppose that we want to compute the variance of our CFT observable, focusing on the RMT statistics of the high-energy spectrum only. Then the corresponding bulk wormhole that captures this variance is constructed by RMT surgery, which is illustrated in Figure \ref{fig:generalizedRMT}. Again, we follow a two-step procedure:
\begin{enumerate}[1)]
    \item Take one edge, $P_1$, and link it with a Wilson loop $P$. We assume that the Wilson loop is non-contractible in $S^3\setminus\Gamma$. A tubular neighborhood of the Wilson loop is a solid torus, which we excise.
    \item Take two copies of the diagram made in Step 1, and glue the torus boundaries (around the Wilson loops) with the torus wormhole $T\times I$ as before.
\end{enumerate}
It is interesting to observe that the above recipe is closely related to a result in 3-manifold topology known as the \emph{JSJ decomposition}. This refers to the theorem that any irreducible 3-manifold can be cut along a unique collection of incompressible tori, into pieces that are either hyperbolic or Seifert-fibered \cite{JacoShalen1978,Johannson1979}\footnote{The original JSJ theorem states ``atoroidal or Seifert-fibered'', but by Thurston's geometrization conjecture, proven by Perelman, we can replace atoroidal by hyperbolic. Moreover, the original theorem refers only to closed 3-manifolds, but a suitable version of the JSJ decomposition for manifolds with boundary was discussed in \cite{Johannson1979}.}. Similarly, RMT surgery builds irreducible 3-manifolds that can be cut along an incompressible torus into pieces that are both hyperbolic. So, topologically, RMT surgery is related to a special case of the JSJ decomposition, similar to how we recognized ETH surgery as a special instance of Heegaard splitting in Section \ref{sec:ETH}.     
 
In order to compute the wormhole partition function for the manifolds constructed in Figure \ref{fig:generalizedRMT}, let us zoom in to a neighborhood of the edges $P_1$ and $P_2$, and use the Verlinde loop operator to unlink the Wilson loops:
\begin{equation}
\begin{tikzpicture}[x=0.75pt,y=0.75pt,yscale=-0.8,xscale=0.8,baseline={([yshift=-0.2ex]current bounding box.center)}]
\draw  (165,150) -- (210,150) ;
\draw  (140,150) -- (160,150) ;
\draw [shift={(210,150)}, rotate = 0] [color={rgb, 255:red, 0; green, 0; blue, 0 }  ][fill={rgb, 255:red, 0; green, 0; blue, 0 }  ][line width=0.75]      (0, 0) circle [x radius= 1.34, y radius= 1.34]   ;
\draw [shift={(140,150)}, rotate = 0] [color={rgb, 255:red, 0; green, 0; blue, 0 }  ][fill={rgb, 255:red, 0; green, 0; blue, 0 }  ][line width=0.75]      (0, 0) circle [x radius= 1.34, y radius= 1.34]   ;
\draw  (165,180) -- (210,180) ;
\draw  (140,180) -- (160,180) ;
\draw [shift={(210,180)}, rotate = 0] [color={rgb, 255:red, 0; green, 0; blue, 0 }  ][fill={rgb, 255:red, 0; green, 0; blue, 0 }  ][line width=0.75]      (0, 0) circle [x radius= 1.34, y radius= 1.34]   ;
\draw [shift={(140,180)}, rotate = 0] [color={rgb, 255:red, 0; green, 0; blue, 0 }  ][fill={rgb, 255:red, 0; green, 0; blue, 0 }  ][line width=0.75]      (0, 0) circle [x radius= 1.34, y radius= 1.34]   ;
\draw  [draw opacity=0] (176.22,153.44) .. controls (175.2,157.56) and (172.5,160.5) .. (169.33,160.5) .. controls (165.28,160.5) and (162,155.69) .. (162,149.75) .. controls (162,143.81) and (165.28,139) .. (169.33,139) .. controls (172.74,139) and (175.6,142.4) .. (176.42,147) -- (169.33,149.75) -- cycle ; \draw  [draw opacity=1 ] (176.22,153.44) .. controls (175.2,157.56) and (172.5,160.5) .. (169.33,160.5) .. controls (165.28,160.5) and (162,155.69) .. (162,149.75) .. controls (162,143.81) and (165.28,139) .. (169.33,139) .. controls (172.74,139) and (175.6,142.4) .. (176.42,147) ;  
\draw  [dash pattern={on 1.5pt off 1.5pt on 1.5pt off 1.5pt}] (129.93,165) .. controls (129.93,140.11) and (150.11,119.93) .. (175,119.93) .. controls (199.89,119.93) and (220.07,140.11) .. (220.07,165) .. controls (220.07,189.89) and (199.89,210.07) .. (175,210.07) .. controls (150.11,210.07) and (129.93,189.89) .. (129.93,165) -- cycle ;
\draw  [draw opacity=0] (176.22,183.77) .. controls (175.2,187.89) and (172.5,190.83) .. (169.33,190.83) .. controls (165.28,190.83) and (162,186.02) .. (162,180.08) .. controls (162,174.15) and (165.28,169.33) .. (169.33,169.33) .. controls (172.74,169.33) and (175.6,172.73) .. (176.42,177.34) -- (169.33,180.08) -- cycle ; \draw  [draw opacity=1 ] (176.22,183.77) .. controls (175.2,187.89) and (172.5,190.83) .. (169.33,190.83) .. controls (165.28,190.83) and (162,186.02) .. (162,180.08) .. controls (162,174.15) and (165.28,169.33) .. (169.33,169.33) .. controls (172.74,169.33) and (175.6,172.73) .. (176.42,177.34) ;  
\draw (141,135) node [anchor=north west][inner sep=0.75pt]  [font=\footnotesize]  {$P_{1}$};
\draw (141,165) node [anchor=north west][inner sep=0.75pt]  [font=\footnotesize]  {$P_{2}$};
\draw (177.86,136.3) node [anchor=north west][inner sep=0.75pt]  [font=\footnotesize]  {$P$};
\draw (177.86,166.3) node [anchor=north west][inner sep=0.75pt]  [font=\footnotesize]  {$P'$};
\end{tikzpicture} = \frac{\mathbb{S}_{PP_1}[\bbi]}{\sker{\bbi}{P_1}{\bbi}}\frac{\mathbb{S}_{P'P_2}[\bbi]}{\sker{\bbi}{P_2}{\bbi}}\,\begin{tikzpicture}[x=0.75pt,y=0.75pt,yscale=-0.8,xscale=0.8,baseline={([yshift=-0.2ex]current bounding box.center)}]
\draw    (160,170) -- (230,170) ;
\draw [shift={(230,170)}, rotate = 0] [color={rgb, 255:red, 0; green, 0; blue, 0 }  ][fill={rgb, 255:red, 0; green, 0; blue, 0 }  ][line width=0.75]      (0, 0) circle [x radius= 1.34, y radius= 1.34]   ;
\draw [shift={(160,170)}, rotate = 0] [color={rgb, 255:red, 0; green, 0; blue, 0 }  ][fill={rgb, 255:red, 0; green, 0; blue, 0 }  ][line width=0.75]      (0, 0) circle [x radius= 1.34, y radius= 1.34]   ;
\draw    (160,200) -- (230,200) ;
\draw [shift={(230,200)}, rotate = 0] [color={rgb, 255:red, 0; green, 0; blue, 0 }  ][fill={rgb, 255:red, 0; green, 0; blue, 0 }  ][line width=0.75]      (0, 0) circle [x radius= 1.34, y radius= 1.34]   ;
\draw [shift={(160,200)}, rotate = 0] [color={rgb, 255:red, 0; green, 0; blue, 0 }  ][fill={rgb, 255:red, 0; green, 0; blue, 0 }  ][line width=0.75]      (0, 0) circle [x radius= 1.34, y radius= 1.34]   ;
\draw  [dash pattern={on 1.5pt off 1.5pt on 1.5pt off 1.5pt}] (149.93,185) .. controls (149.93,160.11) and (170.11,139.93) .. (195,139.93) .. controls (219.89,139.93) and (240.07,160.11) .. (240.07,185) .. controls (240.07,209.89) and (219.89,230.07) .. (195,230.07) .. controls (170.11,230.07) and (149.93,209.89) .. (149.93,185) -- cycle ;
\draw (163,155) node [anchor=north west][inner sep=0.75pt]  [font=\footnotesize]  {$P_{1}$};
\draw (163,185) node [anchor=north west][inner sep=0.75pt]  [font=\footnotesize]  {$P_{2}$};
\end{tikzpicture}
\end{equation}
This equation is valid in VTQFT: to get the gravitational contribution we still have to multiply by the anti-holomorphic counterpart.

Now we glue the torus boundaries labeled by $P,\bar P$ and $P',\bar P'$ using the gluing map $G_\text{T}$. We will use the same microcanonical version of the torus wormhole to perform the gluing. In order to perform the integrals over $P$ and $P'$, recall the fact that $\mathbb{S}\circ G_\text{T}\circ \mathbb{S}=G_\text{T}$, as well as the relation between the matrix elements of $G_\text{T}$ and the spectral correlator $\rho_{0,2}$ in Eq.~\eqref{eq:mel}. Finally, we use that $\sker{\bbi}{P}{\bbi} = \rho_0(P)$. If we denote the VTQFT partition function evaluated on the graph $\Gamma$ with edge $P$ by $Z_\text{Vir}(\Gamma;P)$, then the microcanonical partition function evaluates to
\begin{equation}\label{eq:rmtsurgery}
    \frac{\rho_{0,2}(P_1,\bar P_1,P_2,\bar P_2)}{|\rho_0(P_1)|^2|\rho_0(P_2)|^2} |Z_\text{Vir}(\Gamma;P_1)Z_\text{Vir}(\Gamma;P_2)|^2.
\end{equation}
To go to the canonical partition function, where the boundary moduli are fixed instead of the Liouville momenta of the graph, one has to integrate over all edges, with the appropriate weighing factors of $\rho_0$ and $C_0$, times the conformal blocks in the channel specified by $\Gamma$. In particular, the edges $P_1$ and $P_2$ are weighted by a factor of $\rho_0(P_1)\rho_0( P_2)$, which cancels the denominator of Eq.~\eqref{eq:rmtsurgery}. So, our RMT surgery has effectively replaced \vspace{1mm}
\begin{equation}
    |\rho_0(P_1)|^2|\rho_0(P_2)|^2 \longrightarrow \rho_{0,2}(P_1,\bar P_1,P_2,\bar P_2),\vspace{1mm}
\end{equation}
which is precisely the same effect as taking the connected average $\overline{\rho(h_1,\bar h_1)\rho(h_2,\bar h_2)}^c$ in the boundary CFT. We can again summarize this informally by a slogan:\vspace{1mm}
\begin{equation*}
    \text{``RMT surgery = level repulsion.''}\vspace{1mm}
\end{equation*}

The interested reader can work out the above steps for the genus-two partition function. Its variance is a generalization of the spectral form factor \cite{Belin:2021ibv}. The OPE statistics are already analyzed in Section \ref{sec:ETH}, while the spectral statistics are relevant at post-Thouless times to produce the ramp. The wormhole responsible for this ramp is constructed by RMT surgery: for $\Gamma$, take the sunset diagram, which has 3 internal momenta, and follow the steps given above. Gluing along a single embedded incompressible torus precisely captures the semi-factorized spectral average $\bar\rho \times\bar\rho \times\overline{\rho\rho}^c\times\bar \rho\times\bar\rho$.

\section{Trumpets and 3-manifolds}\label{sec:trumpetgluing}
\begin{figure}
    \centering
    \resizebox{0.85\linewidth}{!}{
   \begin{tikzpicture}[x=0.75pt,y=0.75pt,yscale=-1,xscale=1]
\draw    (324.28,172.57) .. controls (342.39,172.34) and (353.86,188) .. (370.75,187.63) .. controls (387.64,187.25) and (403.75,170.88) .. (404.25,151.63) .. controls (404.75,132.38) and (392.75,113.38) .. (370.75,113.13) .. controls (348.75,112.88) and (339.43,129.57) .. (324.28,129.18) ;
\draw   (313.57,150.87) .. controls (313.57,138.89) and (318.37,129.18) .. (324.28,129.18) .. controls (330.2,129.18) and (334.99,138.89) .. (334.99,150.87) .. controls (334.99,162.86) and (330.2,172.57) .. (324.28,172.57) .. controls (318.37,172.57) and (313.57,162.86) .. (313.57,150.87) -- cycle ;
\draw    (322.82,140.42) .. controls (328.37,143.06) and (328.68,157.81) .. (323.13,162.38) ;
\draw    (324.95,159.98) .. controls (321.98,155.78) and (319.65,149.63) .. (324.5,142.46) ;
\draw  [draw opacity=0] (277.69,129.59) .. controls (278.36,129.32) and (279.07,129.18) .. (279.78,129.18) .. controls (285.7,129.18) and (290.49,138.89) .. (290.49,150.87) .. controls (290.49,162.86) and (285.7,172.57) .. (279.78,172.57) .. controls (279.73,172.57) and (279.68,172.57) .. (279.64,172.57) -- (279.78,150.87) -- cycle ; \draw   (277.69,129.59) .. controls (278.36,129.32) and (279.07,129.18) .. (279.78,129.18) .. controls (285.7,129.18) and (290.49,138.89) .. (290.49,150.87) .. controls (290.49,162.86) and (285.7,172.57) .. (279.78,172.57) .. controls (279.73,172.57) and (279.68,172.57) .. (279.64,172.57) ;  
\draw  [dash pattern={on 1.5pt off 0.75pt on 1.5pt off 0.75pt}]  (278.54,140.77) .. controls (284.09,143.4) and (284.55,157.12) .. (279,161.69) ;
\draw  [dash pattern={on 1.5pt off 0.75pt on 1.5pt off 0.75pt}]  (280.45,159.98) .. controls (277.48,155.78) and (275.15,149.63) .. (280,142.46) ;
\draw   (218.57,150.14) .. controls (218.57,129) and (225.86,111.86) .. (234.86,111.86) .. controls (243.85,111.86) and (251.14,129) .. (251.14,150.14) .. controls (251.14,171.29) and (243.85,188.43) .. (234.86,188.43) .. controls (225.86,188.43) and (218.57,171.29) .. (218.57,150.14) -- cycle ;
\draw    (232.63,131.7) .. controls (241.07,136.35) and (241.54,162.39) .. (233.1,170.45) ;
\draw    (235.87,166.21) .. controls (231.36,158.8) and (227.81,147.95) .. (235.19,135.29) ;
\draw    (240.17,113.86) .. controls (249.75,121.86) and (259.32,131) .. (277.89,129.57) ;
\draw    (238.67,187.39) .. controls (254.52,176.39) and (258.43,172.14) .. (279.78,172.57) ;
\draw  [draw opacity=0][dash pattern={on 1.5pt off 0.75pt on 1.5pt off 0.75pt}] (280.19,172.56) .. controls (280.05,172.57) and (279.92,172.57) .. (279.78,172.57) .. controls (273.87,172.57) and (269.07,162.86) .. (269.07,150.87) .. controls (269.07,140.46) and (272.69,131.76) .. (277.52,129.66) -- (279.78,150.87) -- cycle ; \draw  [dash pattern={on 1.5pt off 0.75pt on 1.5pt off 0.75pt}] (280.19,172.56) .. controls (280.05,172.57) and (279.92,172.57) .. (279.78,172.57) .. controls (273.87,172.57) and (269.07,162.86) .. (269.07,150.87) .. controls (269.07,140.46) and (272.69,131.76) .. (277.52,129.66) ;  
\draw [line width=1.5]    (284.14,124.33) .. controls (294.72,116.84) and (307.7,116.47) .. (319.18,124.41) ;
\draw [shift={(322.33,126.83)}, rotate = 220.36] [fill={rgb, 255:red, 0; green, 0; blue, 0 }  ][line width=0.08]  [draw opacity=0] (6.97,-3.35) -- (0,0) -- (6.97,3.35) -- cycle    ;
\draw [shift={(281,126.83)}, rotate = 318.37] [fill={rgb, 255:red, 0; green, 0; blue, 0 }  ][line width=0.08]  [draw opacity=0] (6.97,-3.35) -- (0,0) -- (6.97,3.35) -- cycle    ;
\draw  [color={rgb, 255:red, 208; green, 2; blue, 27 }  ,draw opacity=1 ] (317.3,150.87) .. controls (317.3,141.67) and (320.42,134.21) .. (324.28,134.21) .. controls (328.14,134.21) and (331.27,141.67) .. (331.27,150.87) .. controls (331.27,160.08) and (328.14,167.54) .. (324.28,167.54) .. controls (320.42,167.54) and (317.3,160.08) .. (317.3,150.87) -- cycle ;
\draw  [color={rgb, 255:red, 208; green, 2; blue, 27 }  ,draw opacity=1 ] (272.8,150.87) .. controls (272.8,141.67) and (275.92,134.21) .. (279.78,134.21) .. controls (283.64,134.21) and (286.77,141.67) .. (286.77,150.87) .. controls (286.77,160.08) and (283.64,167.54) .. (279.78,167.54) .. controls (275.92,167.54) and (272.8,160.08) .. (272.8,150.87) -- cycle ;
\draw (356.65,143.56) node [anchor=north west][inner sep=0.75pt]    {$M$};
\draw (290,122.67) node [anchor=north west][inner sep=0.75pt]  [font=\footnotesize] [align=left] {glue};
\draw (222.5,191.5) node [anchor=north west][inner sep=0.75pt]    {$\tau ,\bar\tau $};
\draw (265.8,178.8) node [anchor=north west][inner sep=0.75pt]  [font=\footnotesize]  {$P,\bar P$};
\draw (309.8,178.8) node [anchor=north west][inner sep=0.75pt]  [font=\footnotesize]  {$P',\bar P'$};
\end{tikzpicture}}
    \caption{Gluing an AdS$_3$ trumpet to a 3-manifold $M$.}
    \label{fig:trumpet}
\end{figure}
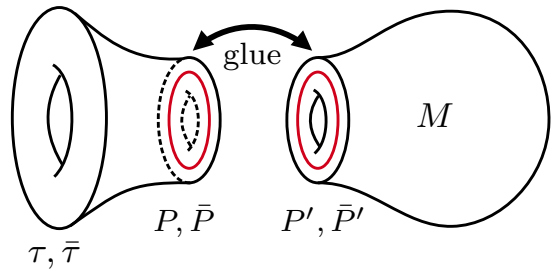

The formalism developed in the previous section can be used to compute another class of off-shell topologies in pure AdS$_3$ gravity. In this section, we provide a method to compute the gravitational partition function for off-shell topologies that are constructed as in Fig.~\ref{fig:trumpet}: we take a finite-volume hyperbolic 3-manifold $M$ with $n$ torus boundaries, and to each boundary we glue the AdS$_3$ trumpet geometry \cite{Cotler:2020ugk}, which is topologically $T\times I$ with one flaring asymptotic boundary. This construction is inspired by JT gravity, where the asymptotically AdS$_2$ trumpet geometry is glued to each circular boundary of a bordered Riemann surface \cite{Saad:2019lba}. 

As will be explained in more detail in a forthcoming work \cite{toappear3}, such topologies are off-shell because of the non-trivial bulk mapping class group at the `cuff' of $T\times I$.\footnote{A different argument showing that these 3-manifolds are off-shell is to follow the analysis in Section 2.1 of \cite{Maloney:2007ud}, which implies that the only \emph{on-shell} topologies whose conformal boundary is a torus are the PSL$(2,\mathbb{Z})$ family of solid tori $D\times S^1$.} The gluing surface is a torus, on which the bulk MCG acts by the same large diffeomorphisms $z\to z + l + m\tau$ that we found in the previous section. However, since we started with a hyperbolic 3-manifold $M$ (whose bulk MCG is finite), the bulk MCG of the manifold after gluing the trumpets is only enhanced by a factor of
$
    (\mathbb{Z}\times\mathbb{Z})^n
$,
where the $i^{\text{th}}$ factor $\mathbb{Z}\times\mathbb{Z}$ acts in a tubular neighborhood of the $i^{\text{th}}$ boundary component. Since this is the same local MCG action as in the previous section, we know how to perform the quotient: we use the gluing map $G_\text{T}$ \eqref{eq:microcan} in each boundary torus Hilbert space $\mathcal{H}_\text{T}$.

For concreteness, here we will explain the case $n=1$. The starting point is a finite-volume, hyperbolic three-manifold $M$ with a single torus boundary; examples include hyperbolic knot complements \cite{Gukov:2003na} or mapping tori $\Sigma_{g,1}\times_\varphi S^1$ with hyperbolic monodromy \cite{toappear3}. Fixing the holonomy coordinates $P,\bar P$ around a cycle on the boundary torus, the gravitational path integral prepares a state
\begin{equation}\label{eq:state5}
    \ket{Z_\text{grav}(M)} = \ket{Z_\text{Vir}(M)}\otimes \ket{Z^*_{\text{Vir}}(M)}\in \mathcal{H}_{\text{T}}\,.
\end{equation}
Expanding in the basis states $\ket{P,\bar P}$ as before, the expansion coefficients are given by two copies of the Virasoro TQFT partition function, which can be computed for arbitrary  values of the Liouville momentum $P$:
\begin{equation}
    Z_{\text{Vir}}[M;P] = \braket{P}{Z_\text{Vir}(M)}.
\end{equation}
The manifold $M$ is hyperbolic, so it admits a complete metric of constant negative sectional curvature; however, the complete metric has a \emph{cusp} torus boundary, meaning that the metric near the boundary behaves as
\begin{equation}
    \dd s^2 = \dd \rho^2 + e^{-2\rho} |\dd z|^2 \quad \text{near } \partial M.
\end{equation}
Here $z \sim z+1 \sim z + \tau_*$ parametrizes the torus with complex structure modulus $\tau_*$. The complete, on-shell metric with cusp boundary corresponds in Virasoro TQFT to taking the limit $P,\bar P \to 0$ \cite{Collier:2024mgv}. Semiclassically, non-zero values of $P,\bar P$ parametrize the deformation space of incomplete hyperbolic structures, which was studied by Neumann and Zagier \cite{NEUMANN1985307} and Thurston \cite{Thurstonlectures}.\footnote{This is analogous to how bordered Riemann surfaces with fixed boundary length $\ell$ are geodesically incomplete in JT gravity.} Leaving a more detailed analysis of this deformation space to \cite{toappear3}, the important upshot is that Virasoro TQFT knows more than just the on-shell limit $P \to 0$: it also correctly computes the  off-shell path integral with $P\neq 0$, which determines the full quantum state \eqref{eq:state5} in  $\mathcal{H}_\text{T}$.

For example, Ref.~\cite{Collier:2024mgv} computed the Virasoro TQFT partition function for the figure-8 knot complement in $S^3$, for arbitrary values of the boundary holonomy:
\begin{equation}\label{eq:fig8partfn}
    Z_\text{Vir}[\mathbf{4}_1;P] = \sqrt{2}\int_{\mathbb{R}-i\epsilon} \!\!\dd x\, S_b(ix + 2iP)S_b(ix-2iP).
\end{equation}
Here $S_b(z)$ is the double sine function, which plays a prominent role in Ponsot and Teschner's solution of the Virasoro crossing kernels \cite{Ponsot:2000mt,Eberhardt:2023mrq}. Additional integral expressions will be obtained in \cite{toappear3} for multi-boundary topologies, such as the complements of the Whitehead link and the Borromean rings.

Now, since the `microcanonical' partition function $Z_\text{grav}[M;P,\bar P]$ with fixed boundary holonomies is computable in Virasoro TQFT, a natural question is how to obtain the `canonical' partition function $Z_\text{grav}[M;\tau,\bar\tau]$ with conformal (i.e.~asymptotically AdS$_3$) boundary conditions, labeled by boundary moduli $\tau,\bar\tau$. The answer is provided by the AdS$_3$ trumpet geometry, depicted on the left-hand side of Fig.~\ref{fig:trumpet}, which evaluates to the product of Virasoro characters, $\langle \tau,\bar\tau | P,\bar P \rangle = |\chi_P(\tau)|^2$ \cite{Cotler:2020ugk}. On the level of the path integral, gluing the trumpet amounts to evaluating the expectation value
\begin{equation}\label{eq:zgrav}
    Z_\text{grav}[M;\tau,\bar\tau] = \mel{\tau,\bar\tau}{G_\text{T}}{Z_\text{grav}(M)}.
\end{equation}
The gluing map $G_\text{T}$ takes into account the quotient by the bulk mapping class group $\mathbb{Z}\times \mathbb{Z}$, as before. Inserting resolutions of the identity, we can also write this as
\begin{equation}
    Z_\text{grav}[M;\tau,\bar\tau] = \int_0^\infty \dd P \dd \bar P\,\rho_\text{grav}(P,\bar P) |\chi_P(\tau)|^2,
\end{equation}
where we have defined
\begin{multline}\label{eq:rhograv3}
    \rho_\text{grav}(P,\bar P) = \\ \int_0^\infty \dd P'\dd \bar P' \mel{P,\bar P}{G_\text{T}}{P',\bar P'} \big|Z_\text{Vir}[M;P'] \big|^2\,.
\end{multline}
In the dual CFT ensemble, this represents an off-shell contribution to the averaged density of primary states $\overline{\rho(h,\bar h)}$. This density is supported above the black hole threshold, as $P,\bar P \geq 0$ corresponds to $h,\bar h \geq \frac{c-1}{24}$. Since the matrix elements of $G_\text{T}$ were computed in Eq.~\eqref{eq:mel}, and $Z_\text{Vir}[M;P']$ can also be computed in principle, the integral in Eq.~\eqref{eq:rhograv3} can be evaluated numerically for any value of $c$. We can also obtain analytic results at large $c$, if we use the \emph{generalized volume conjecture} \cite{Hikami_2007,Hikami2001,Kashaev:1996kc, Dijkgraaf:2009sb, Dijkgraaf:2010ur,EllegaardAndersen:2011vps, andersen2018teichmuller}. As will be shown in Ref.~\cite{toappear3}, the generalized volume conjecture can be stated in the language of Virasoro TQFT as:
\begin{multline}\label{eq:genvolconj}
    \big|Z_\text{Vir}[M;P] \big|^2 = |A|^2 \,e^{-\frac{c}{6\pi}\text{vol}(M)} \\ \times e^{2\pi i \tau_* P^2}\,e^{-2\pi i \bar{\tau}_*\bar{P}^2}(1+ O(\tfrac{1}{c}))\,.
\end{multline}
Here $\text{vol}(M)$ is the hyperbolic volume of $M$, measured in the complete hyperbolic metric, which gives the volume conjecture its name. The generalization is in the dependence of $|Z_\text{Vir}|^2$ on the deformation parameters $(P,\bar P)$, which is through the `Boltzmann weights' $q^{P^2}\bar q^{\bar P^2}$, where the nome $q = e^{2\pi i \tau_*}$ depends on the complex structure modulus $\tau_*$ of $M$ in the complete hyperbolic structure. The factor $|A|^2$ is a one-loop determinant. If we plug in the known value for the cusp modulus of the figure-8 knot complement, $\tau_* = 2i\sqrt{3}$, then one can check numerically that Eq.~\eqref{eq:fig8partfn} is well-approximated by \eqref{eq:genvolconj}.

We can use the large-$c$ approximation \eqref{eq:genvolconj} to obtain an estimate for the off-shell partition function on the single-boundary topology $M$, defined by Eq.~\eqref{eq:zgrav}:
\begin{equation}
    Z_\text{grav}[M;\tau,\bar\tau] \approx e^{-\frac{c}{6\pi}\text{vol}(M)} \frac{|A|^2}{|\eta(\tau)|^2}  \frac{\sqrt{\mathrm{Im}(\tau)\mathrm{Im}(\tau_*)}}{2\pi^2\,|\tau+\tau_*|^2}\,.
\end{equation}
To arrive at this answer, we made use of Eqs.~\eqref{eq:whmicrocanonical} and \eqref{eq:whmicrocanonical2}. Notice that our result is similar to the torus wormhole partition function \cite{Cotler:2020ugk}, where now one of the boundary moduli is evaluated on $\tau_*$. The main new feature is the overall suppression factor $e^{-\frac{c}{6\pi}\text{vol}(M)}$, which makes this off-shell partition function non-perturbatively small in the large-$c$ (i.e.~$G_\text{N}\to 0$) limit. 

The partition function \eqref{eq:zgrav} can be modular completed by summing over the modular images of $\tau,\bar\tau$:
\begin{equation}\label{eq:modsum}
  Z^{\text{m.c.}}_\text{grav}[M;\tau,\bar\tau]  = \sum_{\gamma\in \text{PSL}(2,\mathbb{Z})} Z_\text{grav}[M;\gamma\tau,\gamma\bar\tau]
\end{equation}
where $\gamma$ acts on the boundary modulus by fractional linear transformations \eqref{eq:gamma}.  Interestingly, each term in the modular sum has a geometric interpretation: before gluing the tori in Figure \ref{fig:trumpet}, we apply a relative modular transformation $\gamma$ to obtain a manifold that we call $M_{r,s}$. On the level of the path integral, this modifies the gluing map by the generalized modular kernel $G_\text{T}\circ \mathbb{U}_{r,s}$, as in Sec.~\ref{sec:modcompl}:
\begin{align}
   Z^{\text{m.c.}}_\text{grav}[M;\tau,\bar\tau]  &= \sum_{(r,s)=1}\mel{\tau,\bar\tau}{\mathbb{U}_{r,s}\circ G_\text{T}}{Z_\text{grav}(M)} \\
    &= \sum_{(r,s)=1}\mel{\tau,\bar\tau}{ G_\text{T}}{Z_\text{grav}(M_{r,s})}.
\end{align}
In the last line, we used the commutation relation \eqref{eq:commute}. Unpacking this expression by resolving the identity, 
we see that the contribution to the average spectral density takes a similar form as Eq.~\eqref{eq:rhograv3}, but with $Z_\text{Vir}$ integrated against the modular completed spectral correlator \eqref{eq:modcompleted}.

If we use the large-$c$ approximation \eqref{eq:genvolconj}, we can obtain an explicit expression for the terms in the modular sum:
\begin{align}
     &\braket{P,\bar P}{Z_\text{grav}(M_{r,s})}   = \left|\int_0^\infty \dd P' \,\mathbb{U}(r,s)_{PP'}Z_\text{Vir}[M;P']\,\right|^2 \\
     &\qquad \approx e^{-\frac{c}{6\pi}\text{vol}(M)} \tfrac{ |A|^2}{\sqrt{\beta_*}} \left(\sqrt{\beta_*}\,e^{-2\pi \beta_* E}e^{2\pi i \sigma_* J}\right)\Big\vert_\gamma\,.
\end{align}
Here we wrote the cusp modulus as $\tau_* = \sigma_* + i\beta_*$ and changed variables to $E = P^2+\bar P^2$ and $J= P^2-\bar P^2$. The notation $(\dots)\mid_\gamma$ means that inside the brackets, we replace each $\tau_*$ by its modular image $\gamma\tau_*$\footnote{For example, $(\beta_*)\mid_\gamma \equiv \mathrm{Im}(\gamma\tau_*) = \frac{\mathrm{Im}(\tau_*)}{|s\tau_*+r|^2}$.} So we see that the sum over $\gamma$ produces a sum over generalized Eisenstein series, of the same type studied in Refs.~\cite{Maloney:2007ud,Keller:2014xba} and especially Appendix E of Ref.~\cite{Benjamin:2021ygh}.\footnote{We thank Eric Perlmutter for pointing out this reference.} The upshot is that, at least in the large-$c$ limit, the modular sum \eqref{eq:modsum} can be rendered finite by a natural regularization of the generalized Eisenstein series.

One possible interpretation of the manifolds we have constructed in this section is as follows. The single AdS$_3$ trumpet can be viewed as a solid torus with an internal Wilson loop with hyperbolic holonomy (i.e.~$P \in \mathbb{R}$) \cite{Mertens:2022ujr}. Such a Wilson loop describes a black hole horizon, in Euclidean signature. Gluing the  `internal' manifold $M$ as in Figure \ref{fig:trumpet} could therefore be interpreted as modeling a black hole with non-trivial topology behind the horizon. A similar idea was proposed in chiral 3d gravity \cite{Maloney:2015ina} for $M$ of the form $\Sigma_{g,1}\times S^1$ (see also \cite{Eberhardt:2022wlc}). The figure-8 knot complement \eqref{eq:fig8partfn} is an example of a non-trivially fibered internal manifold $\Sigma_{1,1}\times_\varphi S^1$, with $\varphi = ST^3$ \cite{Collier:2024mgv}. More general fibered 3-manifolds $\Sigma_{g,n}\times_\varphi S^1$ with multiple torus boundaries will be studied in \cite{toappear3}.

To summarize, in this section we studied a class of 3-manifolds whose bulk mapping class group acts non-trivially near each torus end. Using the prescription of the previous section for how to glue tori in 3d gravity, we showed that these manifolds give non-perturbative corrections to the average spectral density and its moments. Moreover, the sum over topologies provides a natural modular completion of these corrections.


\section{Seifert manifolds from Dehn surgery}\label{sec:seifert}

\begin{figure}
    \centering
    \resizebox{0.9\linewidth}{!}{
    \begin{tikzpicture}[x=0.75pt,y=0.75pt,yscale=-1,xscale=1]
\draw [line width=1.5]    (149.75,85.5) -- (190.75,85.16) ;
\draw [shift={(194.75,85.12)}, rotate = 179.52] [fill={rgb, 255:red, 0; green, 0; blue, 0 }  ][line width=0.08]  [draw opacity=0] (8.13,-3.9) -- (0,0) -- (8.13,3.9) -- cycle    ;
\draw [line width=1.5]    (136.5,164) .. controls (165.45,150.97) and (194.4,152.38) .. (225.6,166.76) ;
\draw [shift={(229,168.38)}, rotate = 206.03] [fill={rgb, 255:red, 0; green, 0; blue, 0 }  ][line width=0.08]  [draw opacity=0] (8.13,-3.9) -- (0,0) -- (8.13,3.9) -- cycle    ;
\draw    (60,85.25) -- (130,85.25) ;
\draw [color={rgb, 255:red, 255; green, 255; blue, 255 }  ,draw opacity=1 ][line width=1.5]    (102.93,85.16) -- (107.5,84.98) ;
\draw  [draw opacity=0][line width=1.5]  (85.07,82.38) .. controls (85.64,70.06) and (89.87,60.5) .. (95,60.5) .. controls (100.52,60.5) and (105,71.58) .. (105,85.25) .. controls (105,98.92) and (100.52,110) .. (95,110) .. controls (89.98,110) and (85.82,100.83) .. (85.11,88.88) -- (95,85.25) -- cycle ; \draw  [line width=1.5]  (85.07,82.38) .. controls (85.64,70.06) and (89.87,60.5) .. (95,60.5) .. controls (100.52,60.5) and (105,71.58) .. (105,85.25) .. controls (105,98.92) and (100.52,110) .. (95,110) .. controls (89.98,110) and (85.82,100.83) .. (85.11,88.88) ;  
\draw    (210,86.08) -- (280,86.08) ;
\draw [color={rgb, 255:red, 255; green, 255; blue, 255 }  ,draw opacity=1 ][line width=1.5]    (250,86.08) -- (260,86.08) ;
\draw  [draw opacity=0][line width=6]  (235.07,83.21) .. controls (235.66,71.36) and (239.88,62.17) .. (245,62.17) .. controls (250.52,62.17) and (255,72.87) .. (255,86.08) .. controls (255,99.29) and (250.52,110) .. (245,110) .. controls (239.99,110) and (235.85,101.2) .. (235.11,89.71) -- (245,86.08) -- cycle ; \draw  [line width=6]  (235.07,83.21) .. controls (235.66,71.36) and (239.88,62.17) .. (245,62.17) .. controls (250.52,62.17) and (255,72.87) .. (255,86.08) .. controls (255,99.29) and (250.52,110) .. (245,110) .. controls (239.99,110) and (235.85,101.2) .. (235.11,89.71) ;  
\draw  [draw opacity=0][line width=4.5]  (235.07,83.21) .. controls (235.66,71.36) and (239.88,62.17) .. (245,62.17) .. controls (250.52,62.17) and (255,72.87) .. (255,86.08) .. controls (255,99.29) and (250.52,110) .. (245,110) .. controls (239.99,110) and (235.85,101.2) .. (235.11,89.71) -- (245,86.08) -- cycle ; \draw  [color={rgb, 255:red, 155; green, 155; blue, 155 }  ,draw opacity=1 ][line width=4.5]  (235.07,83.21) .. controls (235.66,71.36) and (239.88,62.17) .. (245,62.17) .. controls (250.52,62.17) and (255,72.87) .. (255,86.08) .. controls (255,99.29) and (250.52,110) .. (245,110) .. controls (239.99,110) and (235.85,101.2) .. (235.11,89.71) ;  
\draw  [fill={rgb, 255:red, 155; green, 155; blue, 155 }  ,fill opacity=0.51 ][line width=0.75]  (60.4,186.25) .. controls (60.4,171.49) and (78.2,159.53) .. (100.17,159.53) .. controls (122.13,159.53) and (139.93,171.49) .. (139.93,186.25) .. controls (139.93,201.01) and (122.13,212.97) .. (100.17,212.97) .. controls (78.2,212.97) and (60.4,201.01) .. (60.4,186.25) -- cycle ;
\draw [fill={rgb, 255:red, 255; green, 255; blue, 255 }  ,fill opacity=1 ]   (88.34,185.84) .. controls (98.06,190.24) and (102.52,190.31) .. (112.68,185.57) ;
\draw [fill={rgb, 255:red, 255; green, 255; blue, 255 }  ,fill opacity=1 ]   (88.34,185.84) .. controls (100.41,180.13) and (104.17,182.51) .. (112.68,185.57) ;
\draw    (85.92,184.49) .. controls (86.14,184.76) and (88.92,186.18) .. (89.14,186.25) ;
\draw    (112.31,185.57) .. controls (112.53,185.84) and (114.14,184.62) .. (114.43,184.42) ;
\draw [color={rgb, 255:red, 208; green, 2; blue, 27 }  ,draw opacity=1 ] [dash pattern={on 1.5pt off 1.5pt on 1.5pt off 1.5pt}]  (101.43,189.14) .. controls (105.14,195.14) and (102.86,205.71) .. (109.14,212) ;
\draw [color={rgb, 255:red, 208; green, 2; blue, 27 }  ,draw opacity=1 ]   (109.14,212) .. controls (113.33,200.92) and (134.14,193.86) .. (133.14,184) .. controls (132.14,174.14) and (123.17,167.25) .. (101.83,165.58) .. controls (80.5,163.92) and (63.86,177.99) .. (67.43,190.29) .. controls (71,202.58) and (86.29,209.43) .. (107.43,201.43) .. controls (128.57,193.43) and (135.17,172.42) .. (100.29,171.43) .. controls (65.4,170.44) and (64.57,215.14) .. (101.43,189.14) ;
\draw    (210.33,186.08) -- (280.33,186.08) ;
\draw [color={rgb, 255:red, 255; green, 255; blue, 255 }  ,draw opacity=1 ][line width=1.5]    (250.33,186.08) -- (260.33,186.08) ;
\draw  [draw opacity=0][line width=6]  (235.4,183.21) .. controls (236,171.36) and (240.22,162.17) .. (245.33,162.17) .. controls (250.86,162.17) and (255.33,172.87) .. (255.33,186.08) .. controls (255.33,199.29) and (250.86,210) .. (245.33,210) .. controls (240.33,210) and (236.18,201.2) .. (235.45,189.71) -- (245.33,186.08) -- cycle ; \draw  [line width=6]  (235.4,183.21) .. controls (236,171.36) and (240.22,162.17) .. (245.33,162.17) .. controls (250.86,162.17) and (255.33,172.87) .. (255.33,186.08) .. controls (255.33,199.29) and (250.86,210) .. (245.33,210) .. controls (240.33,210) and (236.18,201.2) .. (235.45,189.71) ;  
\draw  [draw opacity=0][line width=4.5]  (235.4,183.21) .. controls (236,171.36) and (240.22,162.17) .. (245.33,162.17) .. controls (250.86,162.17) and (255.33,172.87) .. (255.33,186.08) .. controls (255.33,199.29) and (250.86,210) .. (245.33,210) .. controls (240.33,210) and (236.18,201.2) .. (235.45,189.71) -- (245.33,186.08) -- cycle ; \draw  [color={rgb, 255:red, 155; green, 155; blue, 155 }  ,draw opacity=1 ][line width=4.5]  (235.4,183.21) .. controls (236,171.36) and (240.22,162.17) .. (245.33,162.17) .. controls (250.86,162.17) and (255.33,172.87) .. (255.33,186.08) .. controls (255.33,199.29) and (250.86,210) .. (245.33,210) .. controls (240.33,210) and (236.18,201.2) .. (235.45,189.71) ;  
\draw  [draw opacity=0][line width=0.75]  (235.4,183.21) .. controls (236,171.36) and (240.22,162.17) .. (245.33,162.17) .. controls (250.86,162.17) and (255.33,172.87) .. (255.33,186.08) .. controls (255.33,199.29) and (250.86,210) .. (245.33,210) .. controls (240.33,210) and (236.18,201.2) .. (235.45,189.71) -- (245.33,186.08) -- cycle ; \draw  [color={rgb, 255:red, 208; green, 2; blue, 27 }  ,draw opacity=1 ][line width=0.75]  (235.4,183.21) .. controls (236,171.36) and (240.22,162.17) .. (245.33,162.17) .. controls (250.86,162.17) and (255.33,172.87) .. (255.33,186.08) .. controls (255.33,199.29) and (250.86,210) .. (245.33,210) .. controls (240.33,210) and (236.18,201.2) .. (235.45,189.71) ;  
\draw (150.86,68.17) node [anchor=north west][inner sep=0.75pt]  [font=\small] [align=left] {excise};
\draw (42,43.4) node [anchor=north west][inner sep=0.75pt]    {$1)$};
\draw (41.6,140) node [anchor=north west][inner sep=0.75pt]    {$2)$};
\draw (154.47,171.25) node [anchor=north west][inner sep=0.75pt]  [font=\small] [align=left] {glue in};
\draw (153.44,185.27) node [anchor=north west][inner sep=0.75pt]  [font=\small]  {$D\times S^{1}$};
\end{tikzpicture}}
    \caption{Dehn surgery on a link component.}
    \label{fig:Dehnsurgery}
\end{figure}
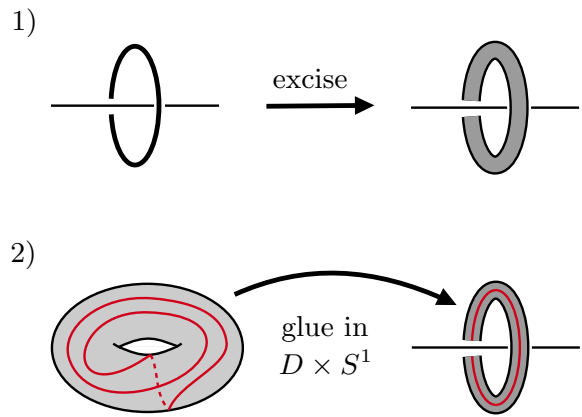

As a last example of a surgery method relevant for the statistical description of 3d gravity, we turn to a class of topologies known as \emph{Seifert manifolds}. It was argued in Ref.~\cite{Maxfield:2020ale} that summing over a class of single-boundary Seifert manifolds can render the average spectral density $\bar\rho$ into a positive distribution in the near-extremal limit. However, so far the evidence for this proposal is rather indirect and uses dimensional reduction to 2-dimensional JT gravity with conical defects.

Here we give a novel and purely 3-dimensional  framework for studying Seifert manifolds, using Dehn surgery. Let us first discuss some basics of Seifert manifolds, referring to Ref.~\cite{JankinsNeumann1983} for further details. A Seifert manifold is a fibration $f: M\to \Sigma_{g,n+1}$ over a Riemann surface, where a tubular neighborhood of each fiber looks like $D \times_\sigma S^1$. Here $D$ is a disk in the base and $\sigma$ is a rotation of the disk by an angle $\frac{2\pi r}{s}$, where $r$ and $s$ are co-prime. When $s=1$ we call the fiber ``regular'', and when $s>1$ ``exceptional''. Exceptional fibers correspond to conical defects on the base surface, with deficit angle $2\pi(1-\frac{1}{s})$.

We will focus on $g=0$. Every Seifert manifold that is fibered over an $n+1$-punctured sphere can be presented by the following Dehn surgery diagram \cite{budney2007jsj}:
\begin{equation}\label{eq:surgerydiagram}
    \mathcal{K}(1,\tfrac{r_1}{s_1},\dots,\tfrac{r_n}{s_n}) = \begin{tikzpicture}[x=0.75pt,y=0.75pt,yscale=-1,xscale=1,baseline={([yshift=-0.5ex]current bounding box.center)}]
\draw  [color={rgb, 255:red, 208; green, 2; blue, 27 }  ,draw opacity=1 ][line width=1.5]  (90.2,141.28) .. controls (90.2,120.23) and (123.57,103.18) .. (164.73,103.18) .. controls (205.88,103.18) and (239.25,120.23) .. (239.25,141.28) .. controls (239.25,162.32) and (205.88,179.38) .. (164.73,179.38) .. controls (123.57,179.38) and (90.2,162.32) .. (90.2,141.28) -- cycle ;
\draw [color={rgb, 255:red, 255; green, 255; blue, 255 }  ,draw opacity=1 ][line width=3]    (127.84,107.53) -- (133.1,107.34) ;
\draw  [draw opacity=0][line width=1.5]  (107.99,112.73) .. controls (107.77,110.74) and (107.65,108.66) .. (107.65,106.52) .. controls (107.65,92.29) and (112.8,80.76) .. (119.15,80.76) .. controls (125.5,80.76) and (130.65,92.29) .. (130.65,106.52) .. controls (130.65,120.75) and (125.5,132.29) .. (119.15,132.29) .. controls (115.11,132.29) and (111.56,127.62) .. (109.51,120.56) -- (119.15,106.52) -- cycle ; \draw  [line width=1.5]  (107.99,112.73) .. controls (107.77,110.74) and (107.65,108.66) .. (107.65,106.52) .. controls (107.65,92.29) and (112.8,80.76) .. (119.15,80.76) .. controls (125.5,80.76) and (130.65,92.29) .. (130.65,106.52) .. controls (130.65,120.75) and (125.5,132.29) .. (119.15,132.29) .. controls (115.11,132.29) and (111.56,127.62) .. (109.51,120.56) ;  
\draw [color={rgb, 255:red, 255; green, 255; blue, 255 }  ,draw opacity=1 ][line width=3.75]    (164.73,103.18) -- (169.98,102.99) ;
\draw  [draw opacity=0][line width=1.5]  (144.97,100.06) .. controls (146.25,88.97) and (150.75,80.78) .. (156.1,80.78) .. controls (162.45,80.78) and (167.6,92.31) .. (167.6,106.54) .. controls (167.6,120.77) and (162.45,132.31) .. (156.1,132.31) .. controls (150.22,132.31) and (145.37,122.43) .. (144.68,109.68) -- (156.1,106.54) -- cycle ; \draw  [line width=1.5]  (144.97,100.06) .. controls (146.25,88.97) and (150.75,80.78) .. (156.1,80.78) .. controls (162.45,80.78) and (167.6,92.31) .. (167.6,106.54) .. controls (167.6,120.77) and (162.45,132.31) .. (156.1,132.31) .. controls (150.22,132.31) and (145.37,122.43) .. (144.68,109.68) ;  
\draw [color={rgb, 255:red, 255; green, 255; blue, 255 }  ,draw opacity=1 ][line width=3.75]    (217.5,114.86) -- (222.36,118.1) ;
\draw  [draw opacity=0][line width=1.5]  (198.2,103.31) .. controls (198.92,90.64) and (203.75,80.84) .. (209.6,80.84) .. controls (215.96,80.84) and (221.1,92.38) .. (221.1,106.61) .. controls (221.1,120.84) and (215.96,132.37) .. (209.6,132.37) .. controls (203.89,132.37) and (199.14,123.02) .. (198.25,110.78) -- (209.6,106.61) -- cycle ; \draw  [line width=1.5]  (198.2,103.31) .. controls (198.92,90.64) and (203.75,80.84) .. (209.6,80.84) .. controls (215.96,80.84) and (221.1,92.38) .. (221.1,106.61) .. controls (221.1,120.84) and (215.96,132.37) .. (209.6,132.37) .. controls (203.89,132.37) and (199.14,123.02) .. (198.25,110.78) ;  
\draw (173.97,120.03) node [anchor=north west][inner sep=0.75pt]    {$\dotsc $};
\draw (92.25,73) node [anchor=north west][inner sep=0.75pt]   {$\frac{r_{1}}{s_{1}}$};
\draw (166.92,73) node [anchor=north west][inner sep=0.75pt]  {$\frac{r_{2}}{s_{2}}$};
\draw (220.58,73) node [anchor=north west][inner sep=0.75pt]   {$\frac{r_{n}}{s_{n}}$};
\draw (158,164.07) node [anchor=north west][inner sep=0.75pt]  [font=\small]  {$1$};
\end{tikzpicture}
\end{equation}

This notation refers to the following two-step procedure, which has been visualized in Figure \ref{fig:Dehnsurgery}:
\begin{enumerate}[1)]
    \item Embed the above diagram into $S^3$. Excise a tubular neighborhood around each ring, thus creating $n+1$ torus boundaries.
    \item For each torus boundary $T_i$, $i=1,\dots,n$, specify a meridian $[m]$. Then glue back a solid torus, identifying the closed curve $r_i[l'] + s_i[m']$ on the boundary of the solid torus with the curve $[m]$ on $T_i$. 
\end{enumerate}
Here $(r_i,s_i)$ are pairs of co-prime integers, presented as rational numbers in Eq.~\eqref{eq:surgerydiagram}, which specify the modular transformation $\gamma$ in \eqref{eq:modtrans} applied to the solid torus before gluing. We leave the torus boundary associated with the distinguished (red) loop unfilled, since our interest lies in single-boundary topologies.

The procedure of excising a solid torus and re-gluing it after some modular transformation is known as \emph{Dehn surgery}. Using the Heegaard splitting of $S^3$ into two solid tori, the result of Step 1 is equivalently described by an $n$-holed disk times $S^1$, shown below for $n=2$:
\begin{equation}\label{eq:solidtorus}
    \begin{tikzpicture}[x=0.75pt,y=0.75pt,yscale=-1,xscale=1,baseline={([yshift=-0.5ex]current bounding box.center)}]
\draw  [fill={rgb, 255:red, 255; green, 255; blue, 255 }  ,fill opacity=1 ] (326.17,142.63) .. controls (326.17,111.35) and (362.06,86) .. (406.33,86) .. controls (450.61,86) and (486.5,111.35) .. (486.5,142.63) .. controls (486.5,173.9) and (450.61,199.25) .. (406.33,199.25) .. controls (362.06,199.25) and (326.17,173.9) .. (326.17,142.63) -- cycle ;
\draw  [color={rgb, 255:red, 74; green, 74; blue, 74 }  ,draw opacity=1 ][fill={rgb, 255:red, 128; green, 128; blue, 128 }  ,fill opacity=0.3 ] (338.77,142.63) .. controls (338.77,118.97) and (369.02,99.79) .. (406.33,99.79) .. controls (443.65,99.79) and (473.89,118.97) .. (473.89,142.63) .. controls (473.89,166.28) and (443.65,185.46) .. (406.33,185.46) .. controls (369.02,185.46) and (338.77,166.28) .. (338.77,142.63) -- cycle ;
\draw  [color={rgb, 255:red, 74; green, 74; blue, 74 }  ,draw opacity=1 ][fill={rgb, 255:red, 255; green, 255; blue, 255 }  ,fill opacity=1 ] (349.26,142.65) .. controls (349.26,124.65) and (374.81,110.06) .. (406.33,110.06) .. controls (437.86,110.06) and (463.41,124.65) .. (463.41,142.65) .. controls (463.41,160.64) and (437.86,175.23) .. (406.33,175.23) .. controls (374.81,175.23) and (349.26,160.64) .. (349.26,142.65) -- cycle ;
\draw  [color={rgb, 255:red, 74; green, 74; blue, 74 }  ,draw opacity=1 ][fill={rgb, 255:red, 155; green, 155; blue, 155 }  ,fill opacity=0.3 ] (363.38,142.63) .. controls (363.38,130.09) and (383.06,119.93) .. (407.34,119.93) .. controls (431.62,119.93) and (451.31,130.09) .. (451.31,142.63) .. controls (451.31,155.16) and (431.62,165.32) .. (407.34,165.32) .. controls (383.06,165.32) and (363.38,155.16) .. (363.38,142.63) -- cycle ;
\draw  [color={rgb, 255:red, 74; green, 74; blue, 74 }  ,draw opacity=1 ][fill={rgb, 255:red, 255; green, 255; blue, 255 }  ,fill opacity=1 ] (373.86,142.62) .. controls (373.86,135.07) and (388.85,128.95) .. (407.34,128.95) .. controls (425.83,128.95) and (440.82,135.07) .. (440.82,142.62) .. controls (440.82,150.18) and (425.83,156.3) .. (407.34,156.3) .. controls (388.85,156.3) and (373.86,150.18) .. (373.86,142.62) -- cycle ;
\draw  [draw opacity=0] (427.21,140.81) .. controls (424.03,145.76) and (416.46,149.24) .. (407.64,149.24) .. controls (398.09,149.24) and (390.01,145.15) .. (387.38,139.54) -- (407.64,135.57) -- cycle ; \draw   (427.21,140.81) .. controls (424.03,145.76) and (416.46,149.24) .. (407.64,149.24) .. controls (398.09,149.24) and (390.01,145.15) .. (387.38,139.54) ;  
\draw  [draw opacity=0] (390.47,143.69) .. controls (392.47,139.73) and (399.13,136.82) .. (407.04,136.82) .. controls (415.22,136.82) and (422.07,139.94) .. (423.81,144.11) -- (407.04,146.19) -- cycle ; \draw   (390.47,143.69) .. controls (392.47,139.73) and (399.13,136.82) .. (407.04,136.82) .. controls (415.22,136.82) and (422.07,139.94) .. (423.81,144.11) ;  
\draw  [draw opacity=0] (412.82,156.13) .. controls (414.48,156.3) and (415.75,158.34) .. (415.67,160.77) .. controls (415.6,163.27) and (414.13,165.24) .. (412.4,165.18) .. controls (412.26,165.17) and (412.12,165.15) .. (411.98,165.12) -- (412.54,160.65) -- cycle ; \draw   (412.82,156.13) .. controls (414.48,156.3) and (415.75,158.34) .. (415.67,160.77) .. controls (415.6,163.27) and (414.13,165.24) .. (412.4,165.18) .. controls (412.26,165.17) and (412.12,165.15) .. (411.98,165.12) ;  
\draw  [draw opacity=0][line width=1.5]  (400.98,174.08) .. controls (400.98,173.97) and (400.98,173.86) .. (400.98,173.76) .. controls (400.98,171.37) and (401.15,169.07) .. (401.47,166.89) -- (413.45,173.76) -- cycle ; \draw  [color={rgb, 255:red, 208; green, 2; blue, 27 }  ,draw opacity=1 ][line width=1.5]  (400.98,174.08) .. controls (400.98,173.97) and (400.98,173.86) .. (400.98,173.76) .. controls (400.98,171.37) and (401.15,169.07) .. (401.47,166.89) ;  
\draw  [draw opacity=0] (414.1,174.98) .. controls (415.49,175.2) and (416.52,177.46) .. (416.44,180.16) .. controls (416.36,182.97) and (415.09,185.21) .. (413.62,185.15) .. controls (413.46,185.14) and (413.3,185.11) .. (413.15,185.05) -- (413.78,180.06) -- cycle ; \draw   (414.1,174.98) .. controls (415.49,175.2) and (416.52,177.46) .. (416.44,180.16) .. controls (416.36,182.97) and (415.09,185.21) .. (413.62,185.15) .. controls (413.46,185.14) and (413.3,185.11) .. (413.15,185.05) ;  
\draw  [line width=1.5]  (343.95,142.62) .. controls (343.95,121.38) and (371.88,104.15) .. (406.33,104.15) .. controls (440.79,104.15) and (468.71,121.38) .. (468.71,142.62) .. controls (468.71,163.87) and (440.79,181.1) .. (406.33,181.1) .. controls (371.88,181.1) and (343.95,163.87) .. (343.95,142.62) -- cycle ;
\draw  [line width=1.5]  (368.53,142.62) .. controls (368.53,132.27) and (385.91,123.88) .. (407.34,123.88) .. controls (428.78,123.88) and (446.16,132.27) .. (446.16,142.62) .. controls (446.16,152.98) and (428.78,161.37) .. (407.34,161.37) .. controls (385.91,161.37) and (368.53,152.98) .. (368.53,142.62) -- cycle ;
\draw  [draw opacity=0][line width=1.5]  (404.67,156.4) .. controls (406.92,152.02) and (410.03,149.31) .. (413.45,149.31) .. controls (420.34,149.31) and (425.93,160.25) .. (425.93,173.76) .. controls (425.93,187.26) and (420.34,198.2) .. (413.45,198.2) .. controls (408.6,198.2) and (404.39,192.77) .. (402.33,184.84) -- (413.45,173.76) -- cycle ; \draw  [color={rgb, 255:red, 208; green, 2; blue, 27 }  ,draw opacity=1 ][line width=1.5]  (404.67,156.4) .. controls (406.92,152.02) and (410.03,149.31) .. (413.45,149.31) .. controls (420.34,149.31) and (425.93,160.25) .. (425.93,173.76) .. controls (425.93,187.26) and (420.34,198.2) .. (413.45,198.2) .. controls (408.6,198.2) and (404.39,192.77) .. (402.33,184.84) ;  
\end{tikzpicture}
\end{equation}
which is a multi-boundary torus wormhole of the type studied in Ref.~\cite{Post:2024itb}. After doing Step 2, the exceptional fibers of the Seifert manifold lie inside the gray solid tori, with a choice of contractible cycle specified by $(r_i,s_i)$.

\subsection{Seifert partition function}
Let us attempt to compute the gravitational path integral for the genus-zero Seifert manifolds in Eq.~\eqref{eq:surgerydiagram}. We will refer to the link in the Dehn surgery diagram as the \emph{keychain link}. Since the keychain link has $n+1$ torus boundaries, the gravitational phase space consists of $n+1$ copies of the torus' Teichm\"uller space $\mathcal{T}\times \overline{\mathcal{T}}$. So, after Step 1, the path integral on the keychain link complement $\mathcal{K}$ defines a state in the tensor product Hilbert space of each torus boundary:
\begin{equation}
    \ket{Z_\text{grav}(\mathcal{K})} \in \mathcal{H}_{\text{T}_0}\otimes \dots \otimes \mathcal{H}_{\text{T}_{n}}\,.
\end{equation}
Its matrix elements in the basis of Virasoro characters are computed by evaluating the path integral with boundary conditions set by fixed holonomies (in the Chern-Simons formulation) around the longitudinal cycles of the boundary tori, which are parametrized by Liouville momenta $P_i$ and their anti-holomorphic counterparts $\bar P_i$:
\begin{multline}\label{eq:keychain2}
    \braket{P_0,\bar P_0; \dots; P_n,\bar P_n}{Z_\text{grav}(\mathcal{K})} = \\[1em]
    Z_\text{grav}\left(\,\begin{tikzpicture}[x=0.75pt,y=0.75pt,yscale=-0.7,xscale=0.7,baseline={([yshift=-0.5ex]current bounding box.center)}]
\draw  [color={rgb, 255:red, 208; green, 2; blue, 27 }  ,draw opacity=1 ][line width=1.5]  (90.2,141.28) .. controls (90.2,120.23) and (123.57,103.18) .. (164.73,103.18) .. controls (205.88,103.18) and (239.25,120.23) .. (239.25,141.28) .. controls (239.25,162.32) and (205.88,179.38) .. (164.73,179.38) .. controls (123.57,179.38) and (90.2,162.32) .. (90.2,141.28) -- cycle ;
\draw [color={rgb, 255:red, 255; green, 255; blue, 255 }  ,draw opacity=1 ][line width=3]    (127.84,107.53) -- (133.1,107.34) ;
\draw  [draw opacity=0][line width=1.5]  (107.99,112.73) .. controls (107.77,110.74) and (107.65,108.66) .. (107.65,106.52) .. controls (107.65,92.29) and (112.8,80.76) .. (119.15,80.76) .. controls (125.5,80.76) and (130.65,92.29) .. (130.65,106.52) .. controls (130.65,120.75) and (125.5,132.29) .. (119.15,132.29) .. controls (115.11,132.29) and (111.56,127.62) .. (109.51,120.56) -- (119.15,106.52) -- cycle ; \draw  [line width=1.5]  (107.99,112.73) .. controls (107.77,110.74) and (107.65,108.66) .. (107.65,106.52) .. controls (107.65,92.29) and (112.8,80.76) .. (119.15,80.76) .. controls (125.5,80.76) and (130.65,92.29) .. (130.65,106.52) .. controls (130.65,120.75) and (125.5,132.29) .. (119.15,132.29) .. controls (115.11,132.29) and (111.56,127.62) .. (109.51,120.56) ;   
\draw [color={rgb, 255:red, 255; green, 255; blue, 255 }  ,draw opacity=1 ][line width=3.75]    (217.5,114.86) -- (222.36,118.1) ;
\draw  [draw opacity=0][line width=1.5]  (198.2,103.31) .. controls (198.92,90.64) and (203.75,80.84) .. (209.6,80.84) .. controls (215.96,80.84) and (221.1,92.38) .. (221.1,106.61) .. controls (221.1,120.84) and (215.96,132.37) .. (209.6,132.37) .. controls (203.89,132.37) and (199.14,123.02) .. (198.25,110.78) -- (209.6,106.61) -- cycle ; \draw  [line width=1.5]  (198.2,103.31) .. controls (198.92,90.64) and (203.75,80.84) .. (209.6,80.84) .. controls (215.96,80.84) and (221.1,92.38) .. (221.1,106.61) .. controls (221.1,120.84) and (215.96,132.37) .. (209.6,132.37) .. controls (203.89,132.37) and (199.14,123.02) .. (198.25,110.78) ;  
\draw (128.25,75) node [anchor=north west][inner sep=0.75pt]  [font=\footnotesize] {$P_1,\bar P_1 \dots$};
\draw (220.58,75) node [anchor=north west][inner sep=0.75pt]  [font=\footnotesize] {$P_n,\bar P_n$};
\draw (146,155.07) node [anchor=north west][inner sep=0.75pt]  [font=\footnotesize]  {$P_0,\bar P_0$};
\end{tikzpicture}\right).
\end{multline}
Let us for the moment postpone the evaluation of the right-hand side and proceed to Step 2 of the Dehn surgery prescription. Filling in $n$ of the $n+1$ boundaries of $\mathcal{K}$ by solid tori, after acting with a modular transformation, is represented in the Hilbert space formalism of the path integral by  the following inner product:
\begin{equation}\label{eq:innpr}
   \big(\bra{\bbi,\bar\bbi}\mathbb{U}_{r_1,s_1}\cdots \bra{\bbi,\bar\bbi}\mathbb{U}_{r_n,s_n}\big)\ket{Z_\text{grav}(\mathcal{K})}.
\end{equation}
The `empty' solid torus that we are gluing in is represented by the vacuum character $\ket{\bbi,\bar \bbi}$ \cite{Collier:2024mgv}, and we have used the generalized modular kernels $\mathbb{U}_{r,s}$ to implement the gluing map. Their matrix elements
\begin{equation}\label{eq:identitykernels}
    \mel{\bbi,\bar\bbi}{\mathbb{U}_{r,s}}{P,\bar P} = \mathbb{U}(r,s)_{\bbi P}\,\mathbb{U}(r,s)^*_{\bbi \bar P}
\end{equation}
are given explicitly in Eq.~\eqref{eq:gen_kernel}. The fact that Dehn surgery is represented in 3d gravity by the identity modular kernels was also used in Ref.~\cite{Collier:2024mgv} to study Dehn fillings of the figure-8 knot complement. 

Having filled in the $n$ torus boundaries associated with the black rings in $\mathcal{K}$, the result of Eq.~\eqref{eq:innpr} is still a state in the single-boundary Hilbert space $\mathcal{H}_\text{T}$ associated with the red ring. We can compute its contribution to the single-boundary partition function $\overline{Z(\tau,\bar \tau)}$ by going to the wavefunction basis $\bra{\tau,\bar\tau}$. This lets us read off the contribution to the average spectral density.

As an example, let us work out $n=1$ in detail. The contribution to the single-boundary partition function is
\begin{align}\label{eq:Z_MWK}
    &\big(\bra{\tau,\bar\tau}\otimes\bra{\bbi,\bar\bbi}\mathbb{U}_{r,s}\big)\ket{Z_\text{grav}(\mathcal{K})} \\
    &= \int_0^\infty\dd^4 P\,|\chi_{P_0}(\tau)|^2 |\mathbb{U}(r,s)_{\bbi P_1}|^2 Z_{\text{grav}}\left(\begin{tikzpicture}[x=0.75pt,y=0.75pt,yscale=-1,xscale=1,baseline={([yshift=-0.5ex]current bounding box.center)}]
\draw  [color={rgb, 255:red, 208; green, 2; blue, 27 }  ,draw opacity=1 ][line width=1.5]  (110.2,135.98) .. controls (110.2,123.43) and (120.14,113.25) .. (132.39,113.25) .. controls (144.65,113.25) and (154.58,123.43) .. (154.58,135.98) .. controls (154.58,148.54) and (144.65,158.71) .. (132.39,158.71) .. controls (120.14,158.71) and (110.2,148.54) .. (110.2,135.98) -- cycle ;
\draw [color={rgb, 255:red, 255; green, 255; blue, 255 }  ,draw opacity=1 ][line width=4.5]    (138.41,113.75) -- (141.25,114.25) ;
\draw  [draw opacity=0][line width=1.5]  (126.07,111.84) .. controls (126.75,104.91) and (129.53,99.72) .. (132.87,99.72) .. controls (136.71,99.72) and (139.82,106.6) .. (139.82,115.09) .. controls (139.82,123.58) and (136.71,130.46) .. (132.87,130.46) .. controls (129.26,130.46) and (126.29,124.38) .. (125.95,116.59) -- (132.87,115.09) -- cycle ; \draw  [line width=1.5]  (126.07,111.84) .. controls (126.75,104.91) and (129.53,99.72) .. (132.87,99.72) .. controls (136.71,99.72) and (139.82,106.6) .. (139.82,115.09) .. controls (139.82,123.58) and (136.71,130.46) .. (132.87,130.46) .. controls (129.26,130.46) and (126.29,124.38) .. (125.95,116.59) ;  
\draw (116.48,139) node [anchor=north west][inner sep=0.75pt]  [font=\scriptsize]  {$P_{0}, \bar P_{0}$};
\draw (141.18,98.67) node [anchor=north west][inner sep=0.75pt]  [font=\scriptsize]  {$\bar P_{1}$};
\draw (110.18,99.67) node [anchor=north west][inner sep=0.75pt]  [font=\scriptsize]  {$P_{1}$};
\end{tikzpicture}\right) \nonumber \\
&= \int_0^\infty\dd^4 P\, \big|\mathbb{U}(r,s)_{\bbi P_1}\mathbb{S}_{P_1P_0}[\bbi]\,\chi_{P_0}(\tau)\big|^2\label{eq:79}
\end{align}
where $\dd^4P\equiv \dd P_0\dd\bar P_0\dd P_1\dd\bar P_1$. In the last equality we have used that the Hopf link evaluates to the modular S-kernel in Virasoro TQFT. To perform the $P_1$ integral, we note that $\mathbb{U}(\gamma)\circ \mathbb{S} = \mathbb{U}(\gamma\circ S)$, since $\mathbb{U}$ is a representation. Multiplying $\gamma$ and $S$, as matrices in $\mathrm{PSL}(2,\mathbb{Z})$, simply sends $(r,s)\mapsto (-s,r)$. So the $P_1,\bar P_1$ integrals in Eq.~\eqref{eq:79} can be done, and  after summing over all Dehn filling parameters $(r,s)$ we find
\begin{equation}
    \overline{Z(\tau,\bar\tau)} \supset \sum_{(r,s)=1} \Big|\int_0^\infty \!\dd P_0\, \mathbb{U}(-s,r)_{\bbi P_0} \,\chi_{P_0}(\tau)\Big|^2.
\end{equation}
This coincides with the partition function of the $\text{PSL}(2,\mathbb{Z})$ family of BTZ black holes, computed and analyzed by Maloney, Witten and Keller (MWK) in \cite{Maloney:2007ud,Keller:2014xba}. It contributes to the average density of primary states as
\begin{equation}
    \overline{\rho(P,\bar P)} \supset \sum_{(r,s)=1} \big |\mathbb{U}(-s,r)_{\bbi P} \big|^2 = \rho_{\text{MWK}}(P,\bar P).
\end{equation}
So, we see that our construction contains as a special case ($n=1$) the sum over all on-shell solutions of the gravitational path integral with a single torus boundary. Every single-boundary Seifert manifold with $n>1$ is necessarily \emph{off-shell}. Indeed, we can see from the figure in Eq.~\eqref{eq:solidtorus}, which has $n=2$, that there is an incompressible torus (surrounding the two gray tubes). More generally, there is an incompressible torus above every non-contractible cycle in the base $\Sigma_{0,n+1}$ of the Seifert fibration, thus violating the criterion for hyperbolicity in the geometrization theorem. 

Before attempting to compute these off-shell contributions with $n>1$, let us discuss when two Seifert manifolds are isomorphic, given the data $\{\frac{r_i}{s_i}\}_{i=1}^n$. This detail is important for determining which set of co-primes should be included in the sum over topologies, ensuring that each topology is counted only once. Fortunately, there is a simple classification theorem: two Seifert manifolds are isomorphic if and only if they have the same base surface $\Sigma_{g,n}$ \textbf{and} they can be related by the following three operations \cite{JankinsNeumann1983}\footnote{There is one more operation in Theorem 1.5 of \cite{JankinsNeumann1983}, which is to replace $(\pm 1,0)$ by $(\mp 1,0)$. However, this is a trivial operation in our case, because we consider $\gamma\in \mathrm{PSL}(2,\mathbb{Z}) = \mathrm{SL}(2,\mathbb{Z})/\{\pm I\}$.}:
\begin{enumerate}[(1)]
    \item Permute the $n$ pairs $(r_i,s_i)$,
    \item Add or remove any pair that has $(r,s) = (0,1)$,
    \item Replace every pair $(r_i,s_i)$ by $(r_i+k_is_i,s_i)$, provided that $\sum_{i=1}^n k_i =0$.
\end{enumerate}
The first operation is obvious: it amounts to sliding the rings in the diagram \eqref{eq:surgerydiagram} underneath each other. The second operation is also clear from the surgery diagram: $(r,s) = (0,1)$ corresponds to $\gamma = 1$, which implements the trivial Dehn filling in VTQFT.
The third operation is realized in VTQFT by noting that a shift of $r\to r+ks$ with $k\in \mathbb{Z}$ is implemented by the Dehn twist $T^k$, so the modular kernels used for Dehn filling are also modified to
\begin{equation}
    \prod_{i=1}^n\mathbb{U}(r_i,s_i)_{\bbi P_i} \longrightarrow \prod_{i=1}^n\mathbb{U}(r_i,s_i)_{\bbi P_i} (\mathbb{T}_{P_i})^{k_i}.
\end{equation}
Multiplying by the anti-holomorphic counterpart, we get an overall factor
$
    \exp(2\pi i \sum_{i=1}^n k_i (P_i^2-\bar P_i^2)).
$
 If the keychain link partition function is spin-diagonal (meaning that it has an overall $\prod_{m\neq n}\delta(J_m-J_n$)), then the constraint $\sum_i k_i = 0$ indeed ensures invariance of the Seifert partition function under the operation (3). While spin-diagonality (before modular completion) is expected based on the maximum ignorance principle discussed in Section \ref{sec:RMT}, it still has to be verified explicitly in Eq.~\eqref{eq:keychain2}, as we will discuss below.

We can use the three operations above to bring the set of surgery data into a canonical form, with $s_i > r_i>0$ for $i\geq 2$. Doing so, the sum over all genus-$0$ Seifert manifolds takes the following form:
\begin{equation}\label{eq:sumoverseifert}
    \overline{\rho(P,\bar P)} \supset \rho_{\text{MWK}}(P,\bar P) + \sum_{n=2}^\infty \frac{1}{n!}\,\rho_{\text{Seifert}}^{(n)}(P,\bar P),
\end{equation}
where we have separated the $n=1$ contribution from the Seifert manifolds with $n>1$ exceptional fibers:
\begin{multline}\label{eq:seif}
    \rho_{\text{Seifert}}^{(n)}(P_0,\bar P_0)\coloneqq \\ \sum_{\mathcal{S}_n} \int_0^\infty \prod_{i=1}^n \dd P_i\dd \bar P_i \,\left|\mathbb{U}(r_i,s_i)_{\bbi P_i}\right|^2 Z_\text{grav}\!\left(\mathcal{K};\mathbf{P},\bar{\mathbf{P}}\right).
\end{multline}
Here $\mathcal{K}$ is the keychain link and $\mathbf{P} = (P_0,P_1,\dots,P_n)$. The set $\mathcal{S}_n$ over which we are summing contains all pairs of co-prime integers $\{(r_1,s_1),\dots, (r_n,s_n)\}$, modulo the constraint that $s_i>r_i>0$ for $i\geq 2$. 

Equations \eqref{eq:sumoverseifert} and \eqref{eq:seif} are a concrete proposal for the sum over Seifert manifolds in 3d gravity from Dehn surgery. However, it remains to evaluate the keychain link partition function on the RHS of Eq.~\eqref{eq:keychain2}, which we refer to as $Z_\text{grav}\!\left(\mathcal{K};\mathbf{P},\bar{\mathbf{P}}\right)$. Naively, we could use VTQFT and the Verlinde loop operator \eqref{eq:TQFTrule} to simplify the keychain link, by successively unlinking the black rings:
\begin{align}\label{eq:vtqft_unlink}
    \!Z_\text{Vir}\!\left(\begin{tikzpicture}[x=0.75pt,y=0.75pt,yscale=-0.6,xscale=0.55,baseline={([yshift=-0.5ex]current bounding box.center)}]
\draw  [color={rgb, 255:red, 208; green, 2; blue, 27 }  ,draw opacity=1 ][line width=1.5]  (90.2,141.28) .. controls (90.2,120.23) and (123.57,103.18) .. (164.73,103.18) .. controls (205.88,103.18) and (239.25,120.23) .. (239.25,141.28) .. controls (239.25,162.32) and (205.88,179.38) .. (164.73,179.38) .. controls (123.57,179.38) and (90.2,162.32) .. (90.2,141.28) -- cycle ;
\draw [color={rgb, 255:red, 255; green, 255; blue, 255 }  ,draw opacity=1 ][line width=3]    (127.84,107.53) -- (133.1,107.34) ;
\draw  [draw opacity=0][line width=1.5]  (107.99,112.73) .. controls (107.77,110.74) and (107.65,108.66) .. (107.65,106.52) .. controls (107.65,92.29) and (112.8,80.76) .. (119.15,80.76) .. controls (125.5,80.76) and (130.65,92.29) .. (130.65,106.52) .. controls (130.65,120.75) and (125.5,132.29) .. (119.15,132.29) .. controls (115.11,132.29) and (111.56,127.62) .. (109.51,120.56) -- (119.15,106.52) -- cycle ; \draw  [line width=1.5]  (107.99,112.73) .. controls (107.77,110.74) and (107.65,108.66) .. (107.65,106.52) .. controls (107.65,92.29) and (112.8,80.76) .. (119.15,80.76) .. controls (125.5,80.76) and (130.65,92.29) .. (130.65,106.52) .. controls (130.65,120.75) and (125.5,132.29) .. (119.15,132.29) .. controls (115.11,132.29) and (111.56,127.62) .. (109.51,120.56) ;  
\draw [color={rgb, 255:red, 255; green, 255; blue, 255 }  ,draw opacity=1 ][line width=3.75]    (164.73,103.18) -- (169.98,102.99) ;
\draw  [draw opacity=0][line width=1.5]  (144.97,100.06) .. controls (146.25,88.97) and (150.75,80.78) .. (156.1,80.78) .. controls (162.45,80.78) and (167.6,92.31) .. (167.6,106.54) .. controls (167.6,120.77) and (162.45,132.31) .. (156.1,132.31) .. controls (150.22,132.31) and (145.37,122.43) .. (144.68,109.68) -- (156.1,106.54) -- cycle ; \draw  [line width=1.5]  (144.97,100.06) .. controls (146.25,88.97) and (150.75,80.78) .. (156.1,80.78) .. controls (162.45,80.78) and (167.6,92.31) .. (167.6,106.54) .. controls (167.6,120.77) and (162.45,132.31) .. (156.1,132.31) .. controls (150.22,132.31) and (145.37,122.43) .. (144.68,109.68) ;  
\draw [color={rgb, 255:red, 255; green, 255; blue, 255 }  ,draw opacity=1 ][line width=3.75]    (217.5,114.86) -- (222.36,118.1) ;
\draw  [draw opacity=0][line width=1.5]  (198.2,103.31) .. controls (198.92,90.64) and (203.75,80.84) .. (209.6,80.84) .. controls (215.96,80.84) and (221.1,92.38) .. (221.1,106.61) .. controls (221.1,120.84) and (215.96,132.37) .. (209.6,132.37) .. controls (203.89,132.37) and (199.14,123.02) .. (198.25,110.78) -- (209.6,106.61) -- cycle ; \draw  [line width=1.5]  (198.2,103.31) .. controls (198.92,90.64) and (203.75,80.84) .. (209.6,80.84) .. controls (215.96,80.84) and (221.1,92.38) .. (221.1,106.61) .. controls (221.1,120.84) and (215.96,132.37) .. (209.6,132.37) .. controls (203.89,132.37) and (199.14,123.02) .. (198.25,110.78) ;  
\draw [font = \small](167.97,120.03) node [anchor=north west][inner sep=0.75pt]    {$\dotsc $};
\draw [font = \scriptsize] (124.25,73) node [anchor=north west][inner sep=0.75pt]   {$P_1$};
\draw [font = \scriptsize] (166.92,73) node [anchor=north west][inner sep=0.75pt]  {$P_2$};
\draw [font = \scriptsize] (218.58,73) node [anchor=north west][inner sep=0.75pt]   {$P_n$};
\draw (158,157.07) node [anchor=north west][inner sep=0.75pt]  [font=\scriptsize]  {$P_0$};
\end{tikzpicture}\right) &= \prod_{i=2}^{n}\frac{\mathbb{S}_{P_iP_0}[\bbi]}{\mathbb{S}_{\bbi P_0}[\bbi]}Z_{\text{Vir}}\!\left(\begin{tikzpicture}[x=0.75pt,y=0.75pt,yscale=-1,xscale=1,baseline={([yshift=-0.5ex]current bounding box.center)}]
\draw  [color={rgb, 255:red, 208; green, 2; blue, 27 }  ,draw opacity=1 ][line width=1.5]  (110.2,135.98) .. controls (110.2,123.43) and (120.14,113.25) .. (132.39,113.25) .. controls (144.65,113.25) and (154.58,123.43) .. (154.58,135.98) .. controls (154.58,148.54) and (144.65,158.71) .. (132.39,158.71) .. controls (120.14,158.71) and (110.2,148.54) .. (110.2,135.98) -- cycle ;
\draw [color={rgb, 255:red, 255; green, 255; blue, 255 }  ,draw opacity=1 ][line width=4.5]    (138.41,113.75) -- (141.25,114.25) ;
\draw  [draw opacity=0][line width=1.5]  (126.07,111.84) .. controls (126.75,104.91) and (129.53,99.72) .. (132.87,99.72) .. controls (136.71,99.72) and (139.82,106.6) .. (139.82,115.09) .. controls (139.82,123.58) and (136.71,130.46) .. (132.87,130.46) .. controls (129.26,130.46) and (126.29,124.38) .. (125.95,116.59) -- (132.87,115.09) -- cycle ; \draw  [line width=1.5]  (126.07,111.84) .. controls (126.75,104.91) and (129.53,99.72) .. (132.87,99.72) .. controls (136.71,99.72) and (139.82,106.6) .. (139.82,115.09) .. controls (139.82,123.58) and (136.71,130.46) .. (132.87,130.46) .. controls (129.26,130.46) and (126.29,124.38) .. (125.95,116.59) ;  
\draw (125.48,145) node [anchor=north west][inner sep=0.75pt]  [font=\scriptsize]  {$P_0$};
\draw (110.18,99.67) node [anchor=north west][inner sep=0.75pt]  [font=\scriptsize]  {$P_{1}$};
\end{tikzpicture}\right) \nonumber\\
&= \frac{1}{\rho_0(P_0)^{n-1}} \prod_{i=1}^n \mathbb{S}_{P_iP_0}[\bbi].
\end{align}
This is a straightforward generalization of the $n=1$ computation given in Eq.~\eqref{eq:Z_MWK}.
When we substitute the above result into Eq.~\eqref{eq:seif}, all the $P_i$-integrals can be done, and we would obtain
\begin{equation}\label{eq:Seifert?}
     \rho_{\text{Seifert}}^{(n)}(P_0,\bar P_0) \stackrel{?}{=} \,\sum_{\mathcal{S}_n}\left | \frac{\prod_{i=1}^n\mathbb{U}(-s_i,r_i)_{\bbi P_0}}{\rho_0(P_0)^{n-1}} \right|^2\,.
\end{equation}
While this answer is appealingly simple, it misses one crucial aspect of the gravitational path integral: we did not take into account the bulk mapping class group.

As before, by \emph{bulk} mapping class group we mean the group of orientation-preserving diffeomorphisms of a 3-manifold $M$ that are not continuously connected to the identity and that act trivially on the boundary (referred to as ``Map$_0(M,\partial M)$'' in Ref.~\cite{Collier:2023fwi}). 
Off-shell 3-manifolds can have infinite bulk mapping class groups, which should be gauged in gravity. Indeed, the Seifert manifolds with $n>1$ constructed in this section have a bulk mapping class group that is generated by \cite{Johannson1979}:
\begin{enumerate}[(1)]
    \item The elements of the (orbifold) mapping class group of the base surface $\Sigma_{g,n+1}$,
    \item $2\pi$ twists along the regular fibers $S^1$, also known as \emph{vertical Dehn twists}.
\end{enumerate}
More precisely, there is a surjective homomorphism from the bulk mapping class group $\text{MCG}(M)$ to the base mapping class group $\text{MCG}_{\text{orb}}(\Sigma_{g,n+1})$ (the subscript ``orb'' stands for orbifold and signifies that the cone points of the base are mapped to cone points of the same order). The kernel of this homomorphism is the first homology group of $\Sigma_{g,n+1}$. These groups fit into a short exact sequence that splits, see Prop.~25.3 in Ref.~\cite{Johannson1979}, so the bulk mapping class group has the structure of a semi-direct product. For the $g=0$ Seifert manifolds considered in this section, with a single boundary and $n$ exceptional fibers, we conclude that the bulk mapping class group is
\begin{equation}\label{eq:seifertMCG}
   \text{MCG}_{\text{orb}}(\Sigma_{0,n+1}) \ltimes \mathbb{Z}^n\,.
\end{equation}
Here we used that the first homology group (over the integers) of the surface $\Sigma_{0,n+1}$ is $\mathbb{Z}^n$. Intuitively, this extra $\mathbb{Z}^n$ is explained by large diffeomorphisms that act as vertical Dehn twists along the $S^1$ fiber for each of the $n$ cycles in a homology basis for $\Sigma_{0,n+1}$. 

The upshot of this discussion is that the bulk MCG in Eq.~\eqref{eq:seifertMCG} is infinite (even for $\Sigma_{0,3}$), hence the VTQFT formalism that was used to arrive at Eq.~\eqref{eq:vtqft_unlink} does not apply to the gravitational path integral on a Seifert manifold. It applies to any 3d TQFT, but it does not take into account the fact that the MCG is gauged in gravity. We should stress, however, that even though we do no currently know how to compute $Z_\text{grav}\!\left(\mathcal{K};\mathbf{P},\bar{\mathbf{P}}\right)$, our Dehn surgery prescription in Eq.~\eqref{eq:seif} still holds true in gravity. 
This is because the bulk MCG of the keychain link $\mathcal{K}$ is the same as that of the associated family of Dehn-filled Seifert manifolds, since large diffeomorphisms restrict to the identity on the exceptional fibers. 

In other words, if we succeed in gauging the MCG for $Z_{\text{grav}}(\mathcal{K})$, which has a \emph{trivial} $S^1$ fibration, then we also obtain the correct partition function for the infinite family of genus-0 Seifert manifolds labeled by the data $\{\frac{r_i}{s_i}\}_{i=1}^n$. Hence, while this section has not fully resolved the problem of computing the sum over Seifert manifolds in 3d gravity, it has reduced it to a slightly simpler problem. In the next section, we will make a prediction for the keychain link partition function, based on a maximum ignorance principle for the boundary dual.    

\subsection{Matrix model ansatz}

Our previous analysis has reduced the problem of computing Seifert manifold partition functions in 3d gravity to computing the gravitational path integral for the keychain link $\mathcal{K}$. The keychain link is, in turn, related to the $n+1$-boundary torus wormhole
$
  Z_\text{grav}(\Sigma_{0,n+1}\times S^1)  
$
by an S-transform on the $P_0,\bar P_0$ link component: 
\begin{multline}\label{eq:strans}
    Z_\text{grav}(\mathcal{K};\mathbf{P},\bar{\mathbf{P}}) = \int_0^\infty \dd P_0'\dd \bar P_0' \Big[\mathbb{S}_{P_0P_0'}[\bbi]\mathbb{S}_{\bar P_0\bar P_0'}[\bbi] \\ Z_\text{grav}(\Sigma_{0,n+1}\times S^1;P_0',\bar P_0',\dots, P_n,\bar P_n)\Big]\,.
\end{multline}
The goal of this subsection is to give a prediction for $Z_\text{grav}(\Sigma_{0,n+1}\times S^1)$, in the microcanonical ensemble, based on a principle of maximum ignorance. 

The principle of maximum ignorance (in the context of holography) is a method to build a statistical model for a microscopic theory, given only coarse-grained, single-boundary input from the semiclassical gravitational path integral \cite{deBoer:2023vsm}. The resulting model maximizes a notion of information entropy and reproduces the single-boundary input on average. It then gives predictions for higher statistical moments, which can be interpreted holographically as multi-boundary wormholes. 

In the case at hand, we take as our single-boundary input the leading spectral density of primary states above the black hole threshold:
\begin{equation}\label{eq:smoothdensity}
    \rho_0(E,J) = \rho_0(\tfrac{E+J}{2})\rho_0(\tfrac{E-J}{2}).
\end{equation}
 Here $\rho_0(h)$ is the universal Cardy-like density introduced before, and we have changed variables from $h,\bar h$ to $E,J$ as in Eq.~\eqref{eq:spectralcorr}, with $E \geq |J|$. This density of primary states comes from a bulk calculation of the path integral on $\Sigma_{0,1}\times S^1$, which is the BTZ black hole. Additionally, we assume that that the spin is quantized, $J\in \mathbb{Z}$, which we view as part of the constraints that the `low-energy observer' can impose on their model. The goal of the maximum ignorance method is to build a statistical model that reproduces the coarse-grained spectral density Eq.~\eqref{eq:smoothdensity} on average, and is otherwise agnostic about the details of the high-energy spectrum. 

The initial step is to separate the random features from the non-random structure fixed by symmetries. First, we separate sectors of different spin. Given the Hamiltonian  (note our convention for the shift of the ground state energy)\vspace{-5mm}
\begin{equation}
    \mathsf{H} = L_0 + \bar L_0 - \frac{c-1}{12},
\end{equation}
 the spin $\mathsf{J} = L_0 - \bar L_0$ commutes with $\mathsf{H}$. Hence we can decompose the Hamiltonian into blocks of fixed integer spin. In each block, we model the spectrum of primaries by a random Hamiltonian $\mathsf{H}_J$ drawn from some distribution $\mu(\mathsf{H}_J)$. Different blocks $\mathsf{H}_J$, $\mathsf{H}_{J'}$ with $J\neq J'$ are taken to be statistically independent. This aligns with the proposal of the matrix/tensor model \cite{Jafferis:2024jkb}.

Second, let us discuss the symmetry class for the random matrix ensemble of each $\mathsf{H}_J$. In the absence of any additional symmetries of $\mathsf{H}_J$, our statistical model is in the universality class of the Gaussian Unitary Ensemble (GUE). However, the microscopic CFT with Hamiltonian $\mathsf{H}$ has an RT symmetry (spatial reflection + time reversal) by the CPT theorem. On the Euclidean torus, it acts as $z\to -z$. The anti-unitary operator $\mathsf{RT}$ that generates this symmetry squares to 1 for bosonic theories, and it commutes with the momentum operator $i(L_0-\bar L_0)$ that we use for the spin-grading. If we want our statistical model to be compatible with this discrete global symmetry, we should take $\mathsf{H}_J$ in the universality class of the Gaussian Orthogonal Ensemble (GOE) \cite{Yan:2023rjh}. Note that this is a choice: the statistical description need not respect all symmetries of the CFT it is supposed to model for each member of the ensemble.\footnote{An analogous situation occurs in JT gravity, where one can choose whether or not to impose time-reversal symmetry $\mathsf{T}$ on each member of the ensemble. In the bulk, this corresponds to a choice of whether or not to \emph{gauge} the global time-reversal symmetry, which, in the gravitational path integral, implies a choice between a sum with or without non-orientable surfaces \cite{Stanford:2019vob,Hsin:2020mfa,Harlow:2023hjb}.} 

The same can be said about modular invariance: we do not have to impose that each member of the statistical ensemble of Hamiltonians is modular invariant. Instead, we assume that modular invariance is implemented by summing over modular images (i.e.~integrating against modular kernels $\mathbb{U}_{r,s}$ and summing over co-primes). Since the modular kernels are non-random, they can be taken outside the average. In particular, we do not build modular invariance into the matrix model. This differs from the matrix/tensor model of Refs.~\cite{Belin:2023efa,Jafferis:2024jkb}, which includes a ``constraint squared'' in the potential that enforces modular S-invariance up to some tolerance parameter $\hbar$. However, doing so has the disadvantage of introducing a double-trace term in the matrix model potential. Therefore, we choose to work with a single-trace matrix model and view the modular completion as an additional rule, of the schematic form \vspace{-1mm}
\begin{multline*}
    \text{Input from single topology} \longrightarrow \\ \text{Statistical model} \longrightarrow \text{Modular completion.}
\end{multline*}
This rule is naturally motivated by the observation that semiclassical gravity (in all known examples) implements modular invariance by summing over topologies. 

As a side note, we remark that since we have built spin quantization into the statistical model, the modular completion is a sum over $\mathrm{PSL}(2,\mathbb{Z})$ modulo the $\mathbb{Z}$ subgroup generated by $T$ transformations.

In summary, our simplified statistical model for the spectrum of primaries above the black hole threshold is a single-trace GOE double-scaled random matrix model for each $\mathsf{H}_J$, with leading average density of states $\rho_0(E,J)$, in which Hamiltonians with different $J$'s are statistically independent. Next, this model gives a \emph{prediction} for connected correlation functions of the form
 \begin{equation}\label{eq:highermom}
    \rho_{(n)}(\mathbf{E},\mathbf{J}) \equiv \overline{\rho(E_1,J_1)\cdots \rho(E_n,J_n)}^{\,\text{c}}
 \end{equation}
 where $\rho(E,J)$ is the spectral density of $\mathsf{H}_J$ at fixed spin $J$ and the overline denotes the matrix model average. For example, for $n=2$ the leading genus-0 contribution gives the universal level repulsion in each spin sector $\delta_{JJ'}\rho_{\text{RMT}}(E-|J|,E'-|J'|)$, where spin-diagonality follows from the statistical independence of different $H_J$'s. As discussed in Sec.~\ref{sec:RMT} and App.~\ref{app:doubletrumpet}, this prediction is confirmed by a gravity calculation on the torus wormhole $T\times I \cong \Sigma_{0,2}\times S^1$. 

To obtain predictions for $n>2$, we need to use the loop equations of the matrix model. For a GUE Hermitian matrix model, the loop equations (in the double-scaling limit) are organized by the \emph{topological recursion} relations of Eynard and Orantin \cite{Eynard:2007kz}. For the GOE universality class, the loop equations are significantly more complicated\footnote{There exists a `refined' version of topological recursion for the $\beta = 1$ GOE ensemble \cite{Chekhov:2006rq}. See also Section 4 in Stanford and Witten \cite{Stanford:2019vob}. For a geometric interpretation of the GOE loop equations, see Ref.~\cite{Stanford:2023dtm}.}, but fortunately, we only need to extract the leading genus-0 prediction of the matrix model for the $n$-boundary partition function on $\Sigma_{0,n}\times S^1$. The genus-0 prediction of the GOE matrix model is related to the genus-0 prediction of the GUE matrix model by a simple multiplicative factor (cf.~Eq.~(5.13) in \cite{Stanford:2019vob}):
\begin{equation}\label{eq:GOEvsGUE}
    \left[\rho^{\text{GOE}}_{(n)} \right]_{g=0} = 2^{n-1} \left[\rho^{\text{GUE}}_{(n)}\right]_{g=0}.
\end{equation}
The factor of 2 is the same $\mathsf{C}_{\text{RMT}}$ factor that is also present in the two-boundary amplitude \eqref{eq:spectralcorr}. For $g \neq 0$, the GOE and GUE predictions differ. From the bulk 3d perspective, this difference can be explained by the fact that there exist smooth orientable Seifert manifolds that are fibered over a \emph{non-orientable} base surface \cite{Johannson1979}. So the topological expansion of the dimensionally reduced theory includes non-orientable surfaces (even if the 3d theory is orientable), which is accounted for by a GOE matrix model expansion \cite{Stanford:2019vob}.   

Since we are only interested in $g=0$, Eq.~\eqref{eq:GOEvsGUE} allows us to compute the higher moments $\rho_{(n)}$ from the GUE topological recursion relations, to which we turn next.

\subsection{Topological recursion}

In order to extract the higher moments in Eq.~\eqref{eq:highermom} from topological recursion (TR), we need to identify the right variables $z_i$ in which the topological recursion relations are formulated.
As a preparation, we define the fixed-spin primary partition function
\begin{multline}
    Z_P^J(\Sigma_{0,n}\times S^1;\beta_1,\dots,\beta_n)\coloneq \\ \int_{0}^1 \prod_{k=1}^n \dd \sigma_k \,e^{2\pi i \sigma_k J_k} \overline{Z_P(\tau_1,\bar\tau_1)\cdots Z_P(\tau_n,\bar\tau_n)}^{\text{c}}_{g=0}
\end{multline}
  where $Z_P(\tau,\bar\tau) = |\eta(\tau)|^2 \Tr e^{-2\pi\beta \mathsf{H}_J}e^{2\pi i \sigma \mathsf{J}}$ is the primary partition function in the \emph{GUE matrix model}, which is multiplied by eta functions to strip off the contribution from the descendants. The superscript $^\text{c}$ denotes the connected average. The subscript $g=0$ means that we take the leading connected contribution in the matrix model expansion. Since the ensemble is diagonal in spin, we get an overall factor of $\prod_{k=1}^n\delta_{J_k,J}$, which is why the LHS is labeled only by $J$.

The input of the topological recursion relations is called the \emph{spectral curve}. The spectral curve is essentially the leading one-point function $\overline{\rho(E,J)}\mid_{g=0} \,= \rho_0(E,J)$ for the spectral density above the threshold $E\geq |J|$, which is related to the primary partition function as
\begin{align}
    Z^J_P(\Sigma_{0,1}\times S^1;\beta) = &\int_{|J|}^\infty \dd E\,\rho_0(E,J) \,e^{-2\pi \beta E} \,.
\end{align}
To define the spectral curve with uniformizing coordinate $z$, we perform the change of variables $E- |J| = -2z^2$,
\begin{equation}
  Z^J_P(\Sigma_{0,1}\times S^1;\beta) = e^{-2\pi \beta |J|} \int_{i\mathbb{R}}\frac{\dd z}{2\pi i} \mathcal{W}^J_{0,1}(z) \,e^{4\pi \beta z^2}
\end{equation}
where we have defined the TR invariant $\mathcal{W}^J_{0,1}(z)$, which specifies the spectral curve as a function of the complex coordinate $z\in \mathbb{P}^1$, to be
\begin{equation}\label{eq:spectralcurve3}
    \mathcal{W}^J_{0,1}(z) = -4\sqrt{2}\pi \sin(2\pi b z)\sin(2\pi b^{-1}z) f_J(z).
\end{equation}
Here we use the convention that $\omega_{0,1}(z) = \mathcal{W}_{0,1}(z)\,\dd z$ denotes the differential form, and we have defined
\begin{equation}
    f_J(z) = 4\sqrt{2}\,\tfrac{\sinh(2\pi b \sqrt{|J|-z^2})\sinh(2\pi b^{-1}\sqrt{|J|-z^2})}{2\sqrt{|J|-z^2}}\,.
\end{equation}
Remark that for $J\neq 0$, $f_J(z)$ does not vanish at $z=0$. Hence, the spectral curve defined by $\mathcal{W}_{0,1}^J(z)$ is of the standard type $\omega_{0,1}(z) = -y(z)\dd x(z)$ with $x(z) = z^2$ and $y(z) \sim z$ near the branch point at $z=0$. So the $J\neq 0$ spectral curve has a square root edge. For $J=0$, the spectral curve has a qualitatively different form near the branch point, namely $y(z) \sim z^2$. Let us for now assume that $J\neq 0$, returning to $J=0$ later.

Similarly, we can read off the two-point function:
\begin{widetext}
\begin{align}\label{eq:two-pointfun}
    Z^J_P(\Sigma_{0,2}\times S^1;\beta_1,\beta_2) = \delta_{J_1J}\delta_{J_2J} &\int_{|J|}^\infty \dd E_1\dd E_2  \,\rho_{\text{RMT}}(E_1-|J|,E_2-|J|)\,e^{-2\pi \beta_1 E_1}e^{-2\pi\beta_2 E_2} \\[1em]
    = \delta_{J_1J}\delta_{J_2J} \,e^{-2\pi(\beta_1+\beta_2)|J|} &\int_{i\mathbb{R}+\epsilon}\frac{\dd z_1}{2\pi i}\frac{\dd z_2}{2\pi i } \,\mathcal{W}^J_{0,2}(z_1,z_2)\,e^{4\pi \beta_1 z_1^2}e^{4\pi \beta_2 z_2^2}
\end{align}
\end{widetext}
where we performed the same change of variables from $E-|J|$ to $-2z^2$ as before, finding\footnote{In \cite{Eynard:2007kz}, the two-point function is  taken to be $\mathcal{W}_{0,2} = \frac{1}{(z_1-z_2)^2}$. However, the output of the topological recursion is unchanged if we take the symmetrized version \eqref{eq:TRtwopoint2} instead.}
\begin{equation}\label{eq:TRtwopoint2}
    \mathcal{W}^J_{0,2}(z_1,z_2) = \frac{1}{2} \left(\frac{1}{(z_1-z_2)^2} + \frac{1}{(z_1+z_2)^2}\right).
\end{equation}
Note that $\mathcal{W}^J_{0,2}$ is the same for each $J$.

Having determined the leading one- and two-point functions, the matrix model predicts the following multi-boundary fixed-spin primary partition functions:\vspace{-1mm}
\begin{multline}\label{eq:fixedspinZ}
    Z^J_P(\Sigma_{0,n}\times S^1;\beta_1,\dots,\beta_n) = \prod_{k=1}^n \left(\delta_{J_k J} \,e^{-2\pi \beta_k |J|}\right)   \\\times
     \int_{i\mathbb{R}+\epsilon}\prod_{k=1}^n \frac{\dd z_k}{2\pi i} \,e^{4\pi \beta_k z_k^2}\,\mathcal{W}^J_{0,n}(z_1,\dots, z_n).
\end{multline}
The functions $\mathcal{W}^J_{0,n}$ can be computed by the topological recursion formula \cite{Eynard:2007kz}, which at genus zero reduces to:
\begin{multline}\label{eq:toporec}
    \mathcal{W}^J_{0,n+1}(z_0,\dots, z_n) = \text{Res}_{z\to 0} \Big[K^J(z_0,z) \\
    \sum_{I_1\cup I_2}' \mathcal{W}^J_{0,|I_1|+1}(z,z_{I_1}) \mathcal{W}^J_{0,|I_2|+1}(-z,z_{I_2})\Big].
\end{multline}
The sum runs over all partitions of the set $\{1,\dots, n\}$, using multi-index notation for $I_1$ and $I_2$. The prime on the sum indicates that terms containing $\mathcal{W}^J_{0,1}$ should be excluded. Finally, the recursion kernel $K^J$ is found to be
\begin{multline}\label{eq:reckernel}
    K^J(z_0,z) = \\ -\frac{1}{(z_0^2-z^2)}\frac{z}{(8\sqrt{2}\pi)\sin(2\pi b z)\sin(2\pi b^{-1}z)f_J(z)}\,.
\end{multline}
Up to the factor $f_J(z)$, this is the same recursion kernel that was found for the matrix model of the Virasoro minimal string \cite{Collier:2023cyw}. This can be understood by the fact that the Virasoro minimal string (VMS) is dual to a \emph{chiral} version of 3d gravity on topologies of the form $\Sigma_{g,n}\times S^1$, for which the spectral density is only the holomorphic copy of the Cardy density $\rho_0(P)$. In our case we have two copies of $\rho_0$, hence the invariants $\mathcal{W}_{0,n}^J$ depend on the spin $J$ through the function $f_J(z)$. The appearance of this function in the recursion kernel \eqref{eq:reckernel} also modifies the $z$-dependence of the the TR invariants $\mathcal{W}_{0,n}^J$ compared to the matrix model of the VMS.

A residue calculation using Eqs.~\eqref{eq:toporec} and \eqref{eq:reckernel} determines the following results for $n=3$ and $n=4$:
\begin{equation}
    \mathcal{W}_{0,3}^J(z_0,z_1,z_2) = \frac{1}{f_J(0)} \underbrace{\frac{-1}{2\sqrt{2}\,(2\pi)^3}\frac{1}{z_0^2z_1^2z_2^2}}_{\text{VMS}}\,,
\end{equation}
\vspace{-6mm}
\begin{multline}
     \mathcal{W}_{0,4}^J(z_0,z_1,z_2,z_3) =\frac{1}{f_J(0)^2}\,\times\\ \frac{1}{(2\pi)^4}\Big(\underbrace{\frac{13-c}{96}+\frac{3}{32\pi^2}\sum_{i=0}^3 \frac{1}{z_i^2}}_{\text{VMS}}+  g_b(J) \Big)\frac{1}{z_0^2z_1^2z_2^2z_3^2}\,.
\end{multline}
We have underscored the terms that are also present in the TR invariants of the Virasoro minimal string, to facilitate the comparison. The $J$-dependent correction in $\mathcal{W}_{0,4}^J$ vanishes in the large-spin limit $J\to \infty$, and can be computed explicitly for any finite $J$:
\begin{multline}
    g_b(J) = \\ \frac{3}{64 \pi^2}\! \left(  \tfrac{2\pi b\coth(2\pi b \sqrt{|J|})}{\sqrt{|J|}} + \tfrac{2\pi b^{-1}\coth(2\pi b^{-1}\sqrt{|J|})}{\sqrt{|J|}}- \tfrac{1}{|J|}\right).
\end{multline}
The most important difference with the Virasoro minimal string is the overall prefactor that scales as an inverse power of $f_J(0) = \rho_0(|J|)$. Since $\log \rho_0(|J|)$ is essentially the Cardy formula at spin $J$, our topological recursion formula predicts the overall scaling
\begin{equation}\label{eq:Jscaling}
    \mathcal{W}_{0,n+1}^J \, \propto \,e^{-(n-1)S_0(|J|)}\,.
\end{equation}
This is the same scaling that was found by Maxfield and Turiaci in the near-extremal limit of 3d gravity \cite{Maxfield:2020ale}. 

With the recursively defined expressions for $\mathcal{W}_{0,n+1}^J$ in hand, we can go back to Eq.~\eqref{eq:fixedspinZ} and evaluate the multi-boundary partition functions. For example, we find
\begin{multline}
    Z_P^J(\Sigma_{0,3}\times S^1;\beta_1,\beta_2,\beta_3) = \\ \frac{1}{\rho_0(|J|)}\prod_{k=1}^3 \left(\delta_{J_kJ}\,\tfrac{\sqrt{2}}{2\pi}\,e^{-2\pi \beta_k |J|} \right) \sqrt{\beta_1\beta_2\beta_3}\,.
\end{multline}
Notice the factors of $e^{-2\pi \beta_k |J|}$, which arise from the non-zero edge of the spectrum $E\geq |J|$, and the Kronecker delta's enforcing the spin diagonality. Similar formulas can be found for higher values of $n$: the structure is always a polynomial in the $\sqrt{\beta_i}$'s, with $J$- and $c$-dependent coefficients. 

Taking into account the factor of $2^{n-1}$ in Eq.~\eqref{eq:GOEvsGUE} for the GOE ensemble, the above matrix model computation gives a prediction for the gravitational path integral on the topology $\Sigma_{0,n}\times S^1$. This prediction is yet to be verified by a first-principles gravity calculation. Moreover, note that our prediction is not yet modular invariant in $\tau_{k},\bar\tau_k$, $k=1,\dots, n$, as is expected from the contribution of a single topology. There are different ways to obtain a modular completion: the most natural option is to sum over the independent modular images of $\tau_{k},\bar\tau_k$. Due to spin diagonality, there is a $\mathbb{Z}^{n}$ redundancy in the $n$-fold modular sum, meaning that (for $n>2$) the modular completion is a sum over $(\text{PSL}(2,\mathbb{Z})/\mathbb{Z})^n$. A second option is to interpret the fixed-spin RMT predictions made in this section as overlaps in a \emph{spectral decomposition} in a basis of eigenfunctions for the Laplacian on the upper-half plane. This modular invariant uplift of random matrix models was called ``RMT$_2$'' in Ref.~\cite{Boruch:2025ilr}, based on earlier works \cite{DiUbaldo:2023qli,Benjamin:2021ygh,Haehl:2023tkr,Haehl:2023xys,Haehl:2023mhf}. 

In order to connect back to the Seifert manifolds that were the focus of this section, recall that our goal was to compute the keychain link partition function in Eq.~\eqref{eq:seif}, or, equivalently, the $(n+1)$-boundary torus wormhole in Eq.~\eqref{eq:strans} as a function of $\mathbf{P},\bar{\mathbf{P}}$. Rewriting the connected $(n+1)$-point correlator in the matrix model as an integral over a spectral density  multiplied by Virasoro characters, our statistical model provides a prediction for the microcanonical partition function (for $P^2_k-\bar P_k^2 \neq 0$):
\begin{align}
    &\frac{1}{2^{n}}\,Z_\text{grav}(\Sigma_{0,n+1},\mathbf{P},\bar{\mathbf{P}}) \stackrel{\text{prediction}}{=}
 \\[1em] &=\sum_{J\in \mathbb{N}_{>0}}\prod_{k=0}^n\left(  \tfrac{\delta( P_k - \sqrt{\bar P_k^2 + J}\,)}{\pi}\right)\mathcal{W}_{0,n+1}^J(i\bar P_0,\dots,i\bar P_n) \nonumber \\ &+ \sum_{J\in \mathbb{N}_{<0}}\prod_{k=0}^n\left(  \tfrac{\delta( \bar P_k - \sqrt{P_k^2 - J}\,)}{\pi}\right)\mathcal{W}_{0,n+1}^J(iP_0,\dots,iP_n).\label{eq:maxigprediction}
\end{align}
where we used that $z = i\sqrt{(E-|J|)/2} = i\, \text{min}(P,\bar P)$.

The above prediction does not encompass the spin-zero case $P_k^2 - \bar P_k^2=0$, because the spectral curve has a qualitatively different form. In fact, since $y(z) \sim z^2$ for $J=0$, the residues in the TR relation \eqref{eq:toporec} with $n>1$ are all zero. In this case, we have to use the full loop equations of the GOE matrix ensemble, which generally involves integrals over the full spectral curve instead of residues at the branch points \cite{Stanford:2019vob}. We currently do not have a physical intuition why the spin-zero contribution seems to have this qualitative difference from the maximum ignorance point of view. 

\subsection{Near-extremal limit}
We would now like to show that the $J\neq 0$ matrix model prediction in Eq.~\eqref{eq:maxigprediction} reduces to the proposal of Maxfield and Turiaci \cite{Maxfield:2020ale} when we take the near-extremal and large spin limit. This limit corresponds to $E\to |J|$ and $|J|\to \infty$,   with the ratio 
\begin{equation}
   \frac{(-1)^{|J|} \,e^{-S_0(|J|)/2}}{4\pi^2 (E-|J|)}
\end{equation}
kept finite. In terms of the Liouville momenta, the limit amounts to $P\to 0$, $\bar P \to \infty$ (or with the role of $P$ and $\bar P$ exchanged; they are treated symmetrically).

In this large-spin near-extremal limit, we first note that the spectral curve in Eq.~\eqref{eq:spectralcurve3} reduces to the square root edge $e^{S_0(|J|)}\sqrt{E-|J|}$, plus terms of order $(E-|J|)^{3/2}$. This is the same behavior near the edge as JT gravity and the matrix model of \cite{Maxfield:2020ale}. Second, the function $\mathcal{W}^J_{0,n+1}$ has a large $J$ scaling given by Eq.~\eqref{eq:Jscaling} and a small $P$ scaling given by $P^{-2n+2}$ (for the most singular term in $z_k = iP_k$). This is the same near-extremal scaling as our naive VTQFT computation in Eq.~\eqref{eq:vtqft_unlink}, since 
\begin{equation}
    \frac{1}{\rho_0(P)^{n-1}\rho_0(\bar P)^{n-1}} \sim \frac{1}{P^{2n-2}}e^{-(n-1)2\pi Q \bar P} 
\end{equation}
as $P\to 0$ and $\bar P\to \infty$. Apart from this qualitative similarity, the matrix model prediction is clearly distinct from the naive Virasoro TQFT answer in Eq.~\eqref{eq:vtqft_unlink} (for a more detailed comparison, see Ref.~\cite{Yan:2025usw}).

Third, we will now show that the Dehn filling prescription in Eq.~\eqref{eq:seif} reduces to the replacement rule
\begin{equation}\label{eq:replacementrule}
    \ell \to 2\pi i\alpha
\end{equation}
in the near-extremal limit. Here $\ell$ is a length variable of one of the boundary tori that is being Dehn filled, and $\alpha$ parametrizes the deficit angle $\theta$ of the conical defect (on the disk slicing the filled-in torus) as $\theta = 2\pi(1-\alpha)$. This replacement rule is well known for JT gravity with sharp conical defects ($\alpha \leq \frac{1}{2}$), which arises as the dimensional reduction of 3d gravity in the large-spin near-extremal limit \cite{Maxfield:2020ale}.

To demonstrate the replacement rule \eqref{eq:replacementrule}, consider one of the $P_i,\bar P_i$ integrals over the generalized modular kernel $|\mathbb{U}(r_i,s_i)_{\bbi P_i}|^2$ in Eq.~\eqref{eq:seif}. Let us pick $P_1,\bar P_1$. Using Eqs.~\eqref{eq:strans} and \eqref{eq:maxigprediction}, the calculation of Eq.~\eqref{eq:seif} boils down to integrals of the form
\begin{multline}\label{eq:boilsdown}
   \int_{\mathbb{R}}\!\frac{\dd P_1}{2} \Big[\mathbb{U}(r_1,s_1)_{\bbi P_1} \mathbb{U}(r_1,s_1)^*_{\bbi \sqrt{P_1^2+|J|}} \\ \times\mathcal{W}^J_{0,n+1}(i P_0, iP_1,\dots, i P_n) \Big]\,.
\end{multline}
Now we note that at large $|J|$, the second generalized modular kernel (cf.~\eqref{eq:gen_kernel}) asymptotes to
\begin{multline}\label{eq:largeJapp}
    \mathbb{U}(r,s)_{\bbi P_1}\mathbb{U}(r,s)^*_{\bbi \sqrt{P_1^2+|J|}} \sim e^{-\frac{2\pi i r}{s} |J|}\,e^{\frac{2\pi Q}{s} \sqrt{|J|}}  \\[1em] \times \frac{4}{s}\,\left(\cosh(\tfrac{2\pi Q }{s}P_1)-e^{\frac{2\pi i r^*}{s}}\cosh(\tfrac{2\pi \hat{Q}}{s}P_1)\right) ,
\end{multline}
with $Q = b+1/b$ and $\hat{Q} = b-1/b$. The $n$ modular kernels therefore contribute an overall spin-dependent factor of $e^{n S_0(|J|)/s}$. Combined with the large-spin scaling of $\mathcal{W}^J_{0,n+1}$, the total prefactor is:
\begin{equation}
    e^{S_0(|J|)} e^{-nS_0(|J|)(1 - \frac{1}{s}))}\,.
\end{equation}
This agrees with the genus-zero, $n$-defect contribution in JT gravity with defects \cite{Maxfield:2020ale,Mertens:2019tcm,Witten:2020wvy}. 

Next, let us define the Laplace transforms $V_{0,n+1}^J$ by:
\begin{multline}
    \mathcal{W}^J_{0,n+1}(z_0,\dots, z_n) =\\ \int_0^\infty \prod_{k=0}^n\dd P'_k \left(-4\sqrt{2}\pi\, P_k' e^{-4\pi z_k P_k'}\right) V^J_{0,n+1}(\mathbf{P}').
\end{multline}
Here $V^J_{0,n+1}(\mathbf{P}')$ is a polynomial in the variables $P_k'$, $k=0,\dots, n$, which has $J$- and $b$-dependent coefficients. Inserting the large-$J$ approximation \eqref{eq:largeJapp} into  Eq.~\eqref{eq:boilsdown}, we see that the integral over $P_1$ gives a Dirac delta function $\delta(P_1' - \frac{iQ}{2s})$ inside the $P_1'$ integral.\footnote{For the $P_1$ integral to converge, we first have to deform the $P_1'$ contour to $\mathbb{R}\pm i\epsilon$, where $\epsilon$ is at least $\frac{Q}{2s}$, and then pick up the contribution from the delta function at $P_1' = \frac{iQ}{2s}$.} Using the parametrization $P_1' = \frac{\ell}{4\pi b}$, familiar from the VMS \cite{Collier:2023cyw}, and using that $Q \sim \frac{1}{b}$ at small $b$ (or large $c$), the Dirac delta becomes proportional to
\begin{equation}
    \delta(\ell - \tfrac{2\pi i}{s}).
\end{equation}
But this is precisely the replacement rule in Eq.~\eqref{eq:replacementrule}, with $\alpha = \frac{1}{s}$. This shows that in the near-extremal limit, with additionally large $c$ and large $J$, our Dehn surgery method applied to the  matrix model ansatz is perfectly consistent with Maxfield and Turiaci's analysis of JT gravity with defects. Moreover, since $s\geq 2$ for the rings $1,\dots, n$, we naturally have $\alpha \leq \frac{1}{2}$, so the deficit angle is sharp, $\theta \geq \pi$ \cite{Eberhardt:2023rzz}.

To summarize, in this section we used Dehn surgery---implemented by generalized modular kernels---to relate every genus-zero Seifert manifold with $n$ singular fibers to the multi-boundary torus wormhole $\Sigma_{0,n+1}\times S^1$. While we do not know a first-principles calculation of the gravitational path integral on the latter topology, we gave a prediction for its partition function using a matrix model based on a principle of maximum ignorance. We showed that this prediction is consistent with the dimensional reduction to JT with defects in the semiclassical, large spin, near-extremal limit. 

Moreover, our framework naturally incorporates modular invariance for the contributions of the Seifert partition functions, by simply summing over the modular images of the ``zeroth'' boundary torus. Geometrically, this corresponds to acting with a modular transformation on the boundary torus before gluing the AdS$_3$ trumpets, as in Section \ref{sec:trumpetgluing}, resulting in the modular sum:
\begin{multline}
    Z^{\text{m.c.}}_\text{grav}(\mathcal{K};\mathbf{P},\bar{\mathbf{P}}) = \sum_{(r,s)=1}\int_0^\infty \dd P_0'\dd \bar P_0' \Big[|\mathbb{U}(r,s)_{P_0P_0'}|^2 \\ Z_\text{grav}(\Sigma_{0,n+1}\times S^1;P_0',\bar P_0',\dots, P_n,\bar P_n)\Big]\,.
\end{multline}
Plugging this modular completion into our formula \eqref{eq:seif} for the Seifert partition function, we obtain a  modular invariant dimensional uplift of the model in Ref.~\cite{Maxfield:2020ale}. It would be very interesting to have a first-principles derivation of $Z_\text{grav}(\Sigma_{0,n+1}\times S^1)$, possibly using a `fixed spin' analog of Mirzakhani's recursion \cite{Mirzakhani:2006eta,Mirzakhani:2006fta}. 

\section{Discussion}

In this paper, we showed how surgery methods in the path integral of 3d gravity with negative cosmological constant can be understood holographically through the universal statistical features of the high-energy sector of 2d CFT’s. The four surgery methods that we introduced—ETH surgery, RMT surgery, trumpet gluing and Dehn surgery—capture different aspects of the statistical theory of the CFT data, including OPE statistics and spectral statistics. Understanding these statistical features and their incarnation in 3-manifold topology is an important step towards establishing an averaged version of the AdS/CFT correspondence for pure gravity in three dimensions.

We will end with some points for further research.

\paragraph{Constrained saddles.}
In this paper, we computed the off-shell partition functions for RMT surgery using cutting-and-gluing rules in the path integral. While these path integrals do not have a saddle-point, one could ask if it is possible to find a constrained saddle, as in Ref.~\cite{Cotler:2020lxj}. That is, can we construct a metric on the topologies constructed by RMT surgery that solves the gravitational equations of motion with constraints? A natural constraint is to fix the trace of the holonomies along the cycles defining the embedded $T\times I$ along which we glue, similar to the fixed length of the minimal geodesic in the double trumpet in JT gravity \cite{Saad:2019lba}.

\paragraph{Generalized RMT surgery.}
Since RMT surgery captures the variance of the spectral density $\overline{\rho\rho}^c$ in the statistics of various CFT observables, it is natural to ask if there is an analogous surgery technique that constructs the bulk dual of higher statistical moments of $\rho$. For example, we could study the cubic cumulant of the sphere four-point function $\mathcal{G}(z,\bar z)$, take the connected contraction of the spectral density
\begin{equation}
    \overline{\rho(E_1,J_1)\rho(E_2,J_2)\rho(E_3,J_3)}^c \times \text{OPE average,}
\end{equation}
and then ask which topology computes this in gravity. A plausible answer involves the following generalization of RMT surgery. While the RMT surgery introduced in this paper involved removing two solid tori and gluing their boundaries via a torus wormhole of topology $T \times I \cong \Sigma_{0,2} \times S^1$, the generalized version would involve excising $n$ disjoint solid tori and gluing them with an $n$-boundary torus wormhole of topology $\Sigma_{0,n} \times S^1$. As discussed in Section \ref{sec:seifert}, a natural candidate for the microcanonical partition function on this geometry—which is needed to carry out the $n$-boundary gluing in the gravitational Hilbert space formalism—is provided by the random matrix ansatz Eq.~\eqref{eq:maxigprediction}. For the cubic cumulant of $\mathcal{G}$, this means that we start with three copies of the sphere with a solid torus removed,
\begin{equation*}
    \begin{tikzpicture}[x=0.75pt,y=0.75pt,yscale=-1.1,xscale=1.1,baseline={([yshift=-0.5ex]current bounding box.center)}]
\draw  [line width=1.5]  (110,200.6) .. controls (110,178.18) and (128.18,160) .. (150.6,160) .. controls (173.02,160) and (191.2,178.18) .. (191.2,200.6) .. controls (191.2,223.02) and (173.02,241.2) .. (150.6,241.2) .. controls (128.18,241.2) and (110,223.02) .. (110,200.6) -- cycle ;
\draw [color={rgb, 255:red, 208; green, 2; blue, 27 }  ,draw opacity=1 ]   (132.29,180) .. controls (142.41,189.38) and (142.41,210.38) .. (132.29,220) ;
\draw [shift={(132.29,220)}, rotate = 181.45] [color={rgb, 255:red, 208; green, 2; blue, 27 }  ,draw opacity=1 ][line width=0.75]    (-2.24,0) -- (2.24,0)(0,2.24) -- (0,-2.24)   ;
\draw [shift={(132.29,180)}, rotate = 87.8] [color={rgb, 255:red, 208; green, 2; blue, 27 }  ,draw opacity=1 ][line width=0.75]    (-2.24,0) -- (2.24,0)(0,2.24) -- (0,-2.24)   ;
\draw [color={rgb, 255:red, 208; green, 2; blue, 27 }  ,draw opacity=1 ]   (168,180) .. controls (157.83,189.38) and (157.83,210.38) .. (168,220) ;
\draw [shift={(168,220)}, rotate = 88.43] [color={rgb, 255:red, 208; green, 2; blue, 27 }  ,draw opacity=1 ][line width=0.75]    (-2.24,0) -- (2.24,0)(0,2.24) -- (0,-2.24)   ;
\draw [shift={(168,180)}, rotate = 182.32] [color={rgb, 255:red, 208; green, 2; blue, 27 }  ,draw opacity=1 ][line width=0.75]    (-2.24,0) -- (2.24,0)(0,2.24) -- (0,-2.24)   ;
\draw  [draw opacity=0][line width=1.5]  (164.12,193.78) .. controls (168.02,195.48) and (170.46,197.89) .. (170.46,200.56) .. controls (170.46,205.69) and (161.44,209.86) .. (150.31,209.86) .. controls (139.19,209.86) and (130.17,205.69) .. (130.17,200.56) .. controls (130.17,197.76) and (132.85,195.24) .. (137.1,193.54) -- (150.31,200.56) -- cycle ; \draw  [line width=1.5]  (164.12,193.78) .. controls (168.02,195.48) and (170.46,197.89) .. (170.46,200.56) .. controls (170.46,205.69) and (161.44,209.86) .. (150.31,209.86) .. controls (139.19,209.86) and (130.17,205.69) .. (130.17,200.56) .. controls (130.17,197.76) and (132.85,195.24) .. (137.1,193.54) ;  
\draw  [draw opacity=0][line width=1.5]  (143.1,191.87) .. controls (145.34,191.47) and (147.77,191.26) .. (150.31,191.26) .. controls (152.86,191.26) and (155.29,191.47) .. (157.53,191.87) -- (150.31,200.56) -- cycle ; \draw  [line width=1.5]  (143.1,191.87) .. controls (145.34,191.47) and (147.77,191.26) .. (150.31,191.26) .. controls (152.86,191.26) and (155.29,191.47) .. (157.53,191.87) ;  
\draw (118.54,176.4) node [anchor=north west][inner sep=0.75pt]  [font=\footnotesize]  {$\clo_1$};
\draw (118.54,214.79) node [anchor=north west][inner sep=0.75pt]  [font=\footnotesize]  {$\clo_1$};
\draw (169.54,175.44) node [anchor=north west][inner sep=0.75pt]  [font=\footnotesize]  {$\clo_2$};
\draw (169.54,214.44) node [anchor=north west][inner sep=0.75pt]  [font=\footnotesize]  {$\clo_2$};
\draw (134.29,210.54) node [anchor=north west][inner sep=0.75pt]  [font=\small]  {$P_i,\bar P_i$};
\end{tikzpicture}\,\,,\quad i=1,\dots,3
\end{equation*}
and then integrate against $Z_\text{grav}(\Sigma_{0,3}\times S^1; \mathbf{P},\bar{\mathbf{P}})$ over the momenta $\mathbf{P} = (P_1,P_2,P_3)$.

This generalization can be understood topologically in terms of the JSJ decomposition: the 3-manifolds constructed in this way are irreducible and can be cut along a set of disjoint tori, such that the pieces are either hyperbolic or trivially fibered Seifert manifolds $\Sigma_{0,n}\times S^1$. This is an extension the RMT surgery discussed in Section \ref{sec:RMT}, where each piece in the JSJ decomposition was assumed to be hyperbolic.

\paragraph{Doubly non-perturbative effects.}

In Section \ref{sec:RMT}, we used the spectral variance $\rho_\text{RMT}(t,t')$ in Eq.~\eqref{eq:spectralcorr} as the leading RMT prediction. This is only true up to corrections that are non-perturbative from the matrix model point of view, but that are \emph{doubly} non-perturbative from the point of gravity. For example, such corrections take the form $\exp(i e^{S})$, with $S$ the Cardy entropy. A well-known example of such behavior appears in the sine kernel of random matrix theory \cite{Saad:2019lba},
\begin{equation}\label{eq:doublynonpert}
\overline{\rho(E_1) \rho(E_2)}^c = -\frac{\sin^2(\pi\omega\varrho(\bar E))}{\pi^2\omega^2} + O(\omega^2),
\end{equation}
which is responsible for the plateau in the spectral form factor \cite{Cotler:2016fpe}.
Writing $\sin^2(x) = 1-\cos^2(x)$, we see that Eqs.~\eqref{eq:spectralcorr} and \eqref{eq:doublynonpert} have the same $\omega\to 0$ behavior up to oscillating terms of the form $\exp(i e^{S})$.
These doubly non-perturbative effects emerge as genuine predictions of the matrix model and lie beyond the reach of its perturbative expansion in the double-scaling limit.

In the case of JT gravity, which is known to be completed by a random matrix integral, similar effects arise and can be understood from the bulk perspective---for example, through the so-called ``universe field theory'' description \cite{Post:2022dfi}. In this framework, such contributions are attributed to target-space D-branes \cite{Altland:2022xqx}. More generally, it is expected that doubly non-perturbative effects involve degrees of freedom—like D-branes—from the UV-complete theory, which have no direct geometric interpretation within the low-energy effective description of the gravitational path integral. Discovering what these UV ingredients would entail for 3d gravity is a very interesting avenue for further research.

\paragraph{Beyond matrix models.}
The random matrix ansatz of Section \ref{sec:seifert} was designed to compute partition functions on topologies of the form $\Sigma_{g,n}\times S^1$. However, in 3d gravity there are more topologies with torus boundaries: for example, there are the mapping tori 
\begin{equation}
    \Sigma_{g,n}\times_\varphi S^1, \quad \varphi \in \mathrm{MCG}(\Sigma_{g,n})
\end{equation}
that were briefly mentioned in Section \ref{sec:trumpetgluing}. The full sum over topologies is therefore not just  the simple genus expansion of RMT: it also includes (at least) a sum over the monodromy maps $\varphi$ for each $g$. Hence, the matrix model only provides a first approximation to the spectral statistics. Our expectation is that this is a good approximation in the large-spin, near-extremal limit, but it would be good to make this precise. 

A related point is that in our simplified statistical description, we either used a ETH-like model for the OPE statistics, or a RMT-like model for the spectral statistics. If we want to combine the two into a matrix-tensor model (in the same way as Refs.~\cite{Jafferis:2022uhu,Belin:2023efa,Jafferis:2024jkb}), then there are necessarily cross-correlations between the matrix and tensor part of the model. In Ref.~\cite{Post:2024itb}, an analysis for the cubic cross-correlation $\overline{\rho CC}^c$ was given, relating it to a wormhole computation in AdS$_3$ gravity. More generally speaking, understanding such cross-correlations is an important next goal in the conjectured correspondence between pure gravity in three dimensions and universal statistics of 2d CFT.

\paragraph{Negativity and the sum over Seifert manifolds.}

Lastly, let us discuss the Dehn surgery presented in Section \ref{sec:seifert}. There, we made a prediction for the genus-0 Seifert partition functions based on a principle of maximum ignorance. If this prediction indeed holds in full 3d gravity, it would suggest a resolution to the long-standing negativity problem of the spectral density. Namely, Maxfield and Turiaci argued that summing over $n$, where $n$ is the number of defects (or in our case the number of exceptional fibers of the genus-0 Seifert manifolds), has the effect of shifting the edge of the black hole threshold by a non-perturbatively small amount \cite{Maxfield:2020ale}, at least in the large-spin, near-extremal limit. Since our procedure in principle works for any $J\neq 0$ and any $E \geq |J|$, it would be very interesting to verify (e.g. numerically) that such a shift appears and cures the negativity in the MWK spectral density. However, to convincingly demonstrate this, we have to show that the modular completion does not introduce new negativities, and, moreover, that the contribution from other off-shell topologies (such as Seifert manifolds fibered over a non-orientable base) does not alter the conclusions. We leave this as an important open problem.

\begin{acknowledgments}
It is our pleasure to thank 
Alex Belin,
Scott Collier,
Gabriele Di Ubaldo,
Hans Jockers,
Alexandros Kanargias,
Diego Li\v{s}ka,
Thomas Mertens,
Eric Perlmutter,
Jiaxin Qiao,
Mosh\'e Rozali,
Julian Sonner and especially Lorenz Eberhardt for many valuable discussions. BP and JdB are supported by the European Research Council under the European Unions Seventh Framework Programme (FP7/2007-2013), ERC Grant agreement ADG 834878. JKK is supported by the National Centre of Competence in Research SwissMAP.  
\end{acknowledgments}

\appendix

\newpage
\section{Variance of the four-point function}\label{app:fourpoint}
 
Let $\mathcal{G}$ be the four-point function, as in Eq.~\eqref{eq:fourpointfunction}, and consider its variance in some statistical ensemble
\begin{equation}\label{eq:appvariance}
    \Delta\mathcal{G}^2 = \overline{\mathcal{G}(z,\bar z)\mathcal{G}(z',\bar z')} - \overline{\mathcal{G}(z,\bar z)}\times \overline{\mathcal{G}(z',\bar z')}.
\end{equation}
We assume that the CFT has a twist gap of order $c$, and a sparse spectrum of fixed, non-random sub-threshold states including $\clo_1$ and $\clo_2$. We expand in conformal blocks in the $t$-channel, so that each $\mathcal{G}$ is a sum over the primary states propagating in the intermediate channel. 

In this sum over states, the contribution of the vacuum state to $\overline{\mathcal{G}\mathcal{G}}$ is trivially factorized, so it drops out of the variance in Eq.~\eqref{eq:appvariance}. Since we assumed the light states to be non-random, their contribution also drops out of $\Delta\mathcal{G}^2$. The remaining sum is over the heavy states $h,h'\geq \frac{c-1}{24}$, which we will treat statistically. 

The statistical moment that we have to compute was denoted by $\mathcal{M}$ in Eq.~\eqref{eq:moment}. In this appendix, we compute the contribution from the spectral statistics only: 
\begin{equation}
    \mathcal{M}\supset \overline{\rho(h,\bar h)\rho(h',\bar h')}^c\times  \overline{|C_{\clo_1\clo_2h}|^2}\times \overline{|C_{\clo_1\clo_2h'}|^2}.
\end{equation}
For the OPE coefficients, we use the leading universal density that follows from sphere crossing symmetry \cite{Collier:2019weq}:
\begin{equation}
    \overline{|C_{\clo_1\clo_2h}|^2} \approx C_0(h_{\clo_1},h_{\clo_2},h)C_0(\bar{h}_{\clo_1},\bar h_{\clo_2},\bar h).
\end{equation}
Here $C_0$ is a smooth meromorphic function of the conformal weights $h = \frac{c-1}{24}+P^2$, given by a ratio of double gamma functions proportional to the DOZZ formula for the Liouville structure constants \cite{Eberhardt:2023mrq}. 

For the spectral correlator, we use the leading random matrix answer Eq.~\eqref{eq:spectralcorr}. We approximate the sum over the heavy states by a continuous integral 
\begin{equation}\label{eq:continuousint}
  \int_{\frac{c-1}{24}}^\infty \dd h \dd\bar h \,\dd h'\dd\bar h'\,\overline{\rho(h,\bar h)\rho(h',\bar h')}^c \Big[\cdots\Big]\vspace{-1mm}
\end{equation}
and then change variables to energy $E=h+\bar h - \frac{c}{12}$ and spin $J = h-\bar h$, in order to use Eq.~\eqref{eq:spectralcorr}. Changing variables to Liouville momenta $P,\bar P$, we get a Jacobian $2(2P)(2\bar P)$. The same holds for $h',\bar h'$. 

Substituting everything into the conformal block expansion of $\Delta \mathcal{G}^2$, we precisely get the expression that was claimed in Eq.~\eqref{eq:bdy_prediction}, with $\rho_{0,2}$ given by
\begin{multline}\label{eq:rho02}
    \rho_{0,2}(P,\bar P, P',\bar P') \coloneqq 4(2P)(2\bar P)(2P')(2\bar P')\times\\[0.6em] \,\delta(J-J') \,\rho_{\text{RMT}}(t_P,t_{P'}),
\end{multline}
where $t_P \equiv P^2 + \bar P^2 - |J| \geq 0$ and $J \equiv P^2-\bar P^2$.

As a small technical remark, note that the RMT kernel $\rho_{\text{RMT}}(t_P,t_{P'})$ has a double pole at $P=P'$, so in order to properly define the double integral over $P,P'$ in Eq.~\eqref{eq:bdy_prediction}, one has to slightly shift the $P$-contour up to $\mathbb{R}+i\epsilon$.

We can refine the above approximation by taking into account spin quantization, inserting a Dirac comb $\delta_{\mathbb{Z}}(J)$ into \eqref{eq:continuousint}. From the bulk, this is reproduced by a sum over Dehn twists, as explained in Section \ref{sec:modcompl}.

\section{Some topology}\label{app:topology}
In this Appendix, we visualize the gluing procedure in Step 2 of the RMT surgery of Section \ref{sec:RMT}. The main observation is to realize that the topology constructed in Step 1 of Figure \ref{fig:RMT_surgery} can be turned ``inside-out'' by an orientation-reversing homeomorphism, see Fig.~\ref{fig:insideout}:
\tikzset{every picture/.style={line width=0.75pt}}
 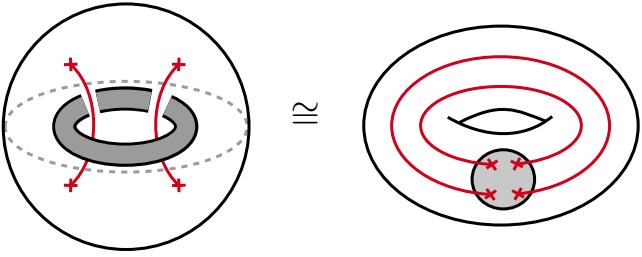
\begin{figure}[h]
    \centering
    \resizebox{\linewidth}{!}{
    \begin{tikzpicture}[x=0.75pt,y=0.75pt,yscale=-1,xscale=1,baseline={([yshift=-0.5ex]current bounding box.center)}]
\draw  [color={rgb, 255:red, 155; green, 155; blue, 155 }  ,draw opacity=1 ][dash pattern={on 1.5pt off 1.5pt on 1.5pt off 1.5pt}] (189.6,80.6) .. controls (189.6,72.32) and (207.51,65.6) .. (229.6,65.6) .. controls (251.69,65.6) and (269.6,72.32) .. (269.6,80.6) .. controls (269.6,88.88) and (251.69,95.6) .. (229.6,95.6) .. controls (207.51,95.6) and (189.6,88.88) .. (189.6,80.6) -- cycle ;
\draw  [line width=0.75]  (189,80.6) .. controls (189,58.18) and (207.18,40) .. (229.6,40) .. controls (252.02,40) and (270.2,58.18) .. (270.2,80.6) .. controls (270.2,103.02) and (252.02,121.2) .. (229.6,121.2) .. controls (207.18,121.2) and (189,103.02) .. (189,80.6) -- cycle ;
\draw [color={rgb, 255:red, 208; green, 2; blue, 27 }  ,draw opacity=1 ]   (211.29,60) .. controls (221.41,69.38) and (221.41,90.38) .. (211.29,100) ;
\draw [shift={(211.29,100)}, rotate = 181.45] [color={rgb, 255:red, 208; green, 2; blue, 27 }  ,draw opacity=1 ][line width=0.75]    (-2.24,0) -- (2.24,0)(0,2.24) -- (0,-2.24)   ;
\draw [shift={(211.29,60)}, rotate = 87.8] [color={rgb, 255:red, 208; green, 2; blue, 27 }  ,draw opacity=1 ][line width=0.75]    (-2.24,0) -- (2.24,0)(0,2.24) -- (0,-2.24)   ;
\draw [color={rgb, 255:red, 208; green, 2; blue, 27 }  ,draw opacity=1 ]   (247,60) .. controls (236.83,69.38) and (236.83,90.38) .. (247,100) ;
\draw [shift={(247,100)}, rotate = 88.43] [color={rgb, 255:red, 208; green, 2; blue, 27 }  ,draw opacity=1 ][line width=0.75]    (-2.24,0) -- (2.24,0)(0,2.24) -- (0,-2.24)   ;
\draw [shift={(247,60)}, rotate = 182.32] [color={rgb, 255:red, 208; green, 2; blue, 27 }  ,draw opacity=1 ][line width=0.75]    (-2.24,0) -- (2.24,0)(0,2.24) -- (0,-2.24)   ;
\draw  [draw opacity=0][line width=6]  (243.12,73.78) .. controls (247.02,75.48) and (249.46,77.89) .. (249.46,80.56) .. controls (249.46,85.69) and (240.44,89.86) .. (229.31,89.86) .. controls (218.19,89.86) and (209.17,85.69) .. (209.17,80.56) .. controls (209.17,77.76) and (211.85,75.24) .. (216.1,73.54) -- (229.31,80.56) -- cycle ; \draw  [line width=6]  (243.12,73.78) .. controls (247.02,75.48) and (249.46,77.89) .. (249.46,80.56) .. controls (249.46,85.69) and (240.44,89.86) .. (229.31,89.86) .. controls (218.19,89.86) and (209.17,85.69) .. (209.17,80.56) .. controls (209.17,77.76) and (211.85,75.24) .. (216.1,73.54) ;  
\draw  [draw opacity=0][line width=6]  (220.2,72.26) .. controls (222.93,71.62) and (226.03,71.26) .. (229.31,71.26) .. controls (232.24,71.26) and (235.03,71.55) .. (237.54,72.06) -- (229.31,80.56) -- cycle ; \draw  [line width=6]  (220.2,72.26) .. controls (222.93,71.62) and (226.03,71.26) .. (229.31,71.26) .. controls (232.24,71.26) and (235.03,71.55) .. (237.54,72.06) ;  
\draw  [draw opacity=0][line width=4.5]  (243.12,73.78) .. controls (246.99,75.48) and (249.41,77.88) .. (249.41,80.54) .. controls (249.41,85.67) and (240.39,89.84) .. (229.27,89.84) .. controls (218.14,89.84) and (209.13,85.67) .. (209.13,80.54) .. controls (209.13,77.72) and (211.83,75.2) .. (216.1,73.5) -- (229.27,80.54) -- cycle ; \draw  [color={rgb, 255:red, 155; green, 155; blue, 155 }  ,draw opacity=1 ][line width=4.5]  (243.12,73.78) .. controls (246.99,75.48) and (249.41,77.88) .. (249.41,80.54) .. controls (249.41,85.67) and (240.39,89.84) .. (229.27,89.84) .. controls (218.14,89.84) and (209.13,85.67) .. (209.13,80.54) .. controls (209.13,77.72) and (211.83,75.2) .. (216.1,73.5) ;  
\draw  [draw opacity=0][line width=4.5]  (220.2,72.26) .. controls (222.87,71.66) and (225.88,71.32) .. (229.06,71.32) .. controls (232.05,71.32) and (234.88,71.62) .. (237.43,72.15) -- (229.06,80.62) -- cycle ; \draw  [color={rgb, 255:red, 155; green, 155; blue, 155 }  ,draw opacity=1 ][line width=4.5]  (220.2,72.26) .. controls (222.87,71.66) and (225.88,71.32) .. (229.06,71.32) .. controls (232.05,71.32) and (234.88,71.62) .. (237.43,72.15) ;  
\draw  [line width=0.75]  (308.11,80.22) .. controls (308.11,62.06) and (328.41,47.33) .. (353.44,47.33) .. controls (378.48,47.33) and (398.78,62.06) .. (398.78,80.22) .. controls (398.78,98.39) and (378.48,113.11) .. (353.44,113.11) .. controls (328.41,113.11) and (308.11,98.39) .. (308.11,80.22) -- cycle ;
\draw    (335.85,77.46) .. controls (348.03,84.49) and (360.89,84.99) .. (370.36,77.21) ;
\draw    (339.69,79.22) .. controls (353.44,72.19) and (358.41,74.95) .. (368.1,78.72) ;
\draw  [fill={rgb, 255:red, 155; green, 155; blue, 155 }  ,fill opacity=0.55 ][line width=0.75]  (343.89,98) .. controls (343.89,92.6) and (348.52,88.22) .. (354.22,88.22) .. controls (359.93,88.22) and (364.56,92.6) .. (364.56,98) .. controls (364.56,103.4) and (359.93,107.78) .. (354.22,107.78) .. controls (348.52,107.78) and (343.89,103.4) .. (343.89,98) -- cycle ;
\draw  [draw opacity=0] (350.47,93.09) .. controls (337.2,92.36) and (326.89,86.88) .. (326.89,80.22) .. controls (326.89,73.07) and (338.78,67.28) .. (353.44,67.28) .. controls (368.11,67.28) and (380,73.07) .. (380,80.22) .. controls (380,86.49) and (370.85,91.72) .. (358.7,92.91) -- (353.44,80.22) -- cycle ; \draw [color={rgb, 255:red, 208; green, 2; blue, 27 }  ,draw opacity=1 ]   (350.47,93.09) .. controls (337.2,92.36) and (326.89,86.88) .. (326.89,80.22) .. controls (326.89,73.07) and (338.78,67.28) .. (353.44,67.28) .. controls (368.11,67.28) and (380,73.07) .. (380,80.22) .. controls (380,86.49) and (370.85,91.72) .. (358.7,92.91) ; \draw [shift={(358.7,92.91)}, rotate = 205.9] [color={rgb, 255:red, 208; green, 2; blue, 27 }  ,draw opacity=1 ][line width=0.75]    (-2.24,0) -- (2.24,0)(0,2.24) -- (0,-2.24)   ; \draw [shift={(350.47,93.09)}, rotate = 235.07] [color={rgb, 255:red, 208; green, 2; blue, 27 }  ,draw opacity=1 ][line width=0.75]    (-2.24,0) -- (2.24,0)(0,2.24) -- (0,-2.24)   ;
\draw  [draw opacity=0] (350.01,103) .. controls (331.69,101.94) and (317.33,92.17) .. (317.33,80.28) .. controls (317.33,67.67) and (333.45,57.46) .. (353.33,57.46) .. controls (373.22,57.46) and (389.33,67.67) .. (389.33,80.28) .. controls (389.33,91.64) and (376.24,101.06) .. (359.09,102.81) -- (353.33,80.28) -- cycle ; \draw [color={rgb, 255:red, 208; green, 2; blue, 27 }  ,draw opacity=1 ]   (350.01,103) .. controls (331.69,101.94) and (317.33,92.17) .. (317.33,80.28) .. controls (317.33,67.67) and (333.45,57.46) .. (353.33,57.46) .. controls (373.22,57.46) and (389.33,67.67) .. (389.33,80.28) .. controls (389.33,91.64) and (376.24,101.06) .. (359.09,102.81) ; \draw [shift={(359.09,102.81)}, rotate = 209.31] [color={rgb, 255:red, 208; green, 2; blue, 27 }  ,draw opacity=1 ][line width=0.75]    (-2.24,0) -- (2.24,0)(0,2.24) -- (0,-2.24)   ; \draw [shift={(350.01,103)}, rotate = 234.78] [color={rgb, 255:red, 208; green, 2; blue, 27 }  ,draw opacity=1 ][line width=0.75]    (-2.24,0) -- (2.24,0)(0,2.24) -- (0,-2.24)   ;
\draw (282.56,71.18) node [anchor=north west][inner sep=0.75pt]    {$\cong $};
\end{tikzpicture}}
    \caption{Two ways of looking at the same topology.}
    \label{fig:insideout}
\end{figure}

The next ingredient is the torus wormhole $T\times I$, which can also be visualized as a solid torus (a doughnut) from whose interior we carve out another solid torus; see Fig.~\ref{fig:doughnut}:
\begin{figure}[h]
    \centering
    \resizebox{0.45\linewidth}{!}{
    \begin{tikzpicture}[x=0.75pt,y=0.75pt,yscale=-1,xscale=1,baseline={([yshift=-0.5ex]current bounding box.center)}]
\draw  [fill={rgb, 255:red, 255; green, 255; blue, 255 }  ,fill opacity=1 ][line width=0.75]  (328.11,100.22) .. controls (328.11,82.06) and (348.41,67.33) .. (373.44,67.33) .. controls (398.48,67.33) and (418.78,82.06) .. (418.78,100.22) .. controls (418.78,118.39) and (398.48,133.11) .. (373.44,133.11) .. controls (348.41,133.11) and (328.11,118.39) .. (328.11,100.22) -- cycle ;
\draw  [color={rgb, 255:red, 128; green, 128; blue, 128 }  ,draw opacity=1 ][line width=6]  (341.59,100) .. controls (341.59,89.51) and (355.85,81) .. (373.44,81) .. controls (391.04,81) and (405.3,89.51) .. (405.3,100) .. controls (405.3,110.5) and (391.04,119.01) .. (373.44,119.01) .. controls (355.85,119.01) and (341.59,110.5) .. (341.59,100) -- cycle ;
\draw    (359.96,99.72) .. controls (371.04,105.14) and (376.13,105.22) .. (387.71,99.39) ;
\draw    (359.96,99.72) .. controls (373.72,92.69) and (378.01,95.62) .. (387.71,99.39) ;
\draw    (357.21,98.05) .. controls (357.46,98.39) and (360.63,100.14) .. (360.87,100.22) ;
\draw    (387.29,99.39) .. controls (387.54,99.72) and (389.38,98.22) .. (389.71,97.97) ;
\draw  [draw opacity=0] (373.9,103.78) .. controls (376.73,104.4) and (378.94,110.72) .. (378.94,118.42) .. controls (378.94,126.23) and (376.66,132.63) .. (373.78,133.08) -- (373.44,118.42) -- cycle ; \draw   (373.9,103.78) .. controls (376.73,104.4) and (378.94,110.72) .. (378.94,118.42) .. controls (378.94,126.23) and (376.66,132.63) .. (373.78,133.08) ;  
\draw  [draw opacity=0][dash pattern={on 1.5pt off 1.5pt on 1.5pt off 1.5pt}] (368.16,114.2) .. controls (368.83,108.14) and (370.94,103.73) .. (373.43,103.73) .. controls (373.44,103.73) and (373.45,103.73) .. (373.46,103.73) -- (373.43,118.42) -- cycle ; \draw  [dash pattern={on 1.5pt off 1.5pt on 1.5pt off 1.5pt}] (368.16,114.2) .. controls (368.83,108.14) and (370.94,103.73) .. (373.43,103.73) .. controls (373.44,103.73) and (373.45,103.73) .. (373.46,103.73) ;  
\draw  [draw opacity=0][dash pattern={on 1.5pt off 1.5pt on 1.5pt off 1.5pt}] (372.56,132.92) .. controls (370.74,132.14) and (369.21,128.95) .. (368.45,124.61) -- (373.44,118.42) -- cycle ; \draw  [dash pattern={on 1.5pt off 1.5pt on 1.5pt off 1.5pt}] (372.56,132.92) .. controls (370.74,132.14) and (369.21,128.95) .. (368.45,124.61) ;  
\draw  [draw opacity=0] (372.87,115.19) .. controls (372.92,115.18) and (372.96,115.18) .. (373,115.18) .. controls (373.85,115.18) and (374.54,116.89) .. (374.54,119.01) .. controls (374.54,121.13) and (373.85,122.84) .. (373,122.84) .. controls (372.84,122.84) and (372.69,122.78) .. (372.55,122.67) -- (373,119.01) -- cycle ; \draw   (372.87,115.19) .. controls (372.92,115.18) and (372.96,115.18) .. (373,115.18) .. controls (373.85,115.18) and (374.54,116.89) .. (374.54,119.01) .. controls (374.54,121.13) and (373.85,122.84) .. (373,122.84) .. controls (372.84,122.84) and (372.69,122.78) .. (372.55,122.67) ;  
\end{tikzpicture}}
    \caption{The topology of the $T\times I$ wormhole.}
    \label{fig:doughnut}
\end{figure}
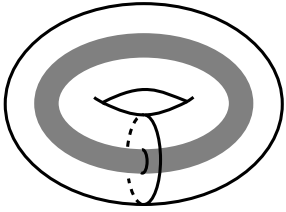

If we now glue the gray torus on the left-hand side in Fig.~\ref{fig:insideout} to the outer torus boundary in Fig.~\ref{fig:doughnut}, and also glue the outer torus on the right-hand side in Fig.~\ref{fig:insideout} to the inner boundary in Fig.~\ref{fig:doughnut} (sitting inside one another like matryoshka dolls) then the result is shown in Fig.~\ref{fig:aftergluing}:
 \begin{figure}[h]
    \centering
    \resizebox{0.5\linewidth}{!}{
    \begin{tikzpicture}[x=0.75pt,y=0.75pt,yscale=-1,xscale=1,baseline={([yshift=-0.5ex]current bounding box.center)}]
\draw  [draw opacity=0][dash pattern={on 1.5pt off 1.5pt on 1.5pt off 1.5pt}] (107.89,86.5) .. controls (129.97,87.37) and (146.74,90.42) .. (148.94,94.16) -- (95,95) -- cycle ; \draw  [color={rgb, 255:red, 155; green, 155; blue, 155 }  ,draw opacity=1 ][dash pattern={on 1.5pt off 1.5pt on 1.5pt off 1.5pt}] (107.89,86.5) .. controls (129.97,87.37) and (146.74,90.42) .. (148.94,94.16) ;  
\draw  [draw opacity=0][dash pattern={on 1.5pt off 1.5pt on 1.5pt off 1.5pt}] (41.29,93.84) .. controls (44.24,90.24) and (60.71,87.34) .. (82.17,86.5) -- (95,95) -- cycle ; \draw  [color={rgb, 255:red, 155; green, 155; blue, 155 }  ,draw opacity=1 ][dash pattern={on 1.5pt off 1.5pt on 1.5pt off 1.5pt}] (41.29,93.84) .. controls (44.24,90.24) and (60.71,87.34) .. (82.17,86.5) ;  
\draw  [color={rgb, 255:red, 155; green, 155; blue, 155 }  ,draw opacity=1 ][dash pattern={on 1.5pt off 1.5pt on 1.5pt off 1.5pt}] (79.64,95) .. controls (79.64,93.49) and (86.52,92.27) .. (95,92.27) .. controls (103.48,92.27) and (110.36,93.49) .. (110.36,95) .. controls (110.36,96.51) and (103.48,97.73) .. (95,97.73) .. controls (86.52,97.73) and (79.64,96.51) .. (79.64,95) -- cycle ;
\draw  [fill={rgb, 255:red, 155; green, 155; blue, 155 }  ,fill opacity=0.47 ] (79.64,95) .. controls (79.64,86.52) and (86.52,79.64) .. (95,79.64) .. controls (103.48,79.64) and (110.36,86.52) .. (110.36,95) .. controls (110.36,103.48) and (103.48,110.36) .. (95,110.36) .. controls (86.52,110.36) and (79.64,103.48) .. (79.64,95) -- cycle ;
\draw [color={rgb, 255:red, 208; green, 2; blue, 27 }  ,draw opacity=1 ]   (131.47,66.6) .. controls (117.69,79.3) and (117.96,111.83) .. (131.73,124.87) ;
\draw [shift={(131.73,124.87)}, rotate = 88.43] [color={rgb, 255:red, 208; green, 2; blue, 27 }  ,draw opacity=1 ][line width=0.75]    (-2.24,0) -- (2.24,0)(0,2.24) -- (0,-2.24)   ;
\draw [shift={(131.47,66.6)}, rotate = 182.32] [color={rgb, 255:red, 208; green, 2; blue, 27 }  ,draw opacity=1 ][line width=0.75]    (-2.24,0) -- (2.24,0)(0,2.24) -- (0,-2.24)   ;
\draw [color={rgb, 255:red, 255; green, 255; blue, 255 }  ,draw opacity=1 ][line width=2.25]    (119.73,99.87) -- (123.47,100.2) ;
\draw [color={rgb, 255:red, 208; green, 2; blue, 27 }  ,draw opacity=1 ]   (104.86,100.26) .. controls (135.24,97.16) and (145.95,108.73) .. (128.32,114.14) ;
\draw [shift={(104.86,100.26)}, rotate = 39.18] [color={rgb, 255:red, 208; green, 2; blue, 27 }  ,draw opacity=1 ][line width=0.75]    (-2.24,0) -- (2.24,0)(0,2.24) -- (0,-2.24)   ;
\draw [color={rgb, 255:red, 208; green, 2; blue, 27 }  ,draw opacity=1 ]   (58.53,66.07) .. controls (72.25,78.77) and (72.52,111.16) .. (58.8,124.2) ;
\draw [shift={(58.8,124.2)}, rotate = 181.45] [color={rgb, 255:red, 208; green, 2; blue, 27 }  ,draw opacity=1 ][line width=0.75]    (-2.24,0) -- (2.24,0)(0,2.24) -- (0,-2.24)   ;
\draw [shift={(58.53,66.07)}, rotate = 87.8] [color={rgb, 255:red, 208; green, 2; blue, 27 }  ,draw opacity=1 ][line width=0.75]    (-2.24,0) -- (2.24,0)(0,2.24) -- (0,-2.24)   ;
\draw [color={rgb, 255:red, 255; green, 255; blue, 255 }  ,draw opacity=1 ][line width=2.25]    (67.6,100.47) -- (70.93,100.6) ;
\draw [color={rgb, 255:red, 208; green, 2; blue, 27 }  ,draw opacity=1 ]   (84.7,100.23) .. controls (79.14,99.88) and (74.32,99.97) .. (70.27,100.4) .. controls (52.45,102.27) and (49.28,110.48) .. (61.85,114.5) ;
\draw [shift={(84.7,100.23)}, rotate = 228.66] [color={rgb, 255:red, 208; green, 2; blue, 27 }  ,draw opacity=1 ][line width=0.75]    (-2.24,0) -- (2.24,0)(0,2.24) -- (0,-2.24)   ;
\draw  [draw opacity=0][dash pattern={on 1.5pt off 1.5pt on 1.5pt off 1.5pt}] (149.03,94.58) .. controls (149.13,94.83) and (149.19,95.09) .. (149.19,95.35) .. controls (149.19,100.85) and (124.93,105.3) .. (95,105.3) .. controls (65.23,105.3) and (41.06,100.89) .. (40.81,95.44) -- (95,95.35) -- cycle ; \draw  [color={rgb, 255:red, 155; green, 155; blue, 155 }  ,draw opacity=1 ][dash pattern={on 1.5pt off 1.5pt on 1.5pt off 1.5pt}] (149.03,94.58) .. controls (149.13,94.83) and (149.19,95.09) .. (149.19,95.35) .. controls (149.19,100.85) and (124.93,105.3) .. (95,105.3) .. controls (65.23,105.3) and (41.06,100.89) .. (40.81,95.44) ;  
\draw  [line width=0.75]  (40,95) .. controls (40,64.62) and (64.62,40) .. (95,40) .. controls (125.38,40) and (150,64.62) .. (150,95) .. controls (150,125.38) and (125.38,150) .. (95,150) .. controls (64.62,150) and (40,125.38) .. (40,95) -- cycle ;
\draw [color={rgb, 255:red, 208; green, 2; blue, 27 }  ,draw opacity=1 ]   (67.14,73.21) .. controls (86.8,70.2) and (101.2,69.8) .. (121.89,73.86) ;
\draw [color={rgb, 255:red, 255; green, 255; blue, 255 }  ,draw opacity=1 ][line width=2.25]    (67.47,89.52) -- (71.11,89.89) ;
\draw [color={rgb, 255:red, 208; green, 2; blue, 27 }  ,draw opacity=1 ]   (84.57,90.02) .. controls (79.6,90.36) and (75.22,90.31) .. (71.44,89.98) .. controls (71,89.94) and (70.56,89.9) .. (70.13,89.85) .. controls (52.31,87.9) and (49.15,79.36) .. (61.71,75.17) ;
\draw [shift={(84.57,90.02)}, rotate = 221.19] [color={rgb, 255:red, 208; green, 2; blue, 27 }  ,draw opacity=1 ][line width=0.75]    (-2.24,0) -- (2.24,0)(0,2.24) -- (0,-2.24)   ;
\draw [color={rgb, 255:red, 255; green, 255; blue, 255 }  ,draw opacity=1 ][line width=2.25]    (119.34,90.05) -- (122.98,90.42) ;
\draw [color={rgb, 255:red, 208; green, 2; blue, 27 }  ,draw opacity=1 ]   (104.73,90) .. controls (135.11,93.22) and (145.82,81.18) .. (128.18,75.55) ;
\draw [shift={(104.73,90)}, rotate = 51.05] [color={rgb, 255:red, 208; green, 2; blue, 27 }  ,draw opacity=1 ][line width=0.75]    (-2.24,0) -- (2.24,0)(0,2.24) -- (0,-2.24)   ;
\draw [color={rgb, 255:red, 208; green, 2; blue, 27 }  ,draw opacity=1 ]   (66.93,116.33) .. controls (85.88,120.43) and (103.28,120.15) .. (122.42,115.76) ;
\end{tikzpicture}}
    \caption{Topology after gluing.}
    \label{fig:aftergluing}
\end{figure}
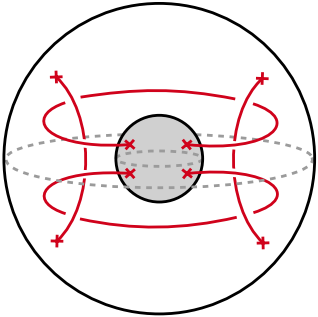

Topologically, this is just a sphere times an interval, with disconnected spherical boundaries. However, the Wilson lines connecting the punctures on the boundary spheres are linked with each other in a non-trivial way.

\section{Virasoro modular kernels}\label{app:kernels}

Here we define the modular kernels that act on the torus Hilbert space $\mathcal{H}_\text{T}$ by $\mathrm{PSL}(2,\mathbb{Z})$ transformations. Since these kernels are holomorphically factorized, \eqref{eq:genmodker}, we will focus on the holomorphic part only; the anti-holomorphic counterpart follows from replacing $P$ by $\bar P$ and taking the complex conjugate. 

Let $\gamma\in \mathrm{PSL}(2,\mathbb{Z})$ be a modular transformation,
\begin{equation}\label{eq:gamma}
    \gamma = \begin{pmatrix}
        a & b \\ s & r
    \end{pmatrix}, \quad \gamma\tau = \frac{a\tau+b}{s\tau + r}\,.
\end{equation}
Then $\gamma$ is (projectively) represented by a unitary operator $\mathbb{U}(\gamma)$ acting on the Hilbert space of torus conformal blocks. The abstract ket vectors $\ket{P}$ with $P\in\mathbb{R}_{\geq 0}$ are related to the non-degenerate Virasoro characters by $\braket{\tau}{P} = \chi_P(\tau)$, where in terms of the nome $q = e^{2\pi i \tau}$,
\begin{equation}
    \chi_P(\tau) = \frac{q^{P^2}}{\eta(\tau)}, \quad \eta(\tau) = q^{\frac{1}{24}}\prod_{n=1}^\infty (1-q^n).
\end{equation}
So we can read off the matrix elements $\mel{P'}{\mathbb{U}(\gamma)}{P}$ by studying the transformation properties of the characters under the $\mathrm{PSL}(2,\mathbb{Z})$ action in Eq.~\eqref{eq:gamma}:
\begin{equation}
    \chi_P(\gamma\tau)= \int_0^\infty \dd P'\mel{P'}{\mathbb{U}(\gamma)}{P}\chi_{P'}(\tau).
\end{equation}
The standard generators
\begin{equation}
    S = \begin{pmatrix}
        0 & -1 \\ 1 & 0
    \end{pmatrix}, \quad T = \begin{pmatrix}
        1 & 1 \\ 0 & 1
    \end{pmatrix}
\end{equation}
are represented by the well-known modular S-kernel and multiplicative phase $\mathbb{T}_P$ \cite{Eberhardt:2023mrq}:
\begin{align}\label{eq:modS}
   \mel{P'}{\mathbb{U}(S)}{P} &= \mathbb{S}_{PP'}[\bbi] = 2\sqrt{2}\cos(4\pi PP'), \\[1em]
   \mel{P'}{\mathbb{U}(T)}{P} &=\delta(P-P') \,e^{2\pi i (P^2-\frac{1}{24})}\,.
\end{align}
When $s\neq 0$, the element $\gamma$ in Eq.~\eqref{eq:gamma} is represented by a generalized modular kernel with matrix elements \cite{Benjamin:2020mfz}:
\begin{multline}\label{eq:mod_ker}
    \mathbb{U}(r,s)_{P_1P_2}\equiv \mel{P_1}{\mathbb{U}(\gamma)}{P_2} =\\[1em] e^{\pi i \theta_{\gamma}} \,2\sqrt{\tfrac{2}{s}}  \cos(\tfrac{4\pi}{s}P_1P_2) \,e^{\frac{2\pi i}{s}(a P_1^2 + r P_2^2)}.
\end{multline}
Here $e^{\pi i\theta_\gamma}$ is a pure phase, which signifies that $\mathbb{U}$ is a projective representation. It is easy to check using the delta-function orthogonality of the cosine that $\mathbb{U}(\gamma)\mathbb{U}(\gamma)^\dagger = \bbi$ and that the special case $(r,s)=(0,1)$ reduces back to the modular S-matrix \eqref{eq:modS}. 

We have adopted the notation $\mathbb{U}(r,s)$ because the constraint $ar-bs=1$ shows that $[a] = [r]^{-1}$ as elements of $(\mathbb{Z}/s\mathbb{Z})^*$ (for $r\neq 0$). So the kernel only depends on $r$ and $s$, up to a phase of the form $e^{2\pi i m P_1^2}$, $m\in \mathbb{Z}$. 

Another special case of the generalized modular kernel is $(r,s) = (1,n)$, which corresponds to the lower triangular matrix $ST^{-n}S$. Indeed, one can show that
\begin{equation}
    \mathbb{U}(1,n)_{P_1P_2} = \int_0^\infty \!\dd P \,\mathbb{S}_{P_1P}[\bbi]\,\mathbb{T}_P^{-n}\,\mathbb{S}_{PP_2}[\bbi].
\end{equation}
This follows from the Gaussian integral identity
\begin{equation}
    \int_0^\infty\!\!\dd P\, \sker{P_1}{P}{\bbi}\,\mathbb{T}_P^{-t}\,\sker{P}{P_2}{\bbi}=\tfrac{e^{\pi i\theta}}{\sqrt{t}}\,\mathbb{T}_{P_1}^{1/t}\,\sker{P_1}{P_2/t}{\bbi}\,\mathbb{T}_{P_2}^{1/t}
\end{equation}
valid for any $t\neq 0$, where $\theta=\frac{1}{4}-\frac{1}{12}(t+\frac{2}{t})$.

There are also generalized modular kernels for degenerate representations of the Virasoro algebra. We will only write the $\mathrm{PSL}(2,\mathbb{Z})$ transformation kernel for the vacuum character \cite{Benjamin:2019stq,Collier:2024mgv}:\vspace{-1mm}
\begin{multline}\label{eq:gen_kernel}
    \mathbb{U}(r,s)_{\bbi P} = e^{\pi i\theta_{\gamma}} e^{\frac{2\pi i}{s}(-a \frac{Q^2}{4}+r P^2)}\\[0.7em] \times  \sqrt{\tfrac{8}{s}}\,\left(\cosh(\tfrac{2\pi Q }{s}P)-e^{\frac{2\pi i a}{s}}\cosh(\tfrac{2\pi \hat{Q}}{s}P)\right)
\end{multline}
where $Q = b+1/b$ and $\hat{Q} = b-1/b$ with $b$ the Liouville coupling. Although the vacuum corresponds to a non-normalizable state in $\mathcal{H}_\text{T}$ at an analytically continued value of the Liouville momentum $P = \frac{iQ}{2}$, the modular kernels for the identity exist nonetheless. They are used in Section \ref{sec:seifert} to perform Dehn surgery. 

As a check, note that Eq.~\eqref{eq:gen_kernel} reduces to the identity S-kernel when we take $\gamma = S$, i.e.~$(r,s) = (0,1)$:
\begin{equation}\label{eq:rho_0}
    \mathbb{S}_{\bbi P}[\bbi] = 4\sqrt{2} \sinh(2\pi b P)\sinh(2\pi b^{-1}P).
\end{equation}
This function is also the universal density of states $\rho_0(P) = \mathbb{S}_{\bbi P}[\bbi] $ derived in Ref.~\cite{Collier:2019weq} and the leading spectral density of the Virasoro minimal string \cite{Collier:2023cyw}. 

\section{Details on the torus wormhole}\label{app:doubletrumpet}
Here we will briefly review the AdS$_3$ double trumpet, computed by Cotler and Jensen, and then derive from it the `microcanonical' torus wormhole Eq.~\eqref{eq:mel}. The gravitational partition function on $T\times I$ (before modular completion), with fixed boundary complex structures and asymptotically AdS$_3$ boundary conditions, reads \cite{Cotler:2020ugk}:
\begin{multline}\label{eq:CJ1}
   Z_{T\times I}(\tau_1,\bar\tau_1;\tau_2,\bar\tau_2) = 2\,\mathsf{C}_\text{RMT} \sqrt{\mathrm{Im}(\tau_1)\mathrm{Im}(\tau_2)}\\ \times \left|\int_0^\infty \dd P\,2P\,\chi_P(\tau_1)\chi_P(\tau_2)\right|^2\,.
\end{multline}
This was derived using a phase space path integral in the Chern-Simons formulation of 3d gravity, interpreting $T\times I$ as an annulus times Euclidean time.
Note that we have multiplied by a factor $\mathsf{C}_{\text{RMT}}$ compared to \cite{Cotler:2020ugk}, which can be 1 (GUE) or 2 (GOE) depending on the statistics of the underlying matrix model.

Writing the moduli as $\tau_{j} = \sigma_{j} + i\beta_{j}$ and $\bar\tau_{j} =\sigma_{j}  - i\beta_{j}$ for $j=1,2$, we will perform a change of variables to energy $E = h+\bar h\! - \!\frac{c-1}{12}=P^2 + \bar P^2$ and spin $J=P^2-\bar P^2$:
\begin{equation}\label{eq:change_of_var2}
   \int_0^\infty \!\!\dd P \,2P \int_0^\infty \!\!\dd \bar P \,2\bar P  \longrightarrow \frac{1}{2}\int_{-\infty}^\infty \!\dd J\int_{|J|}^\infty \!\dd E\,.
\end{equation}
Note that since $P,\bar P\geq 0$, we have $E\geq |J|$. In these variables, the Virasoro characters can be rewritten as
\begin{equation}
    \chi_{P}(\tau_j)\chi_{\bar P}(\bar\tau_j)^* = \frac{1}{|\eta(\tau)|^2} \,e^{-2\pi \beta_j E}e^{2\pi i \sigma_j J}
\end{equation}
so that we can do the $E$-integral in Eq.~\eqref{eq:CJ1}, giving
\begin{equation}
    \int_{|J|}^\infty \dd E\,e^{-2\pi(\beta_1+\beta_2)E} = \frac{e^{-2\pi(\beta_1+\beta_2)|J|}}{2\pi(\beta_1+\beta_2)}\,.
\end{equation}

Next, we use the following double Laplace transform, familiar from JT gravity and random matrix theory \cite{Saad:2019lba}:
\begin{align}\label{eq:laplacetrans}
    &\frac{\mathsf{C}_{\text{RMT}}}{2\pi}\frac{\sqrt{\beta_1\beta_2}}{\beta_1+\beta_2} = \\[0.7em] &\int_{0+i\epsilon}^\infty\!\dd E_1\int_0^\infty \!\!\dd E_2\, \rho_\text{RMT}(E_1,E_2)\,e^{-2\pi \beta_1 E_1}e^{-2\pi\beta_2 E_2}\nonumber
\end{align}
where the spectral two-point function is given by
\begin{equation}
    \rho_{\text{RMT}}(E_1,E_2) = -\frac{\mathsf{C}_{\text{RMT}}}{(2\pi)^2}\frac{E_1+E_2}{\sqrt{E_1E_2}(E_1-E_2)^2}.
\end{equation}
The $E_1$ contour has been shifted by $i\epsilon$ to avoid the double pole at $E_1=E_2$. Substituting Eq.~\eqref{eq:laplacetrans} into the expression for $Z_{T\times I}$, and performing the change of variables $E_1 \to E_1 - |J|$ and $ E_2 \to E_2 - |J|$, we obtain
\begin{widetext}
\begin{equation}\label{eq:CJtwopoint2}
    Z_{T\times I}(\tau_1,\bar\tau_1;\tau_2,\bar\tau_2) = \int_{-\infty}^\infty\! \dd J\,e^{2\pi i (\sigma_1+\sigma_2)J} \int_{
    |J|+i\epsilon}^\infty\!\!\dd E_1\int_{|J|}^\infty \!\!\dd E_2\, \rho_\text{RMT}(E_1-|J|,E_2-|J|)\,\frac{e^{-2\pi \beta_1 E_1}}{|\eta(\tau_1)|^2}
\frac{e^{-2\pi\beta_2 E_2}}{|\eta(\tau_2)|^2}\,.
\end{equation}
\end{widetext}
Next, we rewrite the integral over $J$ as a double integral over $J_1$ and $J_2$ with a delta function $\delta(J_1-J_2)$ setting the spins equal. Finally, we change variables back to Liouville momenta $(P_1,\bar P_1)$ and $(P_2,\bar P_2)$ using the map \eqref{eq:change_of_var2}. Then we have shown that the seed partition function of the $T\times I$ wormhole can be written in the form
\begin{multline}\label{eq:whmicrocanonical}
     Z_{T\times I}(\tau_1,\bar\tau_1;\tau_2,\bar\tau_2) =\\[1em] \int_0^\infty \dd^4 P\,\rho_{0,2}(P_1,\bar P_1,P_2,\bar P_2) \,|\chi_{P_1}(\tau_1)|^2 |\chi_{P_2}(\tau_2)|^2 
\end{multline}
where $\rho_{0,2}$ is precisely the function in Eq.~\eqref{eq:rho02} that came out of our boundary analysis. Interpreting $Z_{T\times I}$ as the matrix elements of some gluing map $G_{\text{T}}$, as in Eq.~\eqref{eq:canonical}, allows us to read off the `microcanonical' torus wormhole partition function to be given by $\rho_{0,2}$, as claimed.

We will now check the modular properties of the seed partition function and translate them to the microcanonical partition function by using modular kernels. First, evaluate the $P$ and $\bar P$ integrals in Eq.~\eqref{eq:CJ1} to get
\begin{equation}\label{eq:whmicrocanonical2}
    Z_{T\times I} = \frac{\mathsf{C}_{\text{RMT}}}{2\pi^2} Z_0(\tau_1)Z_0(\tau_2) \frac{\mathrm{Im}(\tau_1)\mathrm{Im}(\tau_2)}{|\tau_1+\tau_2|^2}
\end{equation}
where $Z_0(\tau)= \big(\sqrt{\mathrm{Im}(\tau)}|\eta(\tau)|^2\,\big)^{-1}$ is the non-compact free boson partition function, which is modular invariant. Now take an element in the modular group $\gamma\in \mathrm{PSL}(2,\mathbb{Z})$ of the form \eqref{eq:gamma},
and use the fact that $\mathrm{Im}(\gamma\tau) = \mathrm{Im}(\tau)/|s\tau+r|^2$ to prove
\begin{equation}\label{eq:property}
    \frac{\mathrm{Im}(\gamma\tau_1)\mathrm{Im}(\tau_2)}{|\gamma\tau_1+\tau_2|^2} = \frac{\mathrm{Im}(\tau_1)\mathrm{Im}(\tilde\gamma\tau_2)}{|\tau_1+\tilde\gamma\tau_2|^2}.
\end{equation}
Here we defined 
\begin{equation}
    \tilde\gamma \coloneqq M\gamma^{-1}M = \begin{pmatrix}
        r & b \\ s & a
    \end{pmatrix}, \quad M = \begin{pmatrix}
        -1 & 0 \\ 0 & 1
    \end{pmatrix}.
\end{equation}
Notice that $M$ acts on $\tau$ as an orientation reversal. Also note that in the special case $\gamma = S$, we have $\tilde S = S$.

Since the $Z_0$ factors are separately modular invariant, Eq.~\eqref{eq:property} shows that 
\begin{equation}\label{eq:TxI}
    Z_{T\times I}(\gamma\tau_1,\gamma\bar\tau_1;\tau_2,\bar\tau_2) = Z_{T\times I}(\tau_1,\bar\tau_1;\tilde\gamma\tau_2,\tilde\gamma\bar\tau_2).
\end{equation}
Geometrically, this reflects the fact that a modular transformation on one boundary, which acts on the $A$- and $B$-cycles, can be smoothly deformed to the other boundary through the bulk $T\times I$.

Translating this property to the microcanonical partition function $\rho_{0,2}$ in Eq.~\eqref{eq:whmicrocanonical}, we use the generalized modular kernels \eqref{eq:mod_ker} to transform the Virasoro characters, finding
\begin{multline}
    \int_0^\infty \!\dd P_1 \dd \bar P_1 \,\mathbb{U}(\gamma)_{P'_1P_1} \mathbb{U}(\gamma)_{\bar P'_1\bar P_1}^*\rho_{0,2}(P_1,\bar P_1,P_2',\bar P_2')=\\  \int_0^\infty \!\dd P_2 \dd \bar P_2\,\rho_{0,2}(P_1',\bar P_1',P_2,\bar P_2)\,\mathbb{U}(\gamma)_{P_2P_2'} \mathbb{U}(\gamma)_{\bar P_2\bar P_2'}^*.
\end{multline}
Here we used that $\tilde \gamma$ is represented by the transposed modular kernel  $\mathbb{U}(\tilde\gamma) = \mathbb{U}(\gamma)^{\mathsf{T}}$ (swap $r$ and $a$ in Eq.~\eqref{eq:mod_ker}). While the above equality looks complicated, it can be elegantly restated as the commutation relation in Eq.~\eqref{eq:commute}.

Finally, let us discuss the modular completion of the $T\times I$ wormhole. For the canonical partition function, the modular completion is a relative Poincar\'e sum: 
\begin{multline}
    Z_{T\times I}^{\text{m.c.}}(\tau_1,\bar\tau_1;\tau_2,\bar\tau_2) =\\ \sum_{\gamma \in \mathrm{PSL}(2,\mathbb{Z})} Z_{T\times I}(\gamma\tau_1,\gamma\bar\tau_1,\tau_2,\bar\tau_2)
\end{multline}
which is modular invariant in both $\tau_1$ and $\tau_2$ thanks to the property \eqref{eq:TxI}. The superscript ``m.c.''~stands for modular completion. We can again translate this to a modular completion of the microcanonical partition function: just take Eq.~\eqref{eq:whmicrocanonical} and add the superscript ``m.c.''~on $Z_{T\times I}$ and $\rho_{0,2}$.
Using the generalized modular kernels \eqref{eq:mod_ker}, we then see that the modular completion of the microcanonical partition function also takes the form of a relative Poincar\'e sum:
\begin{widetext}
    \begin{equation}\label{eq:rhomc}
    \rho_{0,2}^{\text{m.c.}}(P_1,\bar P_1,P_2,\bar P_2) = \sum_{\gamma \in \mathrm{PSL}(2,\mathbb{Z})} \int_0^\infty \dd P_1'\dd\bar P_1' \, \mathbb{U}(\gamma)_{P_1P_1'} \mathbb{U}(\gamma)_{\bar P_1\bar P_1'}^* \,\rho_{0,2}(P_1',\bar P_1',P_2,\bar P_2)\,.
\end{equation}
\end{widetext}
The spectral correlator $\rho^{\text{m.c.}}_{0,2}$ is just the modular completion of the RMT spectral correlator, which was first studied in Refs.~\cite{Cotler:2020ugk,Haehl:2023tkr, Haehl:2023xys, Haehl:2023mhf,DiUbaldo:2023qli}. The Poincar\'e sum can be regularized using standard Eisenstein series techniques. The representation of the spectral correlator in terms of modular kernels \eqref{eq:rhomc} is new. 

We can make the result in Eq.~\eqref{eq:rhomc} conceptually clearer by rewriting it in terms of the gluing map $G_\text{T}$ and the unitary operators $\mathbb{U}_{r,s}$ introduced in Section \ref{sec:modcompl}: 
\begin{multline}\label{eq:rhocompl}
    \rho_{0,2}^{\text{m.c.}}(P_1,\bar P_1,P_2,\bar P_2) = \\\sum_{(r,s)=1} \,\mel{P_1,\bar P_1}{\mathbb{U}_{r,s}\circ G_\text{T}}{P_2,\bar P_2}\,
\end{multline}
where the sum is over all pairs of co-primes $r$ and $s$.

Notice that while $\rho_{0,2}$ is diagonal in spin (there is a $\delta(J_1-J_2)$ in Eq.~\eqref{eq:rho02}), the modular completion $\rho_{0,2}^{\text{m.c}}$ is no longer spin-diagonal. This is a manifestation of the fact that modular invariance always mixes spin sectors.

As a special case, consider the sub-sum over elements of the form $(r,s)=(1,n)$. These correspond to $\mathbb{U}_{1,n} = \mathbb{S}\mathbb{T}^{-n}\mathbb{S}$. If we repeat the derivation of Section \ref{sec:whpartitionfunction} with the more general gluing specified by $(1,n)$, we have to evaluate $\mathbb{S}\,G_\text{T} \,\mathbb{U}_{1,n} \,\mathbb{S} = G_\text{T}\,\mathbb{S}\,\mathbb{U}_{1,n}\, \mathbb{S} = G_\text{T} \mathbb{T}^{-n}$. Summing over $n$ then leads to the quantization of spin, as claimed in Eq.~\eqref{eq:partialsum}.

\bibliography{apssamp}

\end{document}